\documentclass[10pt]{article}

\pdfoutput=1

 \usepackage{graphics}
\usepackage{graphicx}

\usepackage{amssymb, amsmath}

\usepackage[colorlinks=true, pdfstartview=FitV, linkcolor=red, citecolor=blue, urlcolor=blue]{hyperref}

\title{The {RAGE} radiation-hydrodynamic code}

\author{Michael Gittings\footnote{Science Applications International Corp., MS A-1, 10260 Campus Point Drive, San Diego, CA, 92121}\,,  Robert Weaver\footnote{Los Alamos National Laboratory, MS T087,  PO Box 1663, Los Alamos, NM 87545}\,,  Michael Clover$^*$, Thomas Betlach$^*$, \\
 Nelson Byrne$^*$, Robert Coker$^{\dagger}$, Edward Dendy$^{\dagger}$, Robert Hueckstaedt$^{\dagger}$,\\
  Kim New$^{\dagger}$, W Rob Oakes$^{\dagger}$, Dale Ranta$^*$ and Ryan Stefan\footnote{TaylorMade-adidas Golf,  5545 Fermi Court, Carlsbad, CA 92008-7324}
  }

\begin{document}

\maketitle
\begin{abstract}
We describe  RAGE, the ``Radiation Adaptive Grid Eulerian''  radiation-hydrodynamics code, including  its data structures, its parallelization strategy and performance, its hydrodynamic algorithm(s), its (gray) radiation diffusion algorithm, and some of the  considerable amount of verification and validation efforts. The hydrodynamics is a basic Godunov solver, to which we have made significant improvements to increase the advection algorithm's robustness and to converge stiffnesses in the equation of state.  Similarly, the radiation transport is a basic gray diffusion, but our treatment of the radiation-material coupling, wherein we converge nonlinearities in a novel  manner to allow larger timesteps and more robust behavior, can be applied to any multi-group transport algorithm. 
\end{abstract}








\newcommand{\Fig}{Fig.}     
\newcommand{\Figs}{Figs.}
\newcommand{\Eq}{Eq.}
\newcommand{\Eqs}{Eqs.}
\newcommand{\bfvec}[1]{\mbox{\bf #1}}

\subsection*{Introduction}

RAGE, an acronym for Radiation Adaptive Grid Eulerian, 
is a multidimensional (1D, 2D, 3D), multi-material, massively parallel, Eulerian  hydrodynamics code for use in solving the Euler equations coupled with a radiation diffusion equation (gray or multigroup) in a variety of high-deformation flow problems.  It uses computational cells of unit aspect ratio (squares in Cartesian 1D, 2D, 3D; tori in cylindrical 1D and 2D, and shells in spherical 1D).  The mesh is refined (or unrefined) locally every computational cycle by an Adaptive Mesh Refinement (AMR) algorithm that looks at local spatial variation of physical quantities (pressure, density, material boundaries, {\it etc.}) and then bisects or recombines cells so as to maintain a desired degree of resolution.  RAGE is intended for general applications without tuning of algorithms or parameters, and is portable enough to run on a wide variety of platforms, from desktop PC's (Windows, Linux, and Macintosh)  to the latest MMP and SMP supercomputers.

After a brief historical review, the next section will discuss methods of generating meshes from various geometry descriptions and will introduce certain data-structure concepts, while section~\ref{secn:datastruct} will discuss our parallel-centric data structures in greater depth, because they allow us to scale to kilo-pe's without interfering with the physics.  Section~\ref{secn:amr} will discuss the adaption logic briefly, because it uses a very conservative ({\it i.e.} mesh-profligate) algorithm, whose only merit is that all material boundaries and sharp gradients will only be found at the finest level of the mesh.  

Because our Eulerian hydrodynamic algorithm has more of a Lagrangian flavor than most Eulerian codes, section~\ref{secn:hydro} will  describe the relation between the Riemann solution and the advection in detail.  Our equation of state lookup techniques will be briefly discussed in section~\ref{secn:mmeos}, and the logic that is required to modify the Riemann solver to handle self-gravity is covered in section~\ref{secn:gravity}.  

RAGE's novel numerical treatment of the energy coupling between matter and radiation is covered in section~\ref{secn:raddif}.  The exponential differencing of the heating term allows a smooth transition to equilibrium diffusion.  The code uses an iterative solution to the nonlinear terms in this part of the equations, allowing larger timesteps and less computer time.

We conclude with a brief discussion of some of the  work done at Los Alamos National Laboratory to verify  the correctness of the algorithms and validate the appropriateness of those algorithms for a broad range of problems of interest to that community.

A turbulence model and a multifluid hydrodynamic treatment are under development at LANL, and work has also begun on improved AMR methods.  We intend to discuss these and other work ({\it e.g.} multigroup radiation, ionization physics) in future articles.

\subsection*{Historical Background}

The SAGE (SAIC Adaptive Grid Eulerian) code is a multidimensional (1D, 2D, 3D), multimaterial, massively parallel, Eulerian  hydrodynamics code for use in solving the Euler equation (with optional thermal conduction) in a variety of high-deformation flow problems,  begun by M. Gittings in 1990 to support DNA/DTRA work~\cite{MLG92, MLG94}. 
 
The RAGE  code is built on top of  SAGE by adding various radiation packages (gray diffusion and multigroup diffusion packages were written by T. Betlach and N. Byrne as an Internal Research and Development project in 1990-1991~\cite{RNB92}) and is used for problems in which radiative transport of energy is important. 

RAGE  is the result of a long prior evolution from earlier incarnations as a modeling code for  ground-effects experiments at the Nevada Test Site and for underwater blast-effects.  It was brought to Los Alamos National Laboratory in 1995 in order to develop an AMR Eulerian code suitable for problems of interest to that community.   The original code was written for the supercomputers of the early 1990's: Cray vector processors.  We found that these machines were not fast enough to complete 3D simulations (even rudimentary ones) in any reasonable amount of calendar time; based on 2D runs that were done at this time, we estimated that 3D runs would take more than 60 years to complete on the CRAY architecture.

With the advent of the Department of Energy's ASCI (Advanced Simulation and Computing Initiative) program\footnote{now known as the ASC (Advanced Scientific Computing) program.} 
in 1996, LANL and SAIC began a joint effort to convert the original CRAY vector Fortran-77 code to a new paradigm: parallel and modular code via effective software quality engineering.  We chose to employ ``vanilla'' Fortran-90/95 with C-language interfaces for I/O and the explicit use of message passing to effect communication among the processors.  Domain decomposition was used to spread the computational cells of the problem among the processors.

The massively-parallel version of  the codes was designed to run extremely large problems in a reasonable amount of calendar time. Our target is scalable performance to  10,000 processors on a 1 billion  cell problem that requires hundreds of variables per cell, multiple physics packages ({\it e.g.} radiation and hydrodynamics), and implicit matrix solves for each cycle.  Currently, the largest simulations we do are three-dimensional, using over 600 million computation cells and running for literally months of calendar time using  2000 processors.  
 
 SAGE itself provided an early testbed for development of massive parallelism as part of the ASCI program and has been shown to scale well to thousands of processors~\cite{WGC99}.

\section{SETUP and the RAGE Computational Grid}
\label{secn:setup}

The RAGE computational grid is set up by specifying a coarse uniform mesh in the appropriate number of dimensions, $D$,  and then refining that mesh according to various criteria.  Each processor is required to have a multiple of $2^D$ zones at setup, which allows us to  treat our base grid as if it too were composed of adapted cells ({\it i.e.} no special run-time coding is required for this coarsest level of cells).  For efficiency, that multiple should be on the order of thousands, or {\it circa} 50k zones/pe.

\begin{figure}
  \begin{minipage}{0.8\linewidth}
    \includegraphics[width=2.5in,height=2.5in,angle=0]{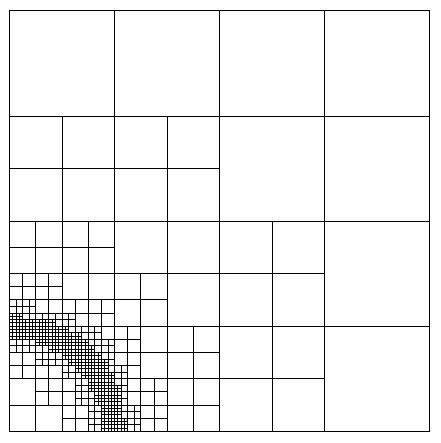}
     \caption{RAGE adaptive mesh refinement is limited to a 2:1 ratio -- adjacent cells may not differ by more than one level.}
 \label{fig:amr_grid}
\end{minipage}
\end{figure}

The RAGE computational grid is an octree decomposition of the model space ({\it i.e.} adaption bisects each zone in each dimension). The tree depth is selected to resolve field gradients within a material to one user-specified resolution and material boundaries to a second user-specified resolution.  Because RAGE allows a mixture of materials in any cell, it is not necessary to subdivide the cells to the size of the smallest geometric features -- if a zone is not mixed on cycle zero, it soon will be.   We refer to the depth of the subdivision required to attain the requested resolution as the ``level'' of the grid. The 2:1 edge length ratio of mother to daughter cells is enforced as the maximum for any adjacent cells, whether or not they have a common ancestor.

The only geometric query that RAGE grid generation requires\footnote{One other query is required to return the material properties -- density, temperature, {\it etc.} -- once the grid is built.} 
is a ``point containment'' test, which determines if a specified region contains a query point.  The user assigns a priority value for each geometric region to be gridded (normally, just the order in which regions are defined).   When the grid generation algorithm queries to determine which geometric region contains a query point, each region is tested in order from highest to lowest (but one) priority. The first region for which the point containment test returns {\em inside} is returned as the containing region. This process assures that all assembly gaps and voids within the model are assigned a material, since region\#1 covers the entire mesh.

The problem domain is subdivided into as many subdomains as the number of processors assigned to the problem. The grid for each subdomain is then generated independently, except for two adjustments that require occasional  interprocessor communication: 1) load balancing, and 2) ensuring the edge length ratio never exceeds  2:1.

With the mesh structure fixed,  we approximate the material properties  within the finest cells (``leaves'') of the octree by using multiple queries in the spirit of a Monte Carlo integral over the distributions of density and temperature (or pressure and temperature, or density and internal energy --  the three alternate ways to specify the initial state of a region).

\subsection{Grid Generation via Geometric Queries}
Since RAGE only requires a point containment test, a wide variety of geometric modeling techniques are feasible.  Geometric support  is currently divided among SAGE primitives, an integrated geometry query library, {\bf Spica}, and a Lagrangian link file query library, {\bf Merak}.  

\subsubsection{SAGE Primitives}
The intrinsic geometry specification support  includes spherical, ellipsoidal, conical frustum, triangular prism and quadrilateral prism regions, as well as perturbed boundary regions (sine and cosine perturbations), and radially swept 2D regions (used, {\it e.g.}, for overlaying a 1-d spherical dump into a 2-d cylindrical mesh).
Use of the intrinsic geometric modeling involves describing a physical problem within the input file and then verifying through RAGE execution that the desired model has been achieved (various graphics options  allow the user to examine the ``cycle zero''  mesh and properties before starting the first physics cycle).  These geometric primitives combined with judicious application of the RAGE region-priority ordering scheme permit a surprising number of complex simulations.

\subsubsection{The Spica Library}
The {\bf Spica} library allows RAGE to generate grids from {\bf OSO} geometric ÔregionsÕ,  {\bf OSO} being a Los Alamos supported geometric modeling program~\cite{F99}. Geometry created in (or imported into)  {\bf OSO} and used for initial grid generation, includes CSG combinations of triangular facetted closed b-rep surfaces (which in turn are imported into {\bf OSO} as stereo lithography or  ``STL''  files), solid representations, and implicit surfaces. {\bf OSO} supports importing data from several geometric modeling and commercial CAD packages.
 
{\bf Spica} currently accepts STL files as input, simply because many packages write that format; many other equivalent formats can easily be translated to STL.  The solid geometry description nodes contain a variety of primitive objects: box, cone, cylinder, sphere, parallelepiped, torus, and a surface of revolution generated by a piecewise linear curve.  These primitives are composed into regions using {\bf OSO}.  While it can be challenging to create extremely detailed models using these primitives, there is a considerable query speed advantage to doing so.  {\bf OSO} also supports STL files as first class primitives that can be translated, rotated, scaled, and combined with the previously specified primitives to compose regions. In practice, combinations of all the geometry description node types are used in computational models.

Separating the region selection from the geometry description is useful because it moves the exact definition of {\em inside} or {\em outside} to the tree, rather than leaving it up to the geometry generation package, with the advantage that all these CAD/CAM descriptions can be used in RAGE at the minimal cost of the query routine.

\begin{figure}[h!]
  \begin{minipage}[l]{.55\linewidth}         	\includegraphics[width=2.9in,height=1.8in]{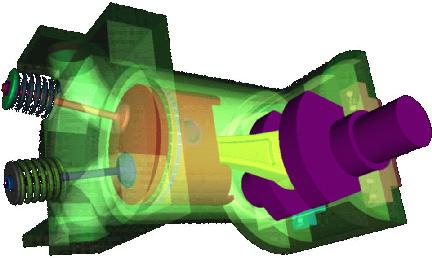}  
   \caption{OSO rendering of an internal combustion engine model created in Pro/E.} \label{fig:setup_1}
  \end{minipage}
  \begin{minipage}[m]{.40\linewidth}
  \end{minipage}
   \begin{minipage}[r]{.45\linewidth}   
  	\includegraphics[width=2.3in,height=1.8in]{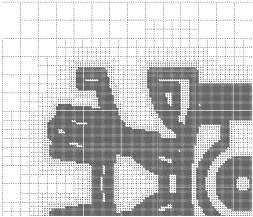} 
  \caption{Close-up of a cross section of a RAGE grid generated from the Pro/E model.}  \label{fig:setup_2}
  \end{minipage}
\end{figure}

Figure~\ref{fig:setup_1} shows  an internal combustion engine model that was created in Pro/ENGINEER,  output as separate STL files for each region of the assembly,  and collected for viewing, scaling, positioning, and rendering by {\bf OSO}.    The {\bf OSO} dataset was then processed via the {\bf Spica} routines while reading the RAGE input file. 

Figure~\ref{fig:setup_2} presents a close-up of a cross section of a RAGE grid generated from the Pro/ENGINEER-{\bf OSO} model. The boundaries of the cylinder block, head, valve springs, and retainers are modeled at level-5. Note the gradual transition from level-1 to level-5 caused by the enforcement of 2:1 edge length ratios between adjacent cells. This grid has been restricted to five levels for purposes of illustration; typically, a well resolved computational grid will descend to level-8. 
 
 One and two-dimensional meshes can be set up from a cross-section of a 3-d model, and 3-d models can be created by rotating cylindrically symmetric 2-d models.



\section{Data Structures and Parallel Implementation}
\label{secn:datastruct}


In a parallel environment, RAGE is a MIMD (Multiple Instruction Multiple Data) code with synchronization at points in the instruction streams where inter-processor communication occurs. Such communication is handled at the lowest level by a machine-specific implementation of the MPI (Message Passing Interface) library, and at the application code level by a set of RAGE routines 
 (the ``token library'') that sits atop the machine-specific library and hides most of the details of inter-processor communication from the application code. Although the code is mainly designed with a distributed-memory multiprocessor architecture in mind, it also runs on shared-memory machines and indeed on uniprocessor machines, including Mac's  
 and PC's.
 

\subsection{Computational Cells and Cell Faces}

Cells are labeled by a single index, $l$, regardless of dimension.  The highest level of refinement may change as the simulation proceeds, and in addition,  cells may be coarsened, {\it i.e.}, a previously refined block of two, four or eight cells may be recombined into a single larger cell. Spatially adjacent cells are not allowed to differ by more than one level of refinement. If changing  conditions, {\it e.g.}, the passage of a shock front, require local refinement by more than one level, the refinement process is made to propagate outward to satisfy this requirement. Also, the refinement algorithm ``looks ahead'' so that it can, for example, anticipate the arrival of a moving shock front;  this tends to put the most intense action into a locally uniform subgrid at the locally highest level of refinement.

At problem setup, blocks of two, four or eight level-1 cells are assigned index values in a linear sequence along a compact (Hilbert)  space-filling curve~\cite{PB94},\cite{PB96} that covers the computational domain, or (by default) in Fortran column-major order (we were surprised to find that on some architectures, this {\it naive} partitioning scaled better than the fancier space-filling curve). 
 Cells are then allocated to processors by means of a simple partitioning of this curve. This produces compact spatial processor-subdomains and at the same time a linear global ordering of the cells. It also approximately minimizes the processor surface-to-volume ratio ({\it i.e.}, the number of cells that share a face with a cell that resides on another processor divided by the number of cells that do not---roughly equivalent to the communication-to-computation ratio). Subsequent cell  refinement, either at problem setup or later during the simulation, proceeds via the creation of new blocks of  daughter cells. The newly created cells are inserted into the existing index sequence following the last cell in the block containing the appropriate mother cell; this preserves a degree of spatial locality in the cell ordering. It also allows us to enforce the rule that a cell and its siblings (cells that share the same mother cell) are always assigned sequential cell indices and that they always reside on the same processor -- during the cycle the daughters are created; there is no guarantee they stay together on subsequent cycles after further adaption. The price that is paid for this locality-preserving indexing is that the cell index-set must be rebuilt at the end of every computational cycle, as part of the cell refinement process. Aside from the specific cases that have just been enumerated, there is no {\em necessary} connection between a cell's index and its spatial location in the mesh, so that a relatively large amount of indirect addressing is required.

{\em Active} or ``leaf'' cells are cells at the finest local level of refinement,
{\it i.e.}, cells with no daughters. The physics algorithms usually
 involve only the active cells, although the self-gravity package's Poisson solver does use inactive cell information to construct its multipole elements (see section~\ref{secn:gravity}).  In addition to the cell arrays, the code constructs face arrays to be used in the implementation of the finite-volume physics algorithms; for example, diffusion coefficients are evaluated on cell faces. Faces are called {\em active} if they connect two active cells. If a cell face lies on a processor boundary, the face is assigned only to one of the two processors, in order to avoid double counting in code that loops over faces. Most cell arrays for physical quantities persist between computational cycles, except for indexing changes due to mesh refinement and load balancing operations, and are written out to restart dump files; face arrays, in contrast, are rebuilt anew at the beginning of each cycle.
 
Load balancing is not done every cycle, but only when the distribution of cells among processors has gotten out of balance by more than some preset threshold.  The load balancing operation occurs via a simple repartition of the global cell index set, followed by the necessary movement of cell data among processors.

\subsection{Communication}\label{Communication}

Communication in RAGE involves movement of data between adjacent cells, between cells and cell-faces, and between cells at different levels (mothers to daughters and daughters to mothers). In a multiprocessor environment, this will involve movement of data between processors. RAGE provides a small set of routines that execute these communication operations in such a way that the details of the inter-processor communication are hidden, as much as possible, from the application code.

Much of the inter-processor communication at the application code level employs the RAGE ``clone" construct. This is a kind of ghost-cell mechanism, where clones are local (on-processor) copies of cells that reside on other processors and share a common face with a cell on the local processor.

The following example shows how the communication process looks from the perspective of the application code. In this example, we compute the divergence of a heat conduction flux, $F=-D\nabla T$, integrated over each active cell, using Gauss's theorem to convert the volume integral to a sum
over faces of the normal component of the flux at each face. In the following pseudocode fragment, $N$ ({\tt Numcell}) is the number of active cells that reside on the local processor, $N_{+}$ ({\tt Numcell\_clone}) is equal to $N$ plus the number of cells cloned from adjacent processors, {\tt T}$\left( 1:N_{+}\right) $ is the temperature array on the local processor, and {\tt divF}$\left( 1:N_{+}\right) $ is an array to hold the computed result, i.e., the integrated divergence of the flux on the local processor. The arrays {\tt T} and {\tt divF} are dimensioned to have $N_{+}$ elements: elements $\left( 1:N\right) $ for local cells (those that reside on the local processor), and elements $\left( N+1:N_{+}\right) $ for the clones (corresponding to cells on adjacent processors). The example contains three typical steps, namely, a gather step ({\tt clone\_get}),  a compute step (the loop over cell-faces), and  a scatter step ({\tt clone\_put}):

\begin{verbatim}
call clone_get(T)            ! Get off-processor temperature data
                             ! i.e. T(Numcell+1:Numcell_clone)
divF(1:Numcell_clone) = 0.0  ! Initialize all elements of divF

do (for all active on-processor faces)   ! in all directions

  ! llo = local index of cell or clone on the lo-side of face
  ! lhi = local index of cell or clone on the hi-side of face
\end{verbatim}
\mbox{\tt \ \  divF(llo) = divF(llo) + D\_face}$\left(
\frac{\mbox{\tt  T(lhi) - T(llo)} }{ \mbox{ {\tt (} $\Delta${\tt x(llo)} + $\Delta${\tt x(lhi)} {\tt )/2} }  } \right)${\tt   A\_face}  \\ 
\mbox{\tt \ \   divF(lhi) = divF(lhi) - D\_face}$\left( 
\frac{\mbox{\tt  T(lhi) - T(llo)} }{ \mbox{ {\tt (} $\Delta${\tt x(llo)} + $\Delta${\tt x(lhi)} {\tt )/2} }  } \right)${\tt   A\_face} 
\begin{verbatim}
enddo
call clone_put("add",divF)  ! Scatter data to adjacent processors
                            ! cumulate into on-processor elements
divF(1:Numcell) = divF(1:Numcell)/Volume(1:Numcell)
\end{verbatim}

Here {\tt llo} and {\tt lhi} refer, respectively, to the cell or clone on the low-side and to the cell or clone on the high-side of the current cell-face in the loop over faces. The map between the active \ local cell faces and the adjacent active cell or clone indices {\tt llo} and {\tt lhi} is built anew at the beginning of each computational cycle, together with the associated machinery used by the two communication routines {\tt clone\_get} and 
{\tt clone\_put}. $D_{face}$ is the diffusion coefficient evaluated at the current cell face via an appropriate interpolation using data from the adjacent cells ({\it e.g.}, a harmonic or geometric average) and  $A_{face}$ is the area of the face.
$\triangle x_{lo}$ and $\triangle x_{hi}$ are the sizes of the cells referred to by {\tt llo} and {\tt lhi}.
Once the off-processor contributions have been cumulated into the on-processor data via the {\tt clone\_put}, they are divided by the volume of the cell in order to ensure that {\tt divF} has the correct dimension to be the divergence of a flux ($\nabla\cdot\bfvec{F}$).
The calculation of the gradient, by using half the sum of the widths of the cells ($\Delta x_{llo}+\Delta x_{lhi}$) allows this expression to be used without modification at the edges of a periodic boundary, whereas using the difference of the positions ($x_{lhi}-x_{llo}$) of the two zones would generate nonsense in that case; otherwise, the two expressions would be equivalent.

 The clone data structure is built by the token library routines at the beginning of the computational cycle and includes a list of the off-processor cells adjacent to the boundary of the local processor, a list of the on-processor cells adjacent to the local processor boundary, and buffers to hold corresponding data to be sent to and received from other processors.

\subsection{Parallel I/O}

When it is necessary to write (or read) restart dump files, RAGE can use either the MPI-IO or specialized bulk I/O libraries at the lowest level.\footnote{The original parallel I/O package was written by Howard Pritchard of SGI for the ASCI Blue-Mountain machine; a more portable version was written by Pat Fay of Intel to also run on the ASCI-Red machine, and it was ported to the ASCI-Q machine by Lori Pritchett of HP.}  At the code developer's level, a subroutine is invoked to write either a scaler or vector which may be the same or different on different processors.  Multidimensional mesh-arrays, {\it e.g.} {\tt cell\_momentum(1:numcell, 1:numdim)}, are written one mesh-array at a time with an outer loop(s) over the non-mesh index(es):
\begin{verbatim}
call pio_io('same', 'write', 0, 'do_hydro', do_hydro, pio_err)
do n = 1,numdim
  call pio_io('diff', 'read', n, 'vel', vel(:,n),numcell,pio_err)
enddo 
\end{verbatim}
The PIO package has both serial and parallel I/O capabilities.  Serial I/O is the default, although a flag allows one to switch to MPI I/O or another parallel I/O package called {\it bulkio} which is based on asynchronous MPI calls that move data to a set of I/O processors.   Both packages can be tuned for optimal performance using input file parameters, {\it e.g.} striping factors, configuration lists, {\it etc}.  The serial I/O package uses MPI sends and receives to move the data from each processor to the I/O processor, overlapping the writing of one processor's data with the receiving of another's.

 When parallel I/O is activated, the lengths of the array on the individual processors are gathered and cumulated so that each processor knows at what location in the dump file\footnote{We choose to write a single disk image, for the convenience of the users, as well as for the flexibility to write with one number of processors and read with a different number.} 
  to begin processing its contribution to the total. All of this data management and the calls to specific I/O library routines occur at a low level that the (physics) code developer never sees. In the case that the data is the `same' or when serial I/O is selected, only the I/O processor (usually pe=0)  does the actual work. Regardless of the type of I/O, a return flag ({\it e.g.} {\tt pio\_err}, above)  indicates whether or not the data has been successfully read into the array or written onto the disk.

Restart dump files are formatted to be valid Scientific Data Sets in the HDF conventional terminology; everything but character strings are written as IEEE 64 bit Reals and characters are written as ASCII character strings. On input, RAGE swaps the bytes on the fly when it detects that the ``endian-ness'' of the machine is different from the big or little ``endian-ness'' of the file it is reading, as well as converting Cray 64 bit Reals to IEEE 64 bit Reals when such dumps are encountered.

\subsection{Parallel Scaling Performance}

\begin{figure}[h!]
  \begin{minipage}[c]{1.0\linewidth}         	
  \includegraphics[width=6.5in,height=3.8in]{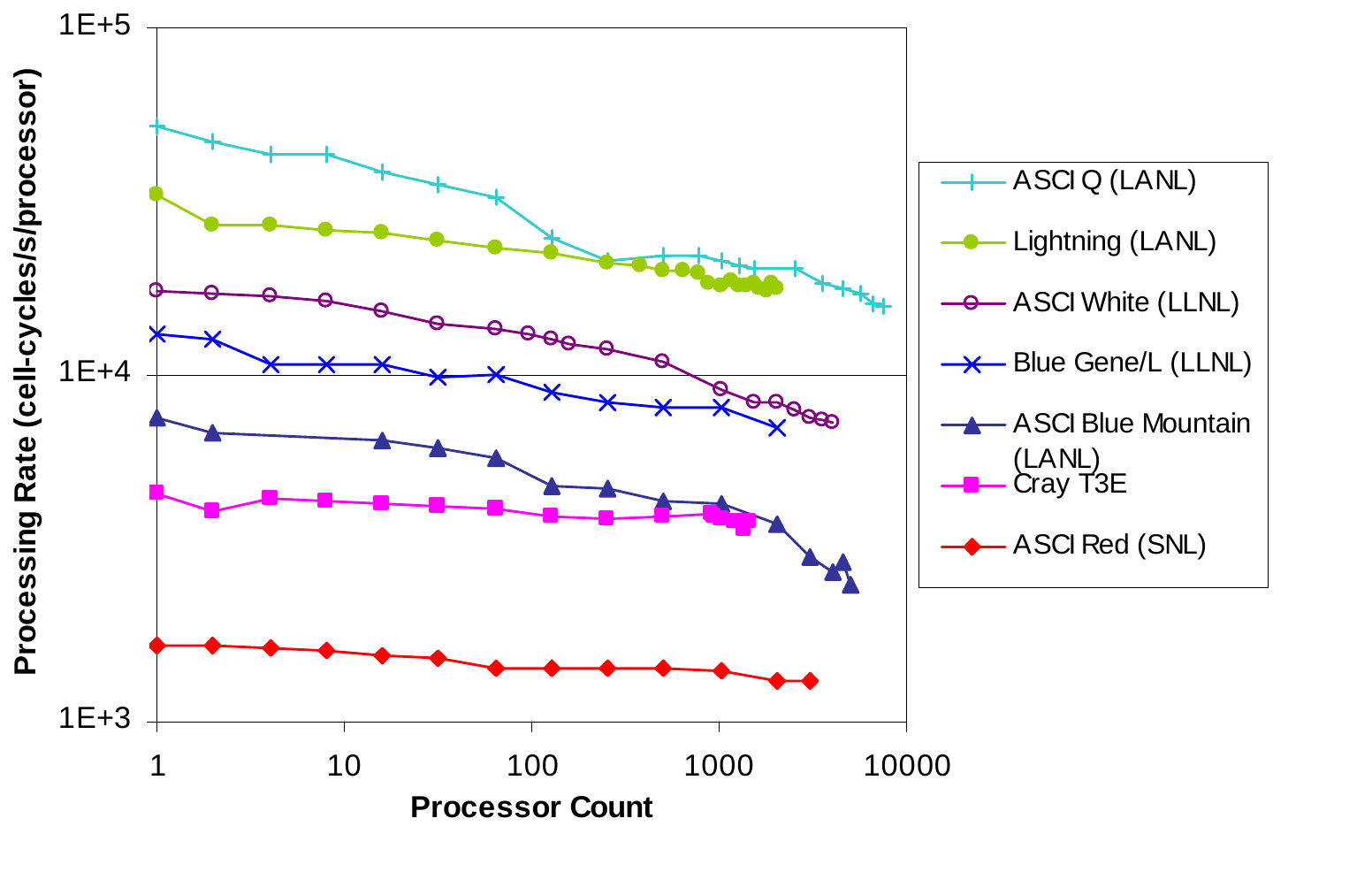}  
   \caption{The scaling behavior of SAGE on a variety of systems in weak-scaling mode (constant work per processor). } \label{fig:scaling}
  \end{minipage}
\end{figure}

In running simple timing problems on various architectures (e.g. hydro-only or  conduction-only on a fixed level mesh\footnote{The scaling problems are 3-d, consisting of three nested blocks of $\gamma=7/5$ ideal gases at different temperatures and pressures ($\rho=0.1,10,0.1$ g/cm$^3$; $T=10^3,0.1,0.1$ eV; $p\sim 10^{13},10^{11}, 10^{9}$ erg/cm$^3$, side length $L_{x,y,z}=30,15,6$ cm respectively), using the same number of fixed zones on each processor.  The same logic is executed in all zones whether the advected quantities will be zero or non-zero.  The problem can be run in hydro only mode (as shown in Figure~\ref{fig:scaling}, or in a pure conduction mode, not shown but giving similar scaling results -- each face calculates  a conductivity, and adds elements to a row and column of the matrix for subsequent inversion).  Because of indirect addressing, the work per zone is the same whether zones are adapted or not, and it would be extremely difficult to force the same amount of adaption on every processor for a scaling study.}
), we have found relatively good scaling with processor count, in the sense of constant work (fixed zones) per processor, as shown in figure~\ref{fig:scaling}.
The metric used for performance is the processing rate in terms of the number of cell-cycles that are processed per second on each processor, and is the inverse of Ôgrind-timeÕ that is often referred to.
Almost a decade of different systems are shown in figure~\ref{fig:scaling}, the oldest being the ASCI Red system installed at Sandia National Laboratory in 1996 and the newest, the Blue Gene/L system, installed at Lawrence Livermore National Laboratory in 2005.
The performance over this time has increased by approximately a factor of 20 resulting from both improvements in processor speeds, and network speeds. 

Ideal scaling performance is a straight horizontal line at the level of the performance of a single processor when a constant amount of work is mapped to each processor; very good scaling performance of SAGE is observed on these large-scale systems.
The performance on almost all of these systems degrades only by about a factor of 2 at 1,000Õs of processors.
The degradation results from the need for communication between logically neighboring processors as well as the need for collective operations. Note that the performance was measured on the largest sized systems installed except for the Blue Gene/L which will be upgraded from the 2K processors indicated in figure 4 to 128K processors during 2005

 
Further details on the performance of SAGE on the ASCI White (IBM), ASCI Blue Mountain (SGI), and T3E (Cray) systems can be found in~\cite{Kerby01}. Details on ASCI Q (Compaq) performance as well as how SAGE was actually used in the optimization of the system performance is described in~\cite{Petrini03}. The performance of Lightning (Linux) has been described in~\cite{Davis04}, and early details on the performance of Blue Gene/L (IBM) is given in~\cite{Davis04a}.




\section{Adaptive Mesh Refinement Overview}
\label{secn:amr}

The adaptive mesh refinement (AMR) method used in RAGE was initially developed to solve multidimensional, time dependent shock hydrodynamics problems. The particular problem of interest at the time required high resolution at a shock front that could be kilometers away from other resolved features in the mesh. In choosing a method, it was desirable to have rectangular (square) cells where integration methods with well understood convergence properties could be used.

 A primary consideration was to ensure global conservation (of mass, momentum and energy);  because the  arbitrarily-orientated patch-based methods such as those of Berger and Oliger~\cite{BO84} make conservation difficult, we chose to have the underlying coarse cell boundaries coincide exactly with the finer mesh boundaries~\cite{BC89}, which greatly simplifies the process of maintaining global conservation. 
 
  The ``patch-based'' AMR of Berger {\it et al.} is not, however, used in RAGE. Our ``cell-based'' AMR method   is most similar to that of Young {\it et al.}~\cite{YMB91} -- it is a block-structured mesh with cell refinement limited to bisection in each dimension where adjacent cells differ by at most one refinement level.

RAGE uses a Continuous Adaptive Mesh Refinement (CAMR) algorithm; it is continuous in space and time in the sense that mesh refinement
occurs on a cell-by-cell and  cycle-by-cycle basis. The mesh is evaluated for refinement on every cycle throughout the entire domain. 
The overhead associated with this CAMR technique has been empirically determined to be about 20\% of the total runtime, which is small compared to the gains of using CAMR instead of a uniform mesh.

The CAMR algorithm has two phases. The first or setup phase performs refinement at the start of a problem and the second or dynamic phase performs refinement during each timestep. It is convenient to separate the algorithm in this way to allow for different refinement criteria in each phase. 

In the setup phase the base grid is refined by checking whether a cell in the grid is composed of a pure material or contains an interface.   It is frequently desirable to allow  material interfaces to be refined to a different maximum level than in the bulk interior of the material ({\it e.g.} during the passage of a shock).  In the case of RAGE, material interfaces are determined either by changes in density or changes in material volume fraction as determined by the user. Additionally, if a cell and its neighbor cell differ significantly in density, pressure, or velocity component then refinement is performed on that cell. The user can also specify levels of refinement at predefined points or regions.

Once the grid has been generated and initialized, the governing equations are integrated one timestep and the entire mesh is evaluated for refinement before any data is moved. This evaluation is a two-part process and occurs on every cycle for every cell, not just the top level cells. The reason for this will be discussed shortly. The first part of the process determines which cells are to be flagged for refinement, coarsening, or neither. The second part of the process handles the addition of new cells, the removal of others, and their associated linkage to neighboring cells. In RAGE, activity tests are performed to refine the mesh ahead of shock waves. These activity tests use small changes in velocity and pressure to trigger refinement. Cells are also refined based on first-order estimates of truncation error in a Taylor series expansion of cell properties. Because error estimates for several cell properties are considered, the maximum error is saved for a cell. This method favors higher refinement in the interest of better accuracy.

When all cells have been evaluated, the second, data-movement,  phase takes place.
Cells are added and removed according to how they are flagged in the first phase.
However, there are two rules that can override that phase: (i) no cell is allowed to be more than one level different from its neighbor (including diagonal neighbors),
and (ii) cells will only be removed if it is unlikely that in the following cycle that cell would be refined.
The last condition is accomplished by basing cell coarsening  on the error estimate in the  mother cell as well as the daughters;
thus the need to update mother cells as mentioned above.
The mother cells are used because they would become the top level cells on the next cycle if the cell was coarsened.
Algorithmically it is easier to update all cells rather than restrict the update operation to top-level cells and their mother cells.
The number of cells unnecessarily updated (``grand-mothers'') is a small fraction of the total. 

When cells are added to the mesh, they are added in groups of known quantity, therefore only a reference to the first daughter cell needs to be stored, since the remaining cells follow in memory.
Inter-processor communications are minimized by inserting the newly created daughter cells immediately after the mother cell in the global cell list.  Load leveling is only done on those cycles where an individual processor's cell count exceeds the average cell count by some tolerance (approximately 10 percent).




\newcommand{\dt}{\Delta t}
\newcommand{\sign}{\hbox{sign}}
\newcommand{\half}{\frac{1}{2}}
\newcommand{\eref}[1]{(\ref{#1})}

\section{Hydro}
\label{secn:hydro}

The hydrodynamic package in RAGE  uses a two-shock or single-intermediate-state approximate Riemann solver to calculate the impulse and work on each face of a cell, using an alternating-direction-explicit (ADX)~\cite{Str68}  technique.  The Riemann solution is used to determine the trajectory in the donor zone along which a Lagrangian mass particle would move through the unperturbed state (and possibly the ``contact'' state), in order to just reach the face at the end of the timestep. This trajectory distance is then used to integrate all quantities that will be advected from the donor zone, and allows us to finesse some details in the typical Eulerian Riemann-solver methods.

For each direction, we test on overall ``volume'' changes within zones, re-evaluating the sound speeds where appropriate and iterating the Riemann solver. This semi-implicitization allows us to handle collapsing bubbles or strongly shocked zones even though the solver itself uses only a ``weak shock'' approximate Riemann solver.  This goes beyond strong-shock or general Riemann solvers to allow a coupling from  logically separated zones through their common neighbor, preventing unphysical results from occurring due to the independent updates.

 We will demonstrate that this method is $2^{nd}$-order accurate in space and time on smooth problems and $1^{st}$-order accurate  on discontinuous or shock problems.  In general, our philosophy has been to develop a method that has a demonstrable convergence rate under mesh refinement, and to then work at reducing the error at a given level of refinement.


\subsection{Conservation Equations}
The hydrodynamical equations solved in the hydro package are:
\begin{eqnarray}
 \frac{\partial\rho}{\partial t} &+ \nabla\cdot  \left({\rho \bfvec{u}}\right)      &=0                                                              \ ,  \label{eq:1a} \\
 \frac{\partial(\rho\bfvec{u})}{\partial t} &+ \nabla\cdot \left( \rho \bfvec{u}\bfvec{u} + \mathbf{P}\right) &=\rho \bfvec{g}    \ , \label{eq:1b} \\
 \frac{\partial( \rho E_m)}{\partial t} &+ \nabla\cdot   \left( \rho E_m \bfvec{u} + \mathbf{P}\cdot  \bfvec{u} \right) &=\rho \bfvec{g}\cdot\bfvec{u}        \ , \label{eq:1c}
\end{eqnarray}
where  RAGE stores  extensive arrays of mass $M=\int_V \rho dV$, momentum $M\bfvec{U} = \int_V \rho \bfvec{u} dV$, and total energy $M E_m = \int_V \rho E_m dV \equiv \int_V \rho (e +\frac{1}{2} u^2)dV$, with $\bfvec{u}$ the specific momentum (velocity), and $E_m$  the specific total energy. The treatment of gravity will be deferred to section~\ref{secn:gravity}; for the  purposes of this section, we will assume $\bfvec{g}=0$.

 RAGE has a strength-of-materials package, which includes a simple failure model and deviatoric strain-dependent stresses ({\it e.g.} Steinberg-Guinan), but for this paper, we will only consider isotropic stresses, $\mathbf{P} = p\mathbf{I}$.
 RAGE also has an operator-split heat conduction package that solves $ \frac{\partial( \rho E_m)}{\partial t}+ \nabla\cdot(-\kappa \nabla T_m) = 0$, but this is infrequently used compared to the radiation diffusion package discussed in section~\ref{secn:raddif}.

The conservation equations can be converted into a Lagrangian form which will be used to calculate $ \bfvec{U}$ and ${P}$ at the half-timestep, for input into the Harten-Lax-van Leer (HLL)~\cite{HLL83}  approximate Riemann solver:
\begin{eqnarray}
 0  = & \frac{d \rho}{dt}       &+ \rho \nabla\cdot\bfvec{U}                                 \ , \label{eq:2a} \\
 0  = & \frac{d\bfvec{U}}{dt} & + \frac{1}{\rho}  \nabla\cdot P                         \ , \label{eq:2b} \\
 0  = & \frac{dE_m}{dt}            &+\frac{1}{\rho}  \nabla\cdot (P \bfvec{U})       
  \ \ \ \ \mbox{or} \ \ \ \ 0 = \frac{de}{dt} + \frac{P}{\rho} \nabla\cdot\bfvec{U}       \ . \label{eq:2c}
\end{eqnarray}
Using  \Eqs~\eref{eq:2a} and \eref{eq:2c} and the thermodynamic identity, $c_s^2 = \left.\frac{dp}{d\rho}\right|_e + \left.\frac{p}{\rho^2}\frac{dp}{de}\right|_{\rho}$,  we can also write:
\begin{eqnarray}
 0  = & \frac{d P}{dt}       &+ \rho c_s^2 \nabla\cdot\bfvec{U}                         \ . \label{eq:2d}
\end{eqnarray}
Aside from  $E_m$, and unless otherwise noted, in this section we will generally use the convention of upper case letters for Lagrangian-frame variables, and lower case for Eulerian-frame variables.  

\subsubsection{Boundary Conditions}

The hydrodynamic package in RAGE has only two real boundary conditions available to the user:  reflecting and outflow; it has a ``treatment'' to allow inflow.

The reflecting boundary condition (BC) is self-explanatory, and was the original BC in the code.  The justification was  that with AMR technology, one could use very large zones to put the boundaries far away from the region of interest, and adapt that region appropriately.  All (normal) derivatives are set to zero in zones adjacent to a vacuum face, and all (normal) components of velocity are set to zero on such faces.

Since then, we have implemented outflow boundary conditions, individually specifiable on the high and low faces of the mesh in each direction.  In this case, zonal values of derivatives normal to the face, instead of being the limited average of the gradient on the far face and zero at the vacuum face ({\it i.e.} therefore zero), are just the value calculated on the face of the zone not adjacent to vacuum.  The velocity of fluid on this face is calculated by the Riemann solver and is not forced to zero, as with reflecting BC's.

In order to meet the demand for some type of in-flow boundary condition, we have developed a crude ``freeze boundary'' condition.  In this case, the user specifies a geometric extent of the problem space in which all properties are to be frozen.  At the top of the timestep, the list of zones falling in this space are recorded, along with the values of all quantities subject to advection (in those zones);  the hydro then operates as usual, and at the end of the timestep, those zones are re-set to their beginning of timestep values.  This has been adequate to handle the class of problems that have a constant inflow, and can be extended to allow simple temporal ramps (albeit only first order accurate in time).

\subsection{Limiters and Linear Reconstructions}\label{secn:limiters}

Although users have the choice of a number of different limiter methods  with which to calculate gradients, the two most commonly used are the {\em minmod} and the modified {\em van Leer}. \   Slopes, $s_{\pm}$, are calculated within a zone by  first linearly interpolating (between neighboring zones) a ``face-value'' of the quantity on the  faces  normal to a given direction on the mesh.  These face-values are then broadcast into cell-centered ``hi'' and ``lo'' arrays,  at which point one can calculate the slope of the quantity between between the high side  and the centroid of mass ($s_{+}$) or between the centroid and the low low side ($s_{-}$).   

The minmod limiter generates the most conservative slopes for reconstruction,
\begin{eqnarray}
s_{\mathrm {minmod}} & \equiv & \mbox{minmod}{(s_+, s_-)} \ \nonumber \\
                         & = & \half[ (\sign(s_-) + \sign(s_+)]\cdot  \min{( |s_-| , |s_+| )} \ , \nonumber
\end{eqnarray}
allowing monotonic discontinuities at a face between zones without introducing any subzonal saw-tooth patterns in the reconstructed field.  This ensures that an approximate Riemann solver will satisfy the correct entropy conditions.   

The van Leer slope and limiter~\cite{LEER4} can be defined in terms of a generalized minmod:
\begin{eqnarray}
s_{\mathrm{vanLeer}} & \equiv & \mbox{minmod}{( \frac{s_+ + s_-}{ 2},\mathcal{F} s_- ,\mathcal{F} s_+)} \ \nonumber \\
 & = &  \half\left(  \sign(s_-) + \sign(s_+) \right) \cdot \left|\sign(s_-) + \sign(\frac{s_+ + s_-}{ 2})\right| \nonumber \\
         & &        \cdot \min{(\left|\frac{s_+ + s_-}{ 2}\right|, |\mathcal{F}s_-| , |\mathcal{F}s_+| )} \ , \nonumber
\end{eqnarray}
where the average, $\frac{s_+ + s_-}{2}$, is the $2^{nd}$-order estimate of the slope between the three zones and van Leer's definition of this limiter has $\mathcal{F}\equiv 2$ ($\mathcal{F}=1$ reduces this to the ordinary {\em minmod}). Our modified version uses $\mathcal{F}=\frac{3}{2}$, which van Leer~\cite{LEER4} has noted gives about the limiting power of his harmonic form, $s_{harm} = \frac{ s_{+}s_{-}}{s_{+}+s_{-}}\, [ (\sign(s_-) + \sign(s_+)]$. 
We have found that the  use of the standard $\mathcal{F}=2$ in this limiter introduces a small amount of ringing behind some shock fronts in problems we have run, while using $\mathcal{F}=\frac{3}{2}$ does not produce any such visible anomalies.
We have also found that the problems associated with stronger-than-minmod limiters ({\it i.e.} the saw-teeth in the reconstruction) is accentuated in two-dimensional problems, even if it is imperceptible in one dimension, and these problems are alleviated by use of  $\mathcal{F}=\frac{3}{2}$.

Our goal is to have a spatially and temporally second order accurate hydrodynamic method, physics permitting.
  For such temporal accuracy, we use a half-step, whole-step method to advance the states for the Riemann solver.  Second order spatial accuracy can be regarded as that accuracy that will exactly preserve linear solutions across the mesh, which means that a minmod limiter will converge to the same order as a van Leer limiter, although in general the van Leer offers  improvements in accuracy at a given level of refinement.

Linear profiles are constructed in each zone with monotonically limited slopes with the zonal average located at the center of mass of the zone.  For example,
\begin{eqnarray}
\rho(r) &=& \overline{\rho} + {(r - r_{cm})}  \nabla\rho\ , \nonumber \\
\Rightarrow & &\int_V \rho(r) dV =   \overline{\rho} V  = M \  , \nonumber \\
 & &\int_V  r  \rho(r) dV =  M \, r_{cm} \ \  \ . \nonumber 
\end{eqnarray}
Kinetic energy is calculated with a 3-point Gaussian quadrature, so that the worst case, $5^{th}$-order polynomial (in 1d spherical geometry), $\int r^2 \rho(r) u^2(r) dr$, is integrated exactly, allowing us to always calculate an ``exact'' internal energy, 
\begin{eqnarray}
e = \left( M \, E_m-\half\int r^2 \rho(r) u^2(r) dr \right)/M \ . \nonumber
\end{eqnarray}
The velocity, $u(r)$, is determined by a two-step method.  Beginning with the specific momentum, $\overline{u}^0 = MU/M$, we calculate its gradient, $\nabla u^0$, and integrate an estimate of the momentum, 
\begin{equation}
\int_V \rho(r) u(r) dV = M\overline{u}^0 + \int_V (r - r_{cm})^2\nabla\rho \nabla u^0 dV \ . \nonumber
\end{equation}
We then recalculate the velocity to minimize the error between this estimate and the known value of momentum,
\begin{equation}
\overline{u} = [M \, U - \int (r - r_{cm})^2\nabla\rho \nabla u^0 dV]/M \ . \nonumber
\end{equation}
We have not found it necessary to converge the velocity  any further, but we do calculate $\nabla \overline{u}$ for use later.

\subsection{Riemann Solver}

\begin{figure}
  \begin{minipage}[t]{.45\linewidth}   
  	\includegraphics[width=2.3in,height=2.0in,angle=0]{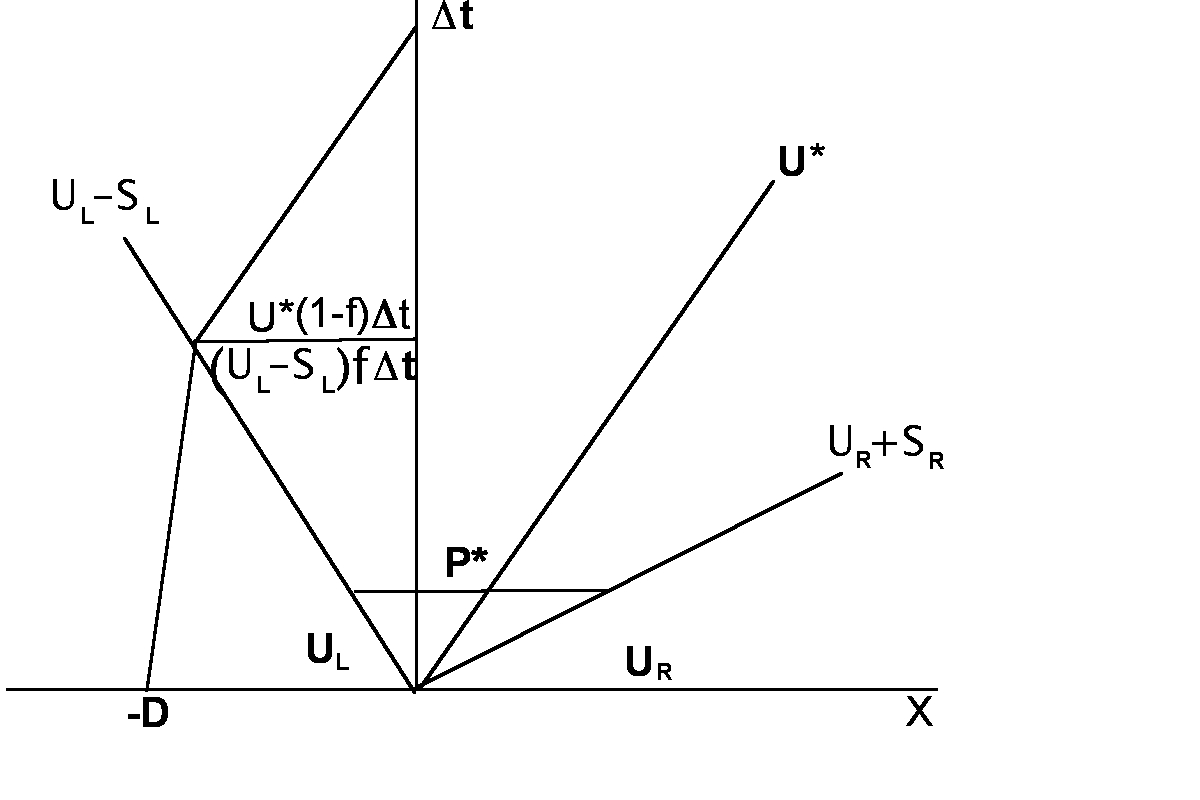} 
        \caption{One characteristic  lies in the Left half plane.  Fluxes  depend on  the Left (donor) and  Intermediate ($^*$) states.}  \label{fig:hydro_1}
 \end{minipage}
 \begin{minipage}[t]{.03\linewidth}
\hspace{.10in}
 \end{minipage}
  \begin{minipage}[t]{.45\linewidth}         	\includegraphics[width=2.3in,height=2.0in,angle=0]{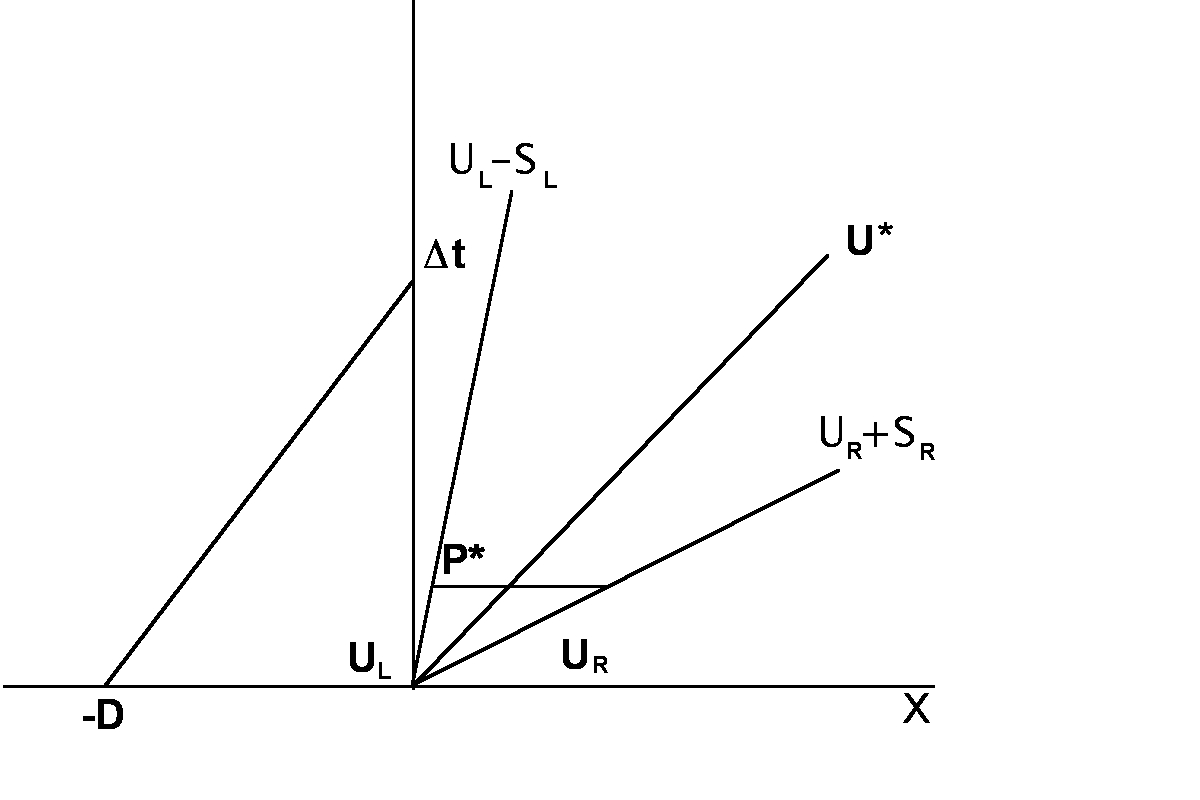}  
   \caption{All characteristics lie in the Right half-plane.   Fluxes only depend on the Left (donor) state. } \label{fig:hydro_2}
  \end{minipage}
\end{figure}

A single-intermediate-state approximate Riemann solution~\cite{HLL83} takes values  that have been reconstructed on the Left and Right sides of a cell's face  ($U_L, P_L, U_R, P_R$), and combines them into the intermediate state, ($U^*,P^*$), shown for example in \Fig~\ref{fig:hydro_1} lying between the lines $U_L - S_L$ and $U_R + S_R$.   This state is a weak ({\it i.e.} integral) solution to the momentum equation on both sides of a face (we use the ``mass'' coordinate, $dm=\rho dx=\rho c_s dt$ in the following formulae):
\begin{eqnarray}
\int_0^{\dt}  dt \int_{- (\rho c)_L t}^0 dm \left( \frac{du}{dt} + \frac{dp}{dm}\right) = 0 = 
 \int_0^{\dt}  dt \int^{ (\rho c)_R t}_0 dm \left( \frac{du}{dt} + \frac{dp}{dm}\right) \ . \nonumber
\end{eqnarray}
On the assumption that a signal propagates from the face at a ``signal speed'' of $\rho c$ into each cell, these two equations in two  unknowns can be solved for the intermediate state:
 \begin{eqnarray}
 U^* &=& \frac{(\rho c)_L U_L + (\rho c)_R U_R - (P_R - P_L)}{(\rho c)_L  + (\rho c)_R} \ , \label{eq:rievel} \\
 P^* &=& \frac{(\rho c)_L P_R + (\rho c)_R P_L - (\rho c)_L (\rho c)_R (U_R - U_L)}{(\rho c)_L + (\rho c)_R} \ , \label{eq:rieprs}
 \end{eqnarray}
 where the impedances, $\rho c$, are the {\em fastest} signal in each zone~\cite{HLL83}, 
 \begin{eqnarray}
(\rho c)^* &=& \sqrt{ (\rho_0 c_0)^2 +\rho_0 \max(0, P^*-P_0) \left(1 +  \half \frac{1}{\rho_0}\frac{dp}{de}\right) } \ , \label{eq:rhoc*}
\end{eqnarray}
found  by iterating \Eqs~\eref{eq:rieprs}  together with the ``L'' and ``R'' versions of \Eq~\eref{eq:rhoc*} to pressure convergence.  
  We have not, however,  found it necessary to perform such iterations, given the ``anti-cavitation'' logic described below in section~\ref{sec:anticav}, and use just the  ``weak-shock'' solution for the intermediate state, solving \Eqs~\eref{eq:rievel} and \eref{eq:rieprs} with the zone-centered, beginning-of-timestep values of $c_L$ and $c_R$.     
We do use $P^*$ in a single pass through \Eq~\eref{eq:rhoc*} before our advection logic,
because this $(\rho c)^*$ can be shown (via Rankine-Hugoniot relations) to be equal to $\rho_0 U_s$,
where $U_s$ is the speed of the shock/rarefaction in the unshocked fluid  (of density $\rho_0$).
The characteristic labeled  ``$U_L - S_L$'' in \Fig~\ref{fig:hydro_1} (and \Fig~\ref{fig:hydro_2}) moves in the lab frame at a speed of $U_L -  U_{s(L)}$ (similarly, the one labeled  ``$U_R+S_R$'' moves at $U_R + U_{s(R)}$).  

We want to ensure that advection is done as accurately as possible, and use certain Lagrangian concepts to effect that end. 
Note that a Lagrangian test particle -- in the absence of any gradients -- will start at the position ``$-D$'' in \Fig~\ref{fig:hydro_1}, and will drift at a constant velocity ($U_L $) until it hits (or is hit by) the characteristic wave that sweeps past it, jerking it to a new drift velocity, $U^*$, with which it will drift to the face of the zone at the end of the timestep. 
If we take our linearly reconstructed density profiles and integrate them between $-D$ and $0$, we will be able to advect the ``exact'' amount of mass that needs to be moved from the donor cell to the donee cell.

The  direct-Eulerian method in RAGE is a hybrid that uses a Lagrangian Riemann solution to infer the correct Eulerian fluxes, in particular, a ``manifestly obvious''  prescription for the multi-material continuity equation's mass fluxes.
Using a Riemann Solver in the solution of any set of (conservative) hyperbolic equations involves four steps~\cite{CG85}:
\begin{itemize}
\item The calculation of interpolated profiles for the dependent variables, 
\item The construction of time-centered left and right states, 
\item The solution of the Riemann problem with those states to generate the evolved state, 
\item the conservative differencing of the fluxes in the original equations.
\end{itemize}

Rather than calculate the left and right states by averaging at $t^n$ over the domain that will be causally connected to the face~\cite{CG85,CW84} during the timestep, we calculate point values on either side of the face:  $p_L + \half \Delta x_L \nabla p_L$ and  $p_R - \half \Delta x_R \nabla p_R$.  As in van Leer~\cite{LEER5}, it is understood that $\half \Delta x$ is actually the distance from the face-center to the centroid of mass.  For T-cell faces, it really is $\Delta\bfvec{x}\cdot{\nabla}p$.  These $(p, u)_{\pm}$ point values  are then advanced to the half time using the  Lagrangian expressions (\Eqs~\eref{eq:2b},\eref{eq:2d}), so that the point value Riemann solutions calculated from them will be second-order accurate in time.

RAGE goes beyond the standard Riemann solver logic to make a further attempt to include gradients of velocity and pressure in its reconstruction of the amount of mass to flux across a face. 
  Clearly, if there were no gradients, the jump in velocity across the leftward traveling characteristic in \Fig~\ref{fig:hydro_1} would always be $U^* - U_L$.  However, if there is a gradient in velocity, then our Lagrangian test particle's initial velocity will nominally be $U_L - D u_x$, where $u_x$ is the gradient of velocity at the beginning of the timestep;  it is reasonable to also assume that the final-state velocity will not be exactly $U^*$ either.  Thus, we regard the Riemann solution as giving us ``point values'' at the half-time rather than time-averaged values, and make  the further assumption that the velocity jump across the left-bound characteristic in \Fig~\ref{fig:hydro_1} is a variable, $v_j(f)$, where $f$ is the fraction of the timestep that elapses before the Lagrangian particle-path hits the characteristic.  We define $v_j$ as:
\begin{eqnarray}
v_j(f) &=& U^*(f) - U_L(f) \ ,  \nonumber \\
         &=& (U^*_0+ U_x^* S_L f \dt)  - ( U_L +  u_x S_L f \dt) \ , \label{eq:ufs} \\
         &=& U^*_0  -  U_L  + f  ( U_x^*   -   u_x)  S_L  \dt \ , \nonumber 
\end{eqnarray}
where we denote the gradient in velocity at the beginning of the timestep with $u_x= U_x$, and the gradient at the end of the timestep ({\it i.e.} the gradient of the Riemann solutions)  with $U^*_x$.  The characteristic Lagrangian speed, $S_L$, is calculated after solving our approximate Riemann problem,
\begin{eqnarray}
S_L &=& \begin{cases}
               -c_L                                                         & \text{if ${p^* \le  p_L} $} , \\
              - (\rho c)^*_L / {\rho_L}                             & \text{otherwise} ,  
              \end{cases} \ , \nonumber 
\end{eqnarray}
with $(\rho c)^*$  as in \Eq~\eref{eq:rhoc*}.  The sign of $S_R$ is positive, meaning that all relevant characteristics (those in a donor cell) move at ``$U+S$''. 

 These expressions for $U(f)$ in  $v_j(f)$ are not immediately obvious:  in principle, the shock or the tail of a rarefaction moves at a speed of $U^* + S_L$, while the head moves at $U_L+c_L$.  In order to be consistent with the single-intermediate state approximation, we assume that the distance from the contact to the tail is  $S_L f \dt$ (measured from the contact moving at $U^*$),  and we assume that the distance from the face to the head of the rarefaction is also $S_L f\dt$ (measured from the face in the unshocked fluid frame moving at $U_L$).  Thus $U^*(f\dt)$ and $U_L(f\dt)$ at \Eq~\eref{eq:ufs} and the jump across them.

If there were no gradients in the problem, then the characteristic in the donor cell will be moving at speed $U_L + S_L$, and the Lagrangian point will have been drifting from the coordinate, $-D$, with a velocity of $U_L$.  Their intersection at a given time gives a constraint,  $-D + U_L f\dt = (U_L + S_L)f\dt$.  It will be noted that the actual velocity of the fluid drops out of this constraint to give,  $-D = S_L f\dt$.  When gradients exist, the characteristic, $U_L+S_L$, ought to show some curvature, but we assume that it  is locally straight at the point of intersection, traveling at some unspecified $\dot{X}+S_L$.  We have a  second constraint  from the definition of the total distance that Lagrangian point moves to get to the face. In general we have
\begin{eqnarray}
-D + \dot{X}f \dt -  ( \dot{X} + S_L )f \dt & = & 0  \ ,                 \label{eq:nr1} \\
 -D + (U_L-Du_x) f \dt + (U_L -Du_x + v_j)(1-f)\dt &=&  0 \ ,  \label{eq:nr2}
\end{eqnarray}
which simplifies to
\begin{eqnarray}
-D  - {S_L f \dt} &=& 0 \ ,                                   \label{eq:snr1} \\
-D(1 + u_x \dt) + U_L \dt + v_j (1-f)\dt &=& 0 \ . \label{eq:snr2}
\end{eqnarray}
Solving the second equation for $D$, and plugging it into the first, we have an
equation for $f$ alone:
\begin{eqnarray}
f &=& \frac{-D}{S_L \dt}  \ \ = \ \  \frac{- (U_L + v_j (1-f))}{S_L ( 1 + u_x \dt)} \ , \nonumber \\
\Rightarrow 0 &=& U_L+ v_j(f) + f[ S_L (1+u_x\dt) - v_j(f)] \ . \label{eq:nr1s}
\end{eqnarray}
 Once this equation is solved for $f$ ({\it e.g.} by Newton-Raphson iteration),  the second equation, \eref{eq:snr2}, can be immediately solved:
\begin{eqnarray}
  D &= & \frac{[U_L+ v_j(f) (1-f)]\dt}{1+u_x\dt} \ . \label{eq:dlimit}
\end{eqnarray}
  Clearly, if we divide $D$ by $\dt$, we will have an average velocity, $\overline{u}$, that should compare to what would have resulted from a more standard Eulerian direct solution, $u^{n+1}$.  Building on that insight,  we construct the analogous value of pressure,
\begin{eqnarray}\label{eq:averages}
\overline{u} & = { D}{\dt}  =&  [ (U_L -D  u_x)  + v_j (1-f) ] \ \ , \label{eq:ubar} \\
\overline{p} & = \ \ \ \ \ \ \ \ &  [ (P_L  - D  p_x)  + p_j (1-f) ] \ \ , \label{eq:pbar}
\end{eqnarray}
where the pressure jump across the characteristic, 
\begin{equation}
p_j(f) =  P^*(f) - P_L(f) = P^*_0  -  P_L  + f  ( P_x^*   -   p_x)  S_L  \dt \ , \nonumber
\end{equation}
is the analog to $v_j(f)$, with $P_x^*$ the derivative of the Lagrangian Riemann pressures across the cell (the ``point'' values on the two faces divided by the size of the cell).

Consider now the case when $U^*>0$ and   $(U_L + S_L)>0$, so that no characteristic reaches into the donor zone  (\Fig~\ref{fig:hydro_2}).  In this (transonic/supersonic) case, ``D'' is determined trivially from the expression for the straight path in space-time,
\begin{eqnarray}
-D &+& (U_L^{n+{1/2}} - D u_{L,x})\Delta t  = 0 \ , \nonumber \\
\Rightarrow  D & = & { \frac{U_L^{n+{1/2}} \Delta t }{1 + u_{L,x} \Delta t} } \ , \label{eq:supersonic}
\end{eqnarray}
where  $u_{L,x} =\left( \frac{du}{dx}\right)^n_L$, and $U _L^{n+{1/2}}$ is the time-advanced value of the velocity on the left-side of the face. 

In order to calculate the value of $\overline{p}$ for the impulse in this supersonic case, recall  that any Lagrangian derivative can be written as an Eulerian derivative:
\begin{eqnarray}
\frac{dP}{dt} & = & \frac{\partial p}{\partial t} +  \overline{u} \frac{\partial p}{\partial x} \ . \nonumber
\end{eqnarray}
 The Eulerian value at  fixed position at the end of a timestep  is given by:
\begin{eqnarray}
p^{n+1} = p^{n} + ( P^{n+1} - P^{n} ) - \Delta t \, \overline{u} \frac{\partial p}{\partial x} \ . \nonumber
\end{eqnarray}
Given that $p_{x} = P_{x}$ to the order of accuracy we need, and since $p^n \equiv P^n$ at the beginning of the timestep, 
\begin{eqnarray}
\overline{p} \equiv p^{n+1} &=& P^{n+1}  - \Delta t \overline{u} \frac{\partial P}{\partial x} \ , \nonumber \\
       &=& P^{n+1}  - D \frac{\partial P}{\partial x} \ , \nonumber
\end{eqnarray}
where the second line follows from the definition at equation~\eref{eq:ubar}.  The average velocity, $\overline{u}$, could have been obtained by similar reasoning,
\begin{eqnarray}
\frac{dU}{dt} = \frac{\partial u}{\partial t} &+& \overline{u} \frac{\partial u}{\partial x} \ \ \ \ 
     = \ \ \  \frac{ u^{n+1}- u^{n}}{\Delta t} + u^{n+1}  \frac{\partial U}{\partial x} \ , \nonumber \\
\Rightarrow \overline{u} \equiv u^{n+1} &=&  \frac{ U^{n+1} }{1 + U_x \Delta t} \ , \nonumber
\end{eqnarray}
and agrees with that from equation~\eref{eq:supersonic}.

\subsubsection{Impulse and Work Terms}

As we have seen, the volume of fluid that is moved from the donor to donee zone is based on the integral from $-D$ to $0$ (for rightward flowing fluid).  In slab geometry, this is $A_{face} \overline{v}\Delta t$, but in cylindrical and spherical geometry, we can calculate the $\pm\left[r^{\delta} \mp (r-D)^{\delta}\right]$ ($\delta = 1,2$) exactly.  Using the average values of $\overline{p}$ and $\overline{u}$ on the face, we calculate the impulse at the face $f_i$ (face normal in the $i$ direction):
\begin{eqnarray}
I_f &=& \overline{p} A_{f_i} \Delta t \ , \nonumber \\
(mv_i)_L^{n+1} &=& (mv_i)^n_L - I_f \ , \nonumber \\
(mv_i)_R^{n+1} &=& (mv_i)^n_R + I_f \ , \nonumber 
\end{eqnarray}
and similarly, the work done on the face is given by,
\begin{eqnarray}
W_f &=& \overline{p} A_{f_i} (r^{\delta} - (r-D)^{\delta}) \ , \nonumber \\
(mE_m)_L^{n+1} &=& (mE_m)^n_L - W_f \ , \nonumber \\
(mE_m)_R^{n+1} &=& (mE_m)^n_R + W_f \ , \nonumber 
\end{eqnarray}
(the area, $A_{f_i} $, has dimensions of solid angle in spherical geometry, length in cylindrical geometry, and  area in slab geometry).  These changes to mass, momentum and energy are temporarily stored as increments in case the anti-cavitation logic requires us to repeat the Riemann step with improved sound speeds.  The state variables are updated before the next directional sweep in the alternating direction explicit (ADX) logic.

\subsubsection{Artificial Viscosity}\label{secn:artvis}

It has been noted  that Godunov solvers in multi-dimensions have to be augmented by some form of artificial viscosity~\cite{WC84} or  by a more dissipative low-order solver~\cite{Quirk94}.  RAGE too has an artificial viscosity, specifically in order to symmetrize numerical errors.

In particular, we want symmetric (comparable) errors  on problems involving shear, independent of the orientation of the shear.  Consider a 2-d box with ({\it  e.g.}, 4) alternating bands of fluid (aligned with mesh boundaries) moving to the left and right with periodic boundary conditions ({\it e.g.}, Kelvin-Helmholtz without the perturbations).  Because the velocity field  contains discontinuities, Strang splitting will no longer be second-order accurate in general; but given our direction-split solver, such bands decouple completely, giving zero numerical error in this orientation.  Now consider the flow field   rotated by 45$^{\circ}$: on each face that straddles the discontinuity, the advection logic on each face will calculate an average velocity of zero, resulting in no change to those border zones; but the intermediate state Riemann logic {\em will} find a non-background pressure (the two velocities are in opposite directions and limiters turn on) on the face, and that will flux some negative momentum into a positive momentum zone (or {\it vice versa}), thereby smearing the discontinuity.  This smearing is  due to the  $1^{st}$-order (slope limited) Riemann solution being Strang-split in this orientation.  

Our approach to such rotational asymmetry is not to attempt to remove the error along the diagonal, but rather to add some error along the axes (but in such a way that in smooth flows, it will have zero magnitude and thus contribute no error).  To that end,  RAGE has developed a tensor-like artificial viscosity of the form
\begin{eqnarray}
\mathcal{Q}_{f|| d} &=& \beta_h \ \max( 0 \, , \, 2 \cos^2 \theta_{|| d} - 1 )\  \frac{ (\rho c_s')_{|| L} (\rho c_s')_{|| R}}{(\rho c_s')_{|| L} + (\rho c_s')_{|| R}}\Delta v_{|| d} \ , \nonumber \\
\cos\theta_{|| d} &=&\frac{  \max( v_{|| L_d}, v_{|| R_d})}{\sum_{c=1}^3  \max( v_{|| L_c}^2, v_{|| R_c}^2 ) } \ , \nonumber \\
\rho c_s' &=& \rho \sqrt{c_s^2 + \Delta v_{|| d}^2} \ , \nonumber \\
\beta_h &=& 0.25 \ , \nonumber
\end{eqnarray}
for each direction, $d\ne f$, parallel to a face $f$.  
  Thus the $y$ and $z$ ($d$) components of velocity-jumps contribute to those components of momentum fluxed across a face normal to the $x$ ($f$) direction. (Note that in smooth flows, $\Delta v_{|| d}\rightarrow 0$ because the (unlimited) slopes will now construct continuous velocities on each side of the face, turning off the viscosity.) The energy fluxed across the face $f$ by these components is calculated  by dotting this component of the tensor  into the velocity parallel to the face, $dW_{\pm f} = A_f \Delta t \, \mathcal{Q}_{f|| d} \cdot v_{|| d}$ where 
\begin{eqnarray} v_{|| d} &=& \frac{(\rho c_s')_{|| L}v_{|| L_d} + (\rho c_s')_{|| R}v_{|| R_d}}{(\rho c_s')_{|| L} + (\rho c_s')_{|| R}} \  . \nonumber
\end{eqnarray}

\begin{figure}\label{figs}
  \begin{minipage}{1.0\linewidth}\includegraphics[width=2.5in,height=2.5in,angle=0.]{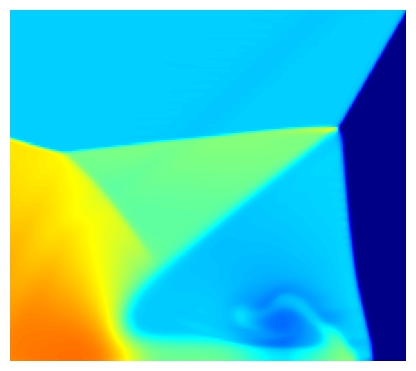}  
                                                  \includegraphics[width=2.5in,height=2.5in,angle=0.]{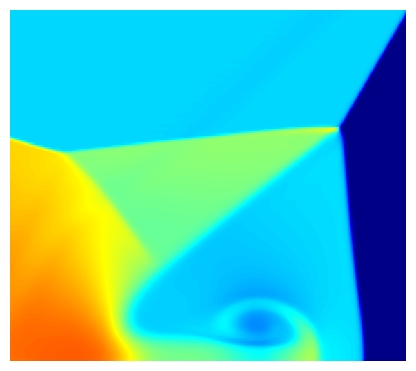}
   \caption{Double Mach reflection test problem, blown up in the vicinity of the Mach stem, showing the effect of the tensor artificial viscosity on the vortical flow ($\beta_h=0.0$, left, $\beta_h=0.25$ (default), right).} \label{fig:hydrobet}
  \end{minipage}
\end{figure}

Figure~\ref{fig:hydrobet} shows an example of the effect of such a viscosity on the vortical flow induced by a jet along along the horizontal axis, in this case for the double Mach reflection test problem\cite{WC84} (960 x 240 mesh, $\Delta x =\Delta y =  1/240$, $t=0.2$).  This same viscosity usually eliminates the ``carbuncle'' growth~\cite{Quirk94} seen at shock fronts along the axes of a multi-dimensional problem.

\subsubsection{Anti-Cavitation Iteration Logic\label{sec:anticav}}

Consider the case that a single, almost empty zone is sandwiched between two high-pressure zones -- for example, a collapsing bubble.  Iterating a standard Riemann solver will ensure that no more than half the zone will fill with material from any face, but on the next cycle -- if the fluid was relatively incompressible, like water, the density in that former bubble might become $1.01$, in which case the over-pressure would cause the now over-dense zone to explode on the next timestep, sending an unphysical shock throughout the fluid.

To prevent this, each face records the effective (1-d) change of volume,  $\overline{u}\,\Delta t \,A_{face}$, which is added and subtracted from the appropriate zone-centered volumes, and a new volume is calculated: $V_{new} = V_{old}^2 / (V_{old}+\Delta V)$.  If the ``zone'' shrinks sufficiently, updated densities and specific energies are used to invoke the equation of state for a new sound speed in those zones (and those zones only), and the Riemann solver is re-invoked (for the entire mesh), until the new volumes' changes are within some small tolerance.

Each such iteration requires a new inter-processor communication to pass the sound speed information, and the Riemann solver has to re-communicate the intermediate state values on the faces back to the zones on either side of the faces (for the $f$-finder logic).  Otherwise, this outer iteration loop costs little more than a typical Riemann iteration loop.  However, by doing the iteration here, we can avoid the over-pressures that occur from causally disconnected faces updating their common zone with too much material. 

van Leer~\cite{LEER5} found that the weak solution is more than adequate for rarefactions, and  Colella~\cite{COL82} found that only 1-2 iterations were necessary in a large class of problems he examined.  In those few cases, we can also  handle strong shocks with our beginning of timestep sound speeds in the weak-shock Riemann solver, since the few zones where a shock occurs will force one of these outer iterations with the shock-modified sound speeds in those zones.  (The biggest problem we have found occurs when a very strong rarefaction would straddle the face if we allowed it.  Then our advection/donor logic will make a large first order error, as seen in \Fig~\ref{fig:hydro_sst} on LeBlanc's shock tube problem.)

\subsection{Advection}

When our problems involve more than one material, it is not enough to just integrate the linearized fractional mass-densities in order to update the equations ({\it e.g.} $\Delta M_j = \int_0^D \rho_j r^{\delta} dr$).
If we wish to ensure that materials in equilibrium remain in equilibrium, we have to advect a certain volume of each material (based on the linearized gradient of volume fraction), multiplied by the partial mass density or partial internal energy density -- for the continuity or energy equation, respectively.
This method works because the partial energies as well as partial volumes of materials are returned by the Multi-Material EOS package (which finds the partial volume that each material has to occupy in order that the weighted partial densities sum to the total density, and the weighted fractional energies sum to the total energy of the zone; see section~\ref{secn:mmeos}, below).

 Because RAGE is a single-fluid code, all specific momenta are identical, and nothing fancy needs be done to update the momentum equation -- it suffices to integrate the bulk momentum profile.
 
 The original RAGE philosophy was to minimize the complexity of the various physics packages in the code.  In the context of advection, this meant using the same slope-limiter logic to advect fractional masses that was used in the hydro to advect momentum and energy; if sharper ``boundaries'' were desired, one should increase the level of adaption, rather than impose a ``subzonal'' model of an interface that wasn't part of the equations being solved.  In the context of an ASCI production code, not all problems run could afford to refine the mesh to the degree required -- even ASCI computers have finite memories -- and the Interface Preserver technique was developed to sharpen contact surfaces without  further adaption.  In a limited number of cases, it was found that even this is not sharp enough, and work is on-going at LANL to implement a Volume of Fluid method~\cite{RK98} that will confine mixed cells to a single interfacial zone.

\subsubsection{The Interface Preserver (IP) Method}

Having taken into consideration the existing data structures and approach to adaptive
mesh refinement in RAGE, an efficient algorithm has been chosen to preserve the steep
slope or profile of a given volume fraction. The approach makes use of the compact
stencil currently used in the code. ÒCompactÓ means that only those mesh cells directly
adjacent to the mesh cell at which a slope is being calculated are used in the
reconstruction. Early renditions of this technique were developed in the late seventies
and early eighties by  Harten~\cite{H78} and subsequently modified by Yang~\cite{Y90}. Although these methods are commonly termed slope steepeners, they were originally called artificial compression methods. The motivation was purely for contact discontinuities.  Here we have adopted the approach of Yang with modifications to capture material  interfaces. For ease and clarity, consider volume fraction $\phi$  distributed on a one dimensional Cartesian mesh. If the volume fraction $\phi_i$ is to be reconstructed using the IP  method, an initial slope $S_i$ is always evaluated using minmod limiting (see section~\ref{secn:limiters}),
 a standard option currently
available in RAGE. With the slope so determined, left and right face values can be
calculated by
 \begin{eqnarray}
\phi_i^L &=& \phi_i - S_i \half \Delta x \ , \nonumber \\
\phi_i^R &=& \phi_i + S_i \half \Delta x \ , \nonumber
 \end{eqnarray}
where $\phi_i^L$ and $\phi_i^R$ are the left and right interpolated cell face values of $\phi_i$.  To modify the initial minmod slope, we first determine the difference in the face values of $\phi$ on either side of the interfaces,
 \begin{eqnarray} 
\delta_i^L  &=& \phi_i^L - \phi_{i-1}^R \ , \nonumber \\
\delta_i^R &=& \phi_{i+1}^L - \phi_i^R \ , \nonumber
 \end{eqnarray}
and the change in slope is then evaluated as
 \begin{eqnarray}
\Delta S_i &=& 2 (33 \,  \alpha_i) \frac{\mbox{minmod} (\delta_i^L, \delta_i^R)}{\Delta x} \ . \label{eq:deltasi}
 \end{eqnarray}
The final, but not necessarily monotone slope, $S^*$, is then
 \begin{eqnarray}
S_i^* &=& S_i + \Delta S_i \ . \nonumber
 \end{eqnarray}
In \Eq~\eref{eq:deltasi}, $\alpha_i$  is a ``discontinuity'' or ``material detector'' that takes on values between 0 and 1 and is evaluated as
 \begin{eqnarray}
\alpha_i &=& \left( \frac{(\phi_{i+1}-\phi_i) - ( \phi_i - \phi_{i-1}) }{|\phi_{i+1}-\phi_i| + |\phi_i - \phi_{i-1}| } \right)^2 \ . \nonumber
 \end{eqnarray}
This ``detector'' is effective in the following manner. At a material interface there will be
a variation in volume fraction over a small number of cells giving a value of $\alpha_i$ of order 1 while away from material interfaces, such as within a pure material or where material fractions vary over a larger number of mesh cells, its value will be near or equal to zero.  The idea at this point is that this detector will not allow a compressing of materials in  regions where material has been either intentionally ÒmixedÓ or the interface is no longer resolvable by the current mesh spacing. The factor of 33 in front of the discontinuity detector is a somewhat empirical constant controlling the onset of the preserver. Notice that if $\alpha_i$ is 1.0, then $\Delta S$  is rather large (due to this factor of 33) and most certainly
makes the final slope non-monotone. This aspect requires a final adjustment to $S^*$ to
ensure that it produces no new extrema relative to the surrounding data. This operation is always performed at the end of the slope calculation. This final limiting gives our final
slope $S_i$ from $S_i^*$.  Although the presentation here is for the simplest of computational grids the actual implementation in RAGE does work with adaptive mesh refinement.   Currently the IP method is used on volume fractions for all materials, all of the time, with $\alpha_i$  being the sole ``switch'' for turning it on or off. 

\subsubsection{Implementation and Integration of IP into RAGE}
The implementation into RAGE is more complicated than the above presentation due to
the need to contend with variable cell sizes, the existence of ÒT-cellsÓ involved with the
AMR, and the use of one and two-dimensional meshes with non-Cartesian symmetry. However
relying on the coding already in place has allowed for a relatively straight forward
implementation of the method.

It turns out that for many applications outside of the area of ideal gases, the IP feeds back detrimentally on
the hydrodynamics and yields unphysical results. We have
implemented a technique where the IP is turned off when a shock is passing through a
cell,  it being preferable to drop the hydro variables to truly first order to alleviate
some of the problems. Unfortunately, it has also been found that turning off the IP even for short amounts of
time will allow enough material diffusion that the material detector in the IP will no longer see the material interface as sharp and never come back on even after the shock has passed. 

Future work will consider additional control switches. The IP is appropriate, in particular, for interfaces undergoing pure compression or expansion. However, interfaces where vorticity is large indicate an amount of physical mixing in which its use would be inappropriate. Thus an additional control switch may need to key off the level of vorticity within a given cell. Also, it has been suggested that as soon as material strength is turned off for a material that IP be turned off as well.

\subsection{Accuracy and Limitations of the Hydro Algorithm\label{secn:hydrovandv}}

RAGE is second-order accurate in space and time when used to calculate ``smooth'' problems (and first-order accurate on discontinuous or shocked problems).  We will demonstrate this by running problems on fixed uniform meshes with fixed timesteps, halving and quartering the zoning and timesteps in order to ensure known conditions (the adaption algorithm is not designed to be self-similar or scale invariant, so refined adapted meshes might not have exactly twice as many zones as coarse adapted meshes).

\begin{figure}
  \begin{minipage}[t]{.9\linewidth}     \hspace{0.9cm}
   \includegraphics[width=4.0in,height=3.0in,angle=0]{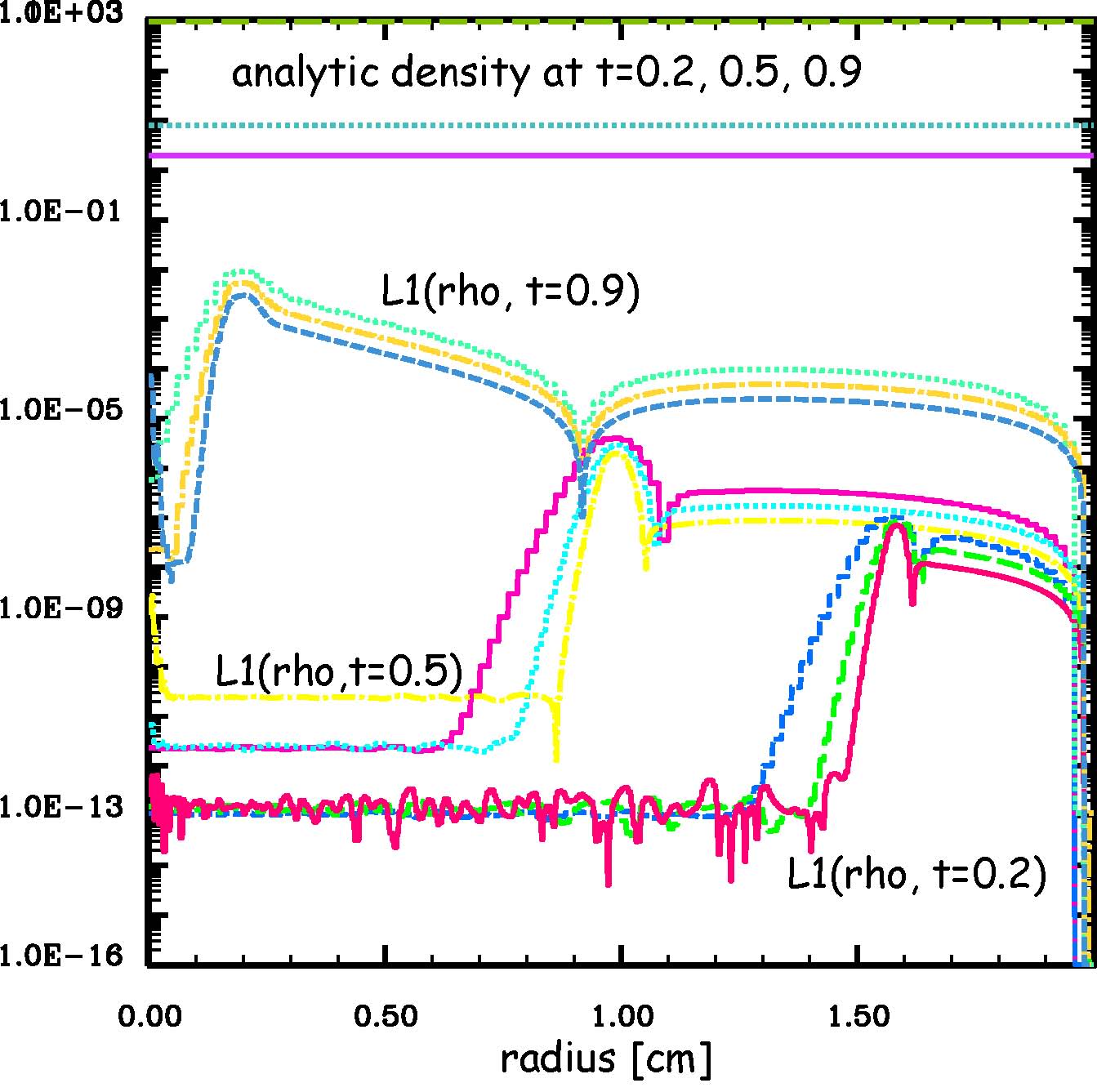}  
   \caption{ Spherical Adiabatic Compression: $L_1$ norm of density as function of position for $\Delta x ,\Delta x/2$ and $\Delta x/4$ at times 0.2, 0.5, and 0.9 (the higher resolutions are the sharper);  analytic values of density for t=0.2, 0.5, 0.9 are also shown at top of figure. A first-order error  can be seen to advect in from the pseudo-inflow boundary condition.} \label{fig:hydro_adcomp}
   \end{minipage}
\end{figure}

Godunov~\cite{GOD99} has shown that conservation of entropy can be obtained by a linear combination of the mass, momentum and energy conservation equations, which means that since RAGE conserves mass, momentum and energy, it also conserves entropy.
In order to demonstrate this, we refer to an adiabatic compression test problem (\Fig~\ref{fig:hydro_adcomp}), where $\bfvec{u}(\bfvec{r}) = -\bfvec{r}$, $(\rho_0=1, \ T_0 \approx 0$), and show that the $L_1$ error of density hovers around machine accuracy independent of mesh resolution (until a first-order error propagates in from the outer boundary's pseudo-inflow condition).
Surprisingly, the internal energy (not shown, not surprisingly) is first-order accurate under mesh refinement.

\begin{figure}
  \begin{minipage}[l]{.50\linewidth}         	
  \includegraphics[width=2.5in,height=2.5in]{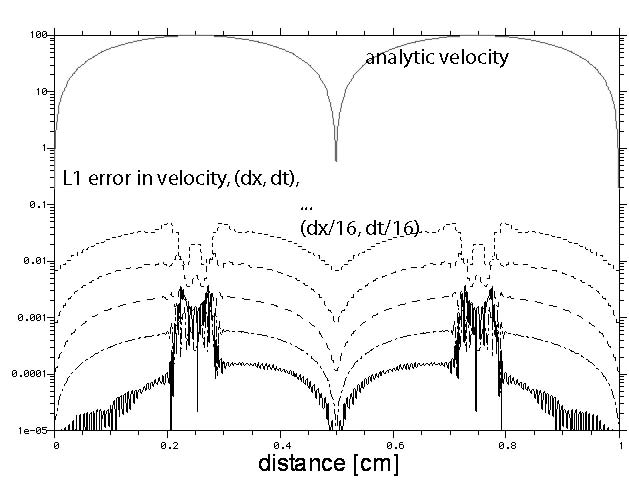}  
   \caption{ Linear Wave: $L_1$ norm of velocity as function of position at time t=0.25 (quarter period) for the perturbative wave problem. Double-humped curve at top is  $|u_{analytic}|$; nonlinear effects begin at $10^{-4}$.
Spatial convergence is $2^{nd}$-order for the 5 $\Delta x$'s shown, although finest is near machine accuracy.} \label{fig:hydro_wibx}
  \end{minipage}
\hspace{0.9cm}
   \begin{minipage}[r]{.50\linewidth}   
  	\includegraphics[width=2.9in,height=2.5in]{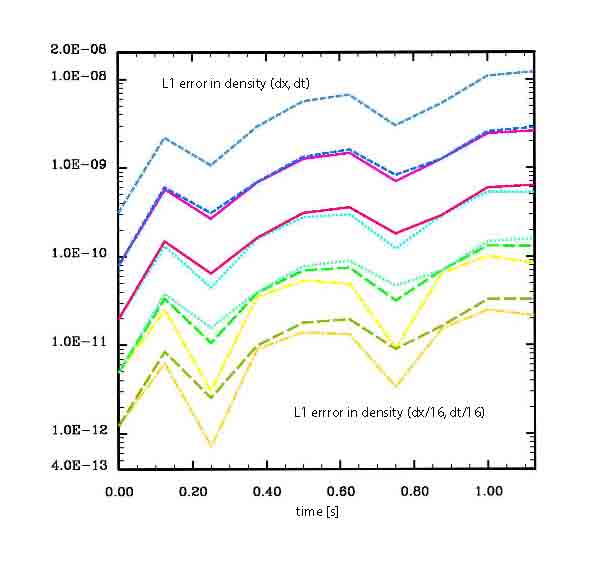} 
  \caption{Linear Wave: Mesh-integrated $L_1$ norm of density as function of time for perturbative wave problem. Spatial convergence is $2^{nd}$-order; overlying curves at fixed $\Delta x$ have $\Delta t, \Delta t/2, \Delta t/4$ timesteps. The two somewhat ``out-lying'' (yellow) curves near bottom were near the Courant limit ($dx/4,8\approx c_s dt/1,2$).}  \label{fig:hydro_wibt}
  \end{minipage}
\end{figure}

\begin{figure}
  \begin{minipage}[l]{.5\linewidth}\includegraphics[width=2.9in,height=2.5in]{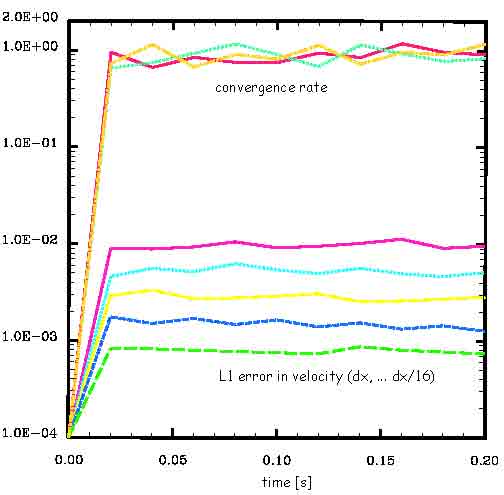}  
   \caption{Log plot of Space-integrated $L_1$ norm of velocity as function of time  for the modified Sod shock tube problem (lower 5 curves). Convergence ratio (upper 4 curves) is close to 1st order (0.75).} \label{fig:hydro_14a}
  \end{minipage}
\hspace{0.9cm}
   \begin{minipage}[r]{.5\linewidth}   
  	\includegraphics[width=2.7in,height=2.4in]{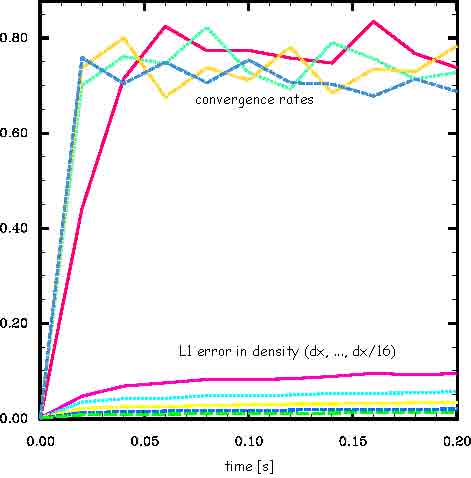} 
  \caption{Linear plot of Space-integrated $L_1$ norm of density  as function of time  for the modified Sod shock tube problem (lower 5 curves). Convergence ratio (upper 4 curves) is close to 1st order (0.75).}  \label{fig:hydro_15a}
  \end{minipage}
\end{figure}

Figures~\ref{fig:hydro_wibx} and \ref{fig:hydro_wibt} show RAGE results on another smooth problem, a ``wave in a bucket'', where a sinusoidal  $10^{-6}$ density perturbation is applied to a unit density fluid initially at rest.  In Figure~\ref{fig:hydro_wibx} the zone size and timestep were both halved for each curve; figure~\ref{fig:hydro_wibt} shows why we usually halve zone size and timestep simultaneously --  the coefficient of the $\Delta x^2$ error term is much larger than that of the $\Delta t^2$ or $\Delta x \, \Delta t$ terms -- it save us a few runs.  Not shown are similar results obtained with 2-d waves in a square bucket, which are also second-order accurate in space and time -- given our Strang splitting and the one dimensional results, this is to be expected. 
Figures~\ref{fig:hydro_14a} and \ref{fig:hydro_15a} show results on a weak (Sod) shock tube problem, and \Figs \ref{fig:hydro_sst} and \ref{fig:hydro_18a} show results on LeBlanc's strong shock tube problem, both of which are demonstrably first-order accurate as expected.

The order of accuracy is determined by reference to the $L_1$ error norms shown in \Figs~\ref{fig:hydro_adcomp}-\ref{fig:hydro_18a}.
If we assume that the theoretical value is given by the computational value plus a truncation error, $\theta = C(\Delta x) + \alpha \Delta x^r + \cdots$,
then the ratio of $L_1$ errors on two different meshes allows us to determine $r$: $|\theta - C(\Delta x)| / |\theta - C(\Delta x/2)| = 2^r$.
The upper portions of \Figs~\ref{fig:hydro_14a}, \ref{fig:hydro_15a}, and \ref{fig:hydro_18a} show the result of calculating $r$,
\begin{eqnarray}
r & =&\frac{  \ln|\theta - C(\Delta x)| - \ln  |\theta - C(\Delta x/2)|}{\ln 2} \ , \nonumber
\end{eqnarray}
 for adjacent sets of $L_1$ errors in the lower parts of the figures (the cycle-0 results have no error, and $r=0$).
 In the problems that involve shocks, we have seen that this order of accuracy, or convergence ratio,  is closer to three quarters than one, for reasons we will suggest presently.
 The same method applied to smooth shockless problems does generate convergence ratios very close to 2 ({\it e.g.} 1.95-1.98), when applied to the point-wise and mesh integrated results of \Figs~\ref{fig:hydro_wibx} and \ref{fig:hydro_wibt}.
 
 As can be seen in \Fig~\ref{fig:hydro_wibt}, the $L_1$ error is sometimes unchanged when $\Delta t$ is refined at a given $\Delta x$.  For this reason, we quote convergence ratios based on runs where both $\Delta t$ and $\Delta x$ are simultaneously halved.  Then the calculation of $r$ is insensitive to whether the leading error term is $\alpha \Delta x^r$, $\beta \Delta x^{r/2}\Delta t^{r/2}$ or $\gamma\Delta t^r$.
 
 \subsubsection{Why Shock tube convergence rates are less than Unity.}
 
The monotonicity constraints of a finite-difference scheme will smear a sharp shock over 3 or 4 zones, no matter what the spatial zone size, so that those zones will always contribute about the same discrepancy, $| X_i - C_i |\gg 0$, (X=exact, C=Code, $i\in(3-4 \mbox{ zones})$ no matter what the zone size.  If this dominates the error, then on mesh refinement, one expects that the new error will be half as much, resulting in a convergence ratio $\approx 1$.  
 
However, it does not seem to be commonly known that  the typical order of accuracy in the treatment of a contact is $\mathcal{O}(\frac{n}{n+1})$\cite{Rider70622}, where $n$ is the order of the difference scheme.  This means that a contact in a shockless (otherwise 2nd-order accurate) problem would be 2/3 order accurate, while a contact (in an otherwise 1st-order accurate) shock tube problem, would be 1/2 order accurate.  While this doesn't dominate the overall error, it does drag our shock tube convergence rates down to approximately 3/4.  Similar results are seen for MUSCL and WENO schemes involving shocks~\cite[Table 5, p. 266 and Table 19, p. 277]{GR04}.
 
\begin{figure}
  \begin{minipage}{.45\linewidth}\includegraphics[width=2.7in,height=2.9in]{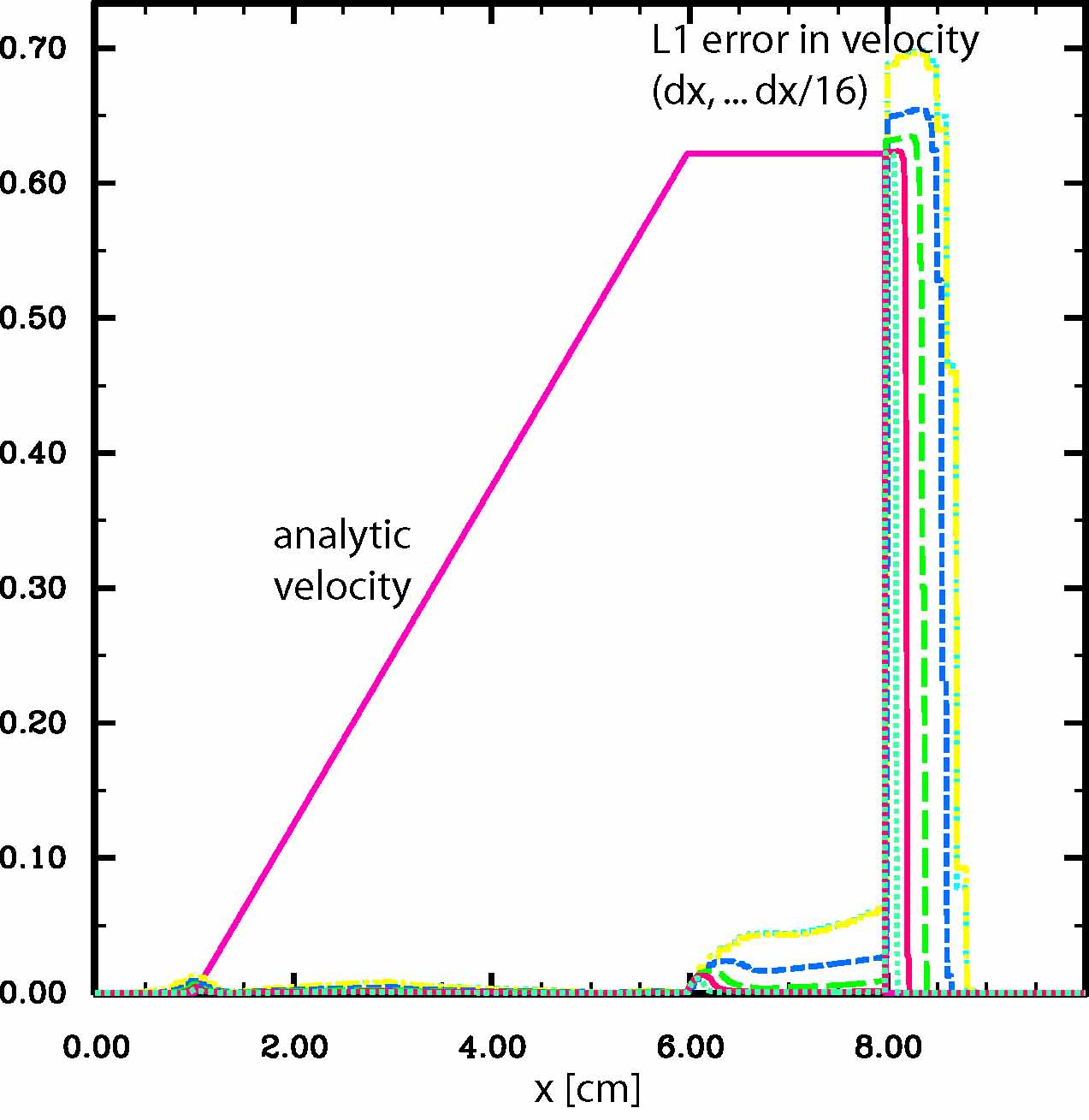}  
   \caption{Pointwise  $L_1$ norm of velocity as function of position at t=0.6 for LeBlanc's strong shock tube  problem.  Trapezoidal curve  is the analytic velocity profile.  Errors at the shock front ($x=8$) dominate this error norm.} \label{fig:hydro_sst}
  \end{minipage}
 \hspace{0.5cm}
  \begin{minipage}{.45\linewidth}\includegraphics[width=3.0in,height=3.0in]{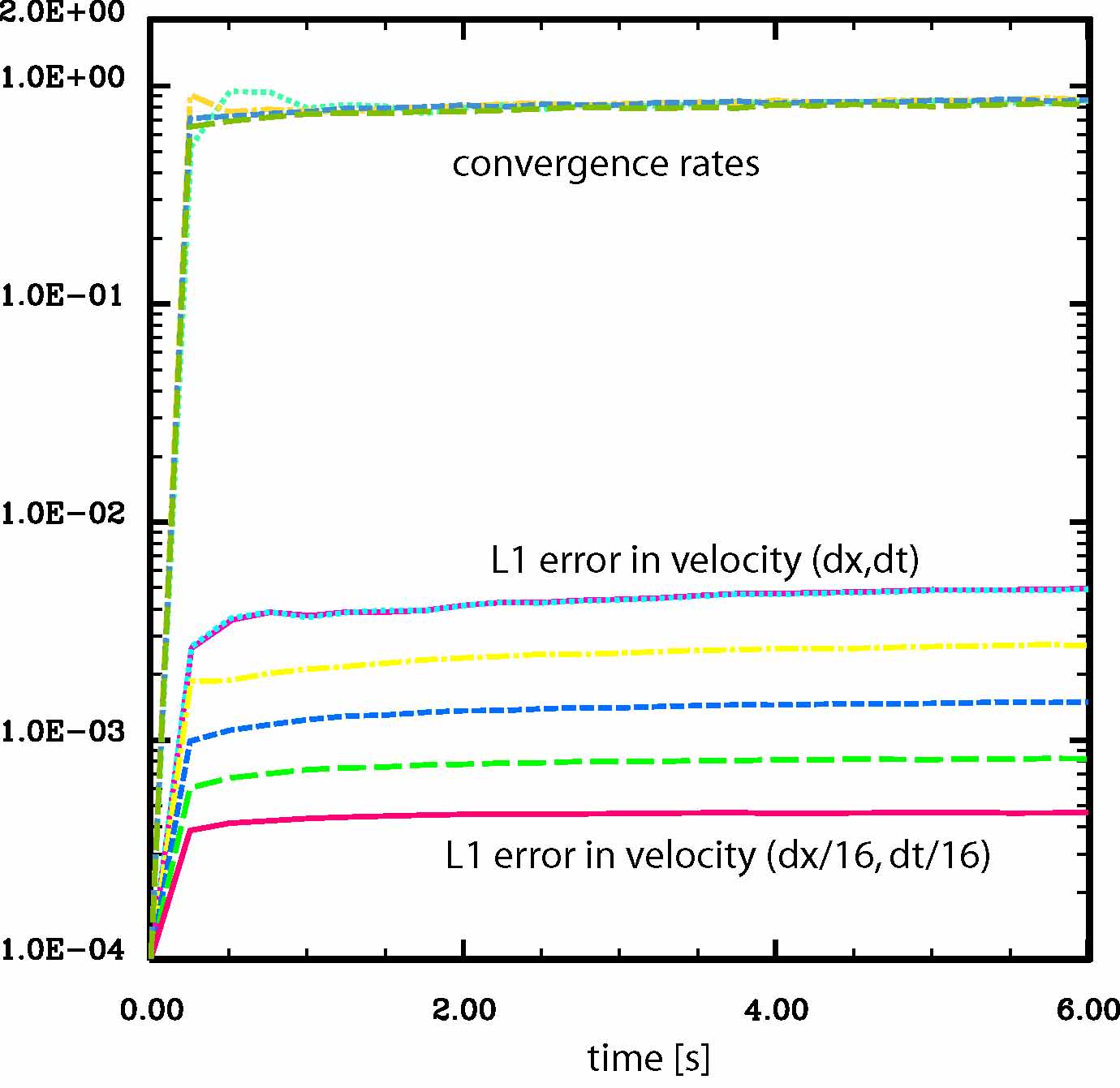}  
   \caption{ Mesh-integrated $L_1$ norm of velocity as function of  time  for LeBlanc's strong shock tube problem (5 lower curves, 4 mesh doublings).  Convergence ratio (upper 4 curves) asymptotes to $\sim 0.85$.} \label{fig:hydro_18a}
  \end{minipage}
  \begin{minipage}{.45\linewidth}\includegraphics[width=2.7in,height=2.3in]{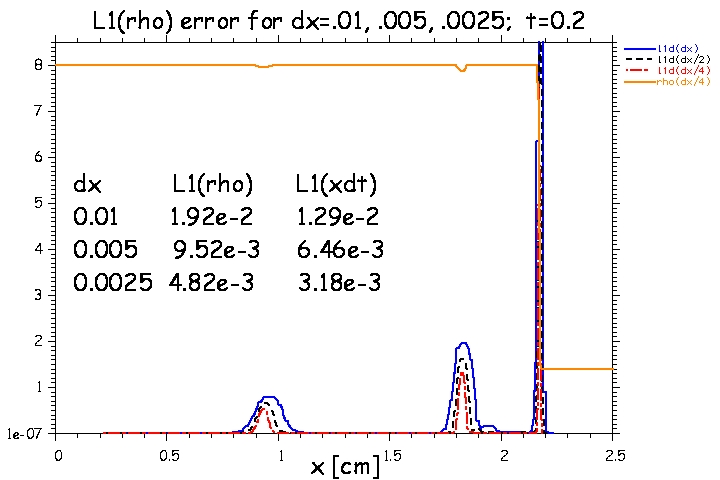}  
   \caption{Pointwise  $L_1$ error in density ($\times 10$) vs. position at t=0.2 for a Mach 10 shock driven from the left into $\rho_0=1.4, \gamma=7/5, p_0=1$ ideal gas.  Gold solid curve is calculated density, showing source errors.} \label{fig:hydro_dmr1d}
  \end{minipage}
 \hspace{0.5cm}
  \begin{minipage}{.55\linewidth}\includegraphics[width=3.6in,height=1.2in,angle=180.]{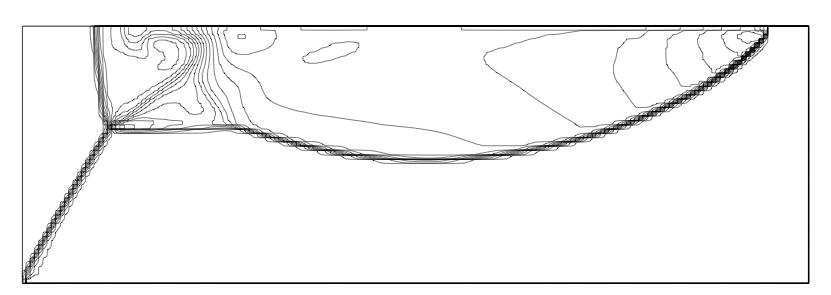}  
                                                 \includegraphics[width=3.6in,height=1.2in,angle=180.]{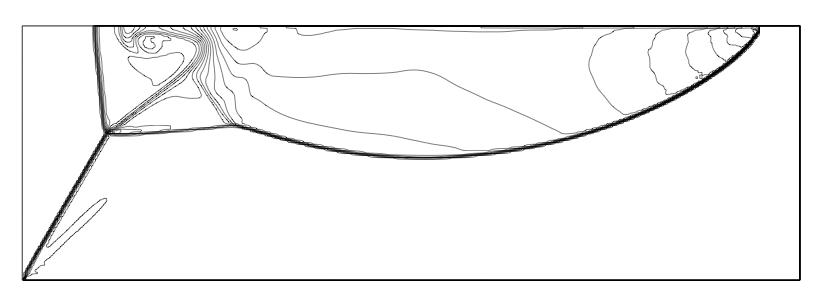}  
   \caption{30 density contours at $t=0.2$ for 2-d double Mach reflection test problem, $\Delta x = \Delta y = 1/60$ (top), $\Delta x = \Delta y = 1/120$ (bottom). {\it Cf.}, \cite[Fig. 9e, bottom 2 plots]{WC84}, \cite[Figure 9, top plot]{SW92}, \cite[Fig. 8, top 2 plots]{RGK07}.} \label{fig:hydro_dmr2d}
  \end{minipage}
\end{figure}

 We have run a further problem, a 1-d Mach 10 shock driven into resting material (based on the 2-d double Mach reflection problem of Woodward and Colella,~\cite{WC84}) to show that a shock without any contacts really is first order accurate in space and time. The table inside Figure~\ref{fig:hydro_dmr1d} shows that the $L_1$ errors in both density and velocity drop almost exactly by factors of 2 under mesh and timestep refinement; the figure itself shows the $L_1$ errors in density (multiplied by 10) compared to the highest resolution's  density profile.  The same error noted by Woodward \& Colella, caused by specifying the time-dependent boundary conditions based on an infinitely sharp shock instead of a numerically broadened one, show up here as two dips in the density behind the shock front.\footnote{One would expect to get similar if not better results with a slab variant of the Noh problem, slamming cold fluid into a wall, and the average convergence ratios for the velocity and pressure are indeed 0.96 and 0.95; the ratio for the density, however,  is only 0.80 -- presumably blamable in this case on a wall-heating effect that acts like a contact.} 

We present in Figure~\ref{fig:hydro_dmr2d} a calculation of the 2-d double Mach reflection test problem from which the 1-d version was abstracted.  Without an analytic representation of the entire density field, it is not possible to do more than compare the position of various features, all of which agree qualitatively with other publications ({\it e.g.}, our linear reconstruction algorithm's 240 x 60 and 480 x 120 mesh results should be compared to the MUSCL scheme in \cite[Figure 9e, lowest 2 plots]{WC84}, or to \cite[Figure 8 bottom plot, Figure 9, top plot]{SW92} or \cite[Figure 8, top 2 plots]{RGK07}, although the last is a PPM (piecewise parabolic) method with $4^{th}$ order steepening, and our scheme is closer to  PLM (piecewise linear method) with $2^{nd}$ order steepening). In this vu-graph norm, one can see the sharpening of contours by a factor of two as expected, although the magnitude of the error is such that, for example, Woodward and Colella's $\Delta x = 1/60$ resolution calculation looks closer to our $\Delta x = 1/120$ resolution.
 
\subsubsection{Grid Imprinting and Grid Seeding}

It may have been remarked ({\it e.g.}, see Figure~\ref{fig:amr_grid}) that the square zones of RAGE do not provide a perfect representation of circular or spherical objects -- the zoning results in a sawtooth representation of what should be a smooth curve (or inclined plane).  There are a number of different cases in which this {\em could} present a problem, and we briefly discuss them. In the case of smooth shockless flow, we have established that the Strang splitting results in a 2nd-order accurate hydro scheme;   thus, while the grid may ``seed'' some fluctuations in volume-fractions around the circumference, the overall motion will be as correct as the hydrodynamics, and Rayleigh-Taylor perturbations in 2-d slabs, for example, do grow at the correct rates.  In the case of shocked flow, there are two cases -- a round shell may be converging (as in an ICF capsule implosion), in which case  no problem arises with a shock going through such a shell and converging at the origin;  however, in the case that a shock reflects off the origin and re-shocks the shell, the shell will exhibit significant instability growth due to the originally grid-seeded perturbations and the fact that gradients of pressure and density are oppositely directed (Rayleigh-Taylor unstable). This growth can be reduced by resolving  the surface -- at some cost -- with finer zoning.  If this re-shock occurs within a small number of zones of the origin ({\it i.e.}, $< 30-50$ zones), then the shock itself will not have achieved self-similarity (roundness) after its bounce, and will hit the (constant radius) shell at different times at different angles.  This can be ameliorated by resolving the section of the mesh at the origin  as well.  In the case that an outbound shock overtakes an outbound shell, there is some Richtmyer-Meshkov instability growth, but it is not as severe a problem since the gradients of pressure and density now tend to be aligned and this configuration is unlikely to be reshocked (or reshocked very strongly) by reflected shocks from larger radii.

\vspace{.1in}



\section{Multi-Material Equations of State and Mixed Cells}
\label{secn:mmeos}

In any Eulerian code, regardless of the degree of adaption, the majority of cells will eventually become mixed with more than one material.   Given that observation, RAGE has developed a multi-material equation of state package (MMEOS) that assumes that {\em all} cells are mixed zones.

Given RAGE's  time-explicit hydrodynamic algorithm,  sound waves will cross the smallest zone in two to  three timesteps, relaxing any subzonal pressure differentials  in that time.   While it is not possible to hand-wave temperature equilibrium in as cavalier a fashion, RAGE nonetheless makes the assumption that all mixed cells are in  pressure and (material) temperature equilibrium.  This assumption has the advantage that there is a unique solution to the question of the mixed-cell temperature and pressure, and has the further advantage of being consistent with the idea that if adaption (based on pressure gradients)  has stopped in a region of zones, there is nothing further to resolve.

\subsubsection{Inverted Pressure-Temperature Tables}
 In order to quickly invert mixed-cell equations of state, a RAGE preprocessor inverts the equations of state for every pure material before beginning a calculation.
 RAGE can do this pre-processing for ($\rho,T$)-based SESAME tables~\cite{sesweb}
 , or a large variety of analytic (either ($\rho,T$)- or $(\rho,e)$-based) equations of state ({\it e.g.} Mie - Gr\"uneisen\cite{Shyue01}, JWL\cite{LFC73}), at which point only the inverted ($P,T$) tables need be saved for use in multiple executions of RAGE problems.
 
 The philosophy of the multi-material equation of state routine(s) is to take these inverted tables -- $v_m(P,T)$ and $e_m(P,T)$, where $v_m\equiv {\rho_m}^{-1}$ is the specific volume and $e_m$ the specific energy of a ``chunk'' of material $m$ --  and construct the total volume and energy by weighting the individual tables by means of the known fractional masses.  In addition to the fractional masses, $M_m$, and their sum, $M$,  the volume of a zone, $V$, and the total energy, $E$, in a zone are also known quantities.
 
 In outline, the user inputs to the preprocessor a range of temperature and pressure on which to build each material's inverted $v_m(P,T)$ and $e_m(P,T)$ tables.
 This is done by making Maxwell constructions in the original form of the tables, if they are not already there, and searching along isotherms (in ($\rho,T$)-based tables) to find the desired pressure.
 The values of $1/\rho$ and $e$ are then saved. In order for tables to be invertible, it is necessary that various thermodynamic derivatives be positive:
\begin{eqnarray}
  \left. \frac{d\rho}{dp}\right|_T > 0 \ \ , \ \  
  \left. \frac{d\rho}{dp}\right|_e > 0 \ \ , \ \ 
  \left. \frac{de}{dT}\right|_p     > 0 \ \ , \ \ 
 \left. \frac{de}{dT}\right|_{\rho} > 0 \ , \nonumber 
\end{eqnarray}
and these are enforced in the preprocessing step that creates the tables.
 
 When RAGE calls the equation of state to find a pressure, it knows the total volume $V$, mass $M$, internal energy $E_{tot}$, and mass fractions, $ x_m \equiv  \frac{M_m}{M}$,  within each cell.  MMEOS begins an iteration process by searching for two adjacent isotherms that bound the specific total internal energy of the cell corresponding to the lowest pressure in the table and calculating the fractional distance between them that would give the correct total energy, such that
 \begin{eqnarray}
\sum_m x_m  e_m(T_{lo},P_{lo})\equiv e_{lo}  &<&   \frac{E_{tot}}{M} \le e_{hi}\equiv \sum_m x_m  e_m(T_{hi},P_{lo})  \ ,  \nonumber \\
\Rightarrow  & s =& \frac{E_{tot}/M- e_{lo}}{e_{hi}-e_{lo}} \ . \nonumber
  \end{eqnarray}
Using the weighted values of various thermodynamic derivatives, MMEOS also calculates
\begin{eqnarray}
\left. \frac{d\rho}{dp}\right|_e &=& \left( \left. \frac{dv}{dT}\right|_p 
                                                                              \cdot \left. \frac{de}{dp}\right|_T 
                                                                               \cdot \left. \frac{de}{dT}\right|_p^{-1} 
                                                                       - \left. \frac{dv}{dp}\right|_t \right) \rho^2 \ , \nonumber 
\end{eqnarray}
so that this can be used to estimate the next pressure in the table at which to repeat the process.  In general, one interpolates along an isotherm at the desired pressure to calculate the internal energy, until one finds the two bounding isotherms.  Although there are various limits designed to ensure that one does not shoot too far through the table too fast, this in broad outline describes the method of finding the point at which a mixed cell exists in pressure and temperature equilibrium.

\subsubsection{Speed and Accuracy}

The procedure outlined above is quite fast, since the expensive table inversion process occurred once during a pre-processing step.
Given the inverted tables, the time to look up a mixed equation of state compared to a pure equation of state is a linear  function of the number of materials, where various table properties have to be mass-weighted (such loops would be short-vector or superscalar loops).

Since some other methods ({\it e.g.}, pairwise balancing of pressures) scale quadratically with materials, they would be significantly inferior applied to  some of the ASCI-scale problems ({\it e.g.}, 50 million zones)  that have upwards of 50 materials in the problem.  Even so, this linear algorithm, on small problems with few materials ( 50k zones, 5 materials) can consume half the execution time for a pure hydro problem.

The accuracy of this process does depend on the models used in the original tables; if some SESAME tables have limited domains, then the extrapolations off those tables to construct the global inverted table will be problematic: densities on the lowest isobar, for example, are overridden so that they will extrapolate to zero at zero pressure for any temperature, while energies are reset to the next higher isobar's value so that $\frac{de}{dp}|_T=0$.  Similarly,  $\frac{de}{dT}|_p>0$ is ensured by requiring that $e'(T_{n-1},P) = \min( 0.999\cdot e(T_n,P), \, e(T_{n-1},P))$, starting at the penultimate temperature and working downward along each isobar.

\subsection{MMEOS Limitations}

The MMEOS algorithm was designed in the context of solid mechanics problems, and works equally well in the  higher-temperature (LTE) regime at Los Alamos.\footnote{By this we mean that if one of the chunks in a mixed zone is in a state ``under'' its vapor dome, MMEOS is going to return a single mean density, and not separate densities and volume fractions for the two phases.} 
 This does mean, however, that the algorithm is not easily adapted to the ``2-T'' regime of separate ion and electron temperatures -- one would have to construct  3-dimensional arrays, {\it e.g.} $v_m(T_i, T_e, P_{tot})$, assuming that it is the total pressure of each chunk of material that equilibrates.  Moreover, one would need 4 such arrays: $v_m$, $e_m^{ion}$, $e_m^{elec}$ and $g_m^{elec} \equiv {p_m^{elec}(v_m, T_e)}/{P_{tot}}$, in order to satisfy all the energy and volume constraints, as well as to partition $pdV$ work among the ions and electrons.  This is a subject of research at present.




\section{Gravity}  
\label{secn:gravity}
             
Because gravity is a long range force, the large spatial scales   characteristic of astrophysical systems make necessary its inclusion in   many problems.  Three different types of gravity have been incorporated into  the RAGE code: constant acceleration, analytic, and  self-gravity.  They may be used in any combination.  The simplest is a constant acceleration, which may be applied in any coordinate direction. The magnitude and direction of the acceleration are specified by the user.  The analytic gravity routines accessible through user inputs are spatially varying but constant in time.
They include single and binary star potentials.  Also included is a power-law mass distribution to model the potential due to a star cluster.  The gravity routines are written to allow users to add other analytic potentials with little effort.  

\subsection{Self-Gravity}
For problems involving self-gravity, the potential and acceleration are determined through a global consideration of masses.   For a matter distribution characterized by  a spatial function of density $\rho (x,y,z)$, the specific gravitational potential ($\phi$) is   determined everywhere in space by solving Poisson's equation,
\begin{equation}
\nabla^{2}\phi  = 4{\pi}G\rho (x,y,z) \ , \nonumber
\end{equation}

where $G$ is the gravitational constant, $G=6.67259\times{10}^{-8} \mathrm{cm}^3\mathrm{g}^{-1}\mathrm{s}^{-2}$.  The acceleration is in turn given by the negative gradient of the potential, $g=-\nabla \phi$.  The equations of hydrodynamics are hyperbolic equations and can be solved by a local (explicit) method, but Poisson's equation is an elliptic equation which must be solved by a global (implicit) method.   For this reason, the calculation of gravitational energies and accelerations typically requires as much computing time as
all other hydrodynamic processes combined.  In order to model self-gravitating systems with sufficient spatial resolution within reasonable run times, a fast gravity solver is desired.   

While many methods for solving Poisson's equation exist,  the adaptive grid algorithm used by RAGE restricts the   possibilities.  The chosen method is the multipole method of Salmon, Warren, \& Winckelmans~\cite{SWW94}.  The SWW method was developed for N-body particle codes,
but the hierarchical tree structure employed is also compatible with the unstructured mesh of RAGE.  
At any moment in time, the discrete integral   representation of the Poisson equation is written as
\begin{equation}
\phi (x)= - \sum_q \frac{m_q}{\parallel x-x_q \parallel} \ . \label{poisson}
\end{equation}
This is a direct sum over $q$ cells, each of mass $m_q$, in order to determine the potential at a chosen location $x$.  To solve equation (\ref{poisson}) on a grid containing N cells requires $\mathcal{O}(N^2)$ operations.  The multipole method  is designed to lower this number to approximately $\mathcal{O}(N\log N)$.

The key to the multipole method lies in combining distant cells into groups and then solving for the potential due to the group rather than each individual cell.  As SWW point out, this is analogous to treating the Earth as one extended object rather than a collection of individual atoms when calculating the gravitational force beyond its surface.   The right hand side of equation (\ref{poisson}) is approximated by a multipole   expansion for a group of $q$ cells within the region $S$:
\begin{eqnarray}
\sum_{q\in S} \frac{m_q}{\parallel x-x_q\parallel} & = &  
\frac{M_S}{\parallel x-x_{cm} \parallel}  + \frac{1}{2}\frac{Q_{ij}(x-x_{cm})_i(x-x_{cm})_j}
{\parallel x-x_{cm} \parallel^5}  + ...  \label{expan}
\end{eqnarray}
In this expression, $M_S$ is the total mass contained in $S$, $x_{cm}$ the center of mass of $S$, and $Q_{ij}$ the quadrupole moment of $S$ about its center of mass.  The dipole term of the expansion vanishes when taken about the center of mass.  The quadrupole moment $Q_{ij}$,  is determined by
\begin{eqnarray}
Q_{ij} & = & \sum_{q\in S} m_q\left[ 3(x_q-x_{cm})_i(x_q-x_{cm})_j
                                     -\delta_{ij}(x_q-x_{cm})\cdot (x_q-x_{cm}) \right] \ , \label{moment}
\end{eqnarray}  
where $\delta_{ij}$ is the Kronecker delta function, and repeated indices indicate a summation.  A multipole acceptability criterion (MAC) is required in order to  quantify the error introduced through the use of equation (\ref{expan}).  The multipole expansion requires the use of data on all cells, but RAGE typically computes data on leaf cells only.  (A leaf cell is one that is not sub-divided into daughter cells.)  So before the gravity solver can begin, the mass, center of mass, and quadrupole moment for all mother cells must be determined through the aggregation of data from the daughter cells. Once the needed data are in place, cell-centered values of $\phi$ and $-\nabla \phi$ are determined for each leaf cell with a treewalk through the grid hierarchy.  The treewalk starts with a loop over the cells on the coarsest grid level.  If the cell in question  is a leaf cell, its contribution to the potential is computed from equation (\ref{poisson}).  If the cell is a mother cell, its MAC is determined.  Should the MAC prove to be satisfied, equation (\ref{expan}) is used to calculate the contribution
to the potential.  If the approximation error exceeds the allowable value, the source cell is divided into its daughters and the test is repeated.  This continues until either the MAC is satisfied or only leaf cells remain.  Note that in the case that AMR is not used, this procedure becomes the direct method embodied by equation (\ref{poisson}). 

In the manner of SWW, we consider the MAC to be satisfied should the MAC error  be lower than a user specified limit $e$:
 \begin{equation}
\frac{1}{d^2 \left(1-\frac{b}{d}\right)^2}
            \left( (p+2)\frac{B_{p+1}}{d^{p+1}}-(p+1)\frac{B_{p+2}}{d^{p+2}}
                  \right) \leq e \ . \label{err}
\end{equation}  
In this expression, $p$ represents the highest order retained in the multipole expansion, $d=\parallel x-x_{cm}\parallel $, and $b=\max_{q\in S}\parallel x_q-x_{cm} \parallel$. The moments $B_n$ are defined as
\begin{equation}
B_n= \sum_{q\in S} \parallel x_q-x_{cm} \parallel^n m_q \ . \label{bmom}
\end{equation}   
Embodied in the MAC is the notion that the multipole approximation is more accurate if the distance to the measurement point is large (large $d$),   the sources are scattered over a small region (small $b$), higher order approximations are used (large $p$), and the multipoles which have been  
neglected are small (small $B_{p+1}$).  For the RAGE implementation, the  quadrupole approximation is used, so $p=2$. 

As expressed by equation (\ref{err}), the MAC is an  absolute criterion.  The error tolerance must be set according to the smallest contribution of $\nabla \phi$. However, applying the most stringent
error tolerance everywhere on the computational grid will result in a higher accuracy than needed and a significant slow down of the code.  The solution to this dilemma is to turn the MAC into a relative error criterion.  This is accomplished by noting that the expression in  equation (\ref{err}) has the dimensions of a mass divided by distance squared.   Therefore, a dimensionless relative MAC is defined by dividing the left hand side of equation (\ref{err}) by $M_s/d^{2}$ to get (for $p=2$),
\begin{equation}
\frac{1}{M_s \left(1-\frac{b}{d}\right)^2}
\left( 4\frac{B_{3}}{d^{3}}-3\frac{B_{4}}{d^{4}}\right) \leq \mbox{\tt mac\_tol} \ . \label{relerr}
\end{equation}
This new relative MAC (named {\tt mac\_tol} in RAGE) represents the relative size of the error in using the quadratic approximation instead of the direct sum.  Typically a value of  {\tt mac\_tol}=.01 proves sufficient.

It is useful to recast the relative MAC into a form in which it becomes a property of the source cell~\cite{WS93}.  The MAC can then be calculated for all mother cells prior to the treewalk.  A critical radius $r_c$ is defined by solving equation (\ref{relerr}) for $d$.    This problem is simplified by noting that by the definition of the moments in equation (\ref{bmom}),  $B_4$ is always greater than zero.  Therefore, that term can be dropped from  equation (\ref{relerr}) without violating
the inequality.  The equation can then be re-written to produce a cubic expression for $d$:
\begin{equation}
d^3-2bd^2+b^2d-\frac{4B_3}{M_s \; \mbox{\tt mac\_tol}} > 0.
\end{equation} 
When treated as an equation, this expression can be solved analytically.  The result is the critical radius $r_c$, defined as the closest distance  to the source for which the quadrupole approximation meets  the criterion set by  {\tt mac\_tol}.  Note that the computational simplicity gained by dropping the $B_4$ term  comes at the price of a critical radius which is larger than is   strictly necessary.  

\subsection{Integration of Gravity into RAGE}

Once the gravitational potential and acceleration have been determined, the results must be incorporated into RAGE.  The conservation equations for mass, momentum, and energy take the modified form:
\begin{eqnarray}
 \frac{\partial\rho}{\partial t} &+ \nabla\cdot  \left({\rho \bfvec{u}}\right)      &=0                                                              \ ,  \label{eq:cont} \\
 \frac{\partial(\rho\bfvec{u})}{\partial t} &+ \nabla\cdot \left( \rho \bfvec{u}\bfvec{u} + p\right) &=-\rho \nabla \phi    \ , \label{eq:mome} \\
 \frac{\partial( \rho E)}{\partial t} &+ \nabla\cdot  \left( \rho E \bfvec{u} + p  \bfvec{u} \right) &=-\rho \bfvec{u} \cdot \nabla \phi        \ , \label{eq:ener}
\end{eqnarray}
where $\phi$ is the gravitational potential, and $\nabla \phi \equiv \bfvec{g}$ the gravitational acceleration.

Adding gravity into a Godunov code is a multi-step process.
During each direction-sweep through a hydro cycle, gravity affects the  dynamics both before and after the Riemann solver.
Before being passed to the Riemann solver, the gravitational acceleration is combined with the acceleration due to the pressure gradient to form half-time values of  $u_L$ and $u_R$.
After the Riemann solution is formed and the state variables are advected,  the energy and momentum are then modified by  the changes in gravitational potential and acceleration such that
\begin{eqnarray}
\Delta{E} =&                \Delta{E}            &+ \phi \;\Delta{m} \ , \label{eng} \\
\Delta (m \bfvec{u}) =& \Delta (m \bfvec{u}) &+ m\; g  \Delta{t} \ , \label{mtm}
\end{eqnarray}
for a cell of mass $m$ with a change in mass due to advection of $\Delta{m}$.

\subsection{Limitations}

A significant decrease in the run time of the self-gravity solver can be achieved by taking advantage of the manner in which RAGE indexes cells among different grid levels.  Setting up the best combination of grid size, AMR levels, and gravity input parameters requires some vigilance on the part of the user.

As of this writing, the RAGE self-gravity solver operates only for three-dimensional problems with non-periodic boundary conditions. Two-dimensional solutions and periodic boundary conditions are under development.


\newcommand{\energy}[1]{e_#1}

\section{Radiation-Matter Energy Exchange in RAGE}\label{secn:raddif}

Many authors\cite{nb:676,nb:765,nb:789,nb:766,nb:627,nb:787,nb:652,nb:262,nb:254,nb:686,nb:413,nb:572,nb:776,nb:431,nb:272,nb:534,nb:624,nb:533,nb:587,nb:261,nb:629,nb:736,nb:319,nb:756} have addressed the topic of handling the coupling of energy and radiation at high temperatures. Our approach has three interesting features:
1) It produces an exponential relaxation of the difference between these two, which is more accurate in general. We explain this in detail and compare it to alternative treatments. 2) It can handle arbitrary variation of $C_{V}$ and ``Planck'' opacity within a timestep, and 3) it  uses a novel technique to solve our resultant set of nonlinear equations.

\subsection{Nomenclature}

The primary dependent variables in RAGE are $E$, $\rho$, $ E_m$, and $\mathbf{u}$,
where $E$ represents the radiation energy density, $\rho$ the mass density, $\rho E_m$ the material total energy density, and $\mathbf{u}$ the material velocity.
The material specific internal energy  is defined by
\begin{equation}
 e \equiv E_{m}-\frac{1}{2} \mathbf{u}^{2} \ . \nonumber
\end{equation}
The temperature $\theta$, pressure $p$,  and specific heat $C_{V}$ are given in terms of $e$ and $\rho$ by an (inverted) equation of state, which may be tabular or analytic.
The radiation coupling $\mu=\rho\kappa_{abs}$ and mean free path $\lambda=1/\rho\kappa_{tot}$ are also given by opacity tables or analytic functions.

The quantities $\Phi$ and $T_{R}$ will appear; they are given by
\begin{eqnarray}
\Phi  \equiv  a{\theta}^{4}  \ , \label{nb:Phidef}  \\
T_{R} \equiv  \sqrt[4]{E/a} \nonumber \ ,
\end{eqnarray}
where $a$ is the radiation constant. $T_{R}$ is called the radiation temperature, but we do not mean to imply by this that the radiation has any particular spectrum; $T_{R}$ is just a measure of radiation energy density.

\subsection{Energy Equations: General}

Although RAGE has options that enable more ambitious physics treatments, 
here we assume local thermodynamic equilibrium, with electrons and ions each described locally by the same temperature $\theta$.

RAGE is a multifrequency code, but this article will treat radiation in the gray diffusion approximation, for which only its energy density $E$ is of concern. Its spectrum is of interest only insomuch as it affects the calculation of $\mu$ and $\lambda$; it is not calculated by the equations described herein. The spectrum may be the result of a simultaneous multifrequency-gray calculation (as described by Winslow and others\cite{nb:624,nb:526,nb:772,nb:797}) of which these equations form the gray pass. Or it may be assumed to be that of a Planckian at temperature $\theta$, or even at temperature $T_{R}$.
 The treatment described herein applies to all of these cases, with small suitable modifications.

RAGE is not a relativistic code, but it does consider terms in $u/c$ to first order~\cite{nb:624,nb:533}. All the quantities involved with radiation are taken to be evaluated in the rest frame of whatever fluid element they reside in at the moment, the co-moving frame. To ${\mathcal O}(u/c)$, there is a simple transformation between the radiant fields in a comoving frame and those of an inertial frame~\cite{nb:624}.  Finally, by making the $P_1$ approximation ($I = \frac{1}{4\pi}(cE + 3 \Omega\cdot\mathbf{F})$ or equivalently, $p_r = E/3$), the hierarchy of moments of the transport equation close at the momentum equation, so that the relativistically correct equations (to ${\mathcal O}(u/c)$) become
\begin{eqnarray}
\frac{\partial E}{\partial t} + \nabla \cdot \left( [E + p_r] \mathbf{u} + \mathbf{F} \right)
            &=& -c \mu ( E - \Phi )
                   - \mathbf{u} \cdot \frac{\mathbf{F}}{\lambda c}
                   - 2\frac{ \mathbf{u}}{c^2}\cdot \frac{\partial \mathbf{F}}{\partial t} \ , \label{eq:p1e} \\
\frac{1}{c^2}\frac{\partial \mathbf{F}}{\partial t} + \nabla \cdot \left(\frac{\mathbf{F}}{c^2}\mathbf{u} + p_r\right)
            &=& -\frac{\mathbf{F}}{\lambda c}
                  - \frac{1}{c^2} (\mathbf{F}\cdot\nabla) \mathbf{u}
                   -\frac{\mathbf{u}}{ c^2}\frac{\partial p_r}{\partial t}  \ . \label{eq:p1f} 
\end{eqnarray}

\subsubsection{Gray Radiation Hydrodynamics}

We now make the diffusion approximation, $\frac{\partial \mathbf{F}}{\partial t} = 0$, relieving us of the burden of solving the radiation momentum equation, \eref{eq:p1f}, and dropping the last term from the radiation energy equation, \eref{eq:p1e}, above.  We further assume that velocity-dependent terms do not contribute to the steady-state flux so that we can maintain $\mathbf{F} \propto  \nabla p_r$.  The resulting radiation { diffusion} equations for the energy $E$ and flux $\mathbf{F}$ are
\begin{eqnarray}
\frac{\partial E}{\partial t} + \nabla \cdot \mathbf{F} & =& 
-c\mu \left( {E - \Phi } \right)\nonumber \\
&&- \nabla \cdot (E \mathbf{u} )\nonumber \\
& &-\nabla\cdot (p_r \mathbf{u})-  \mathbf{u}\cdot\frac{\mathbf{F}}{c \lambda}  \ , \label{nb:uoverc} \\
 \hspace{1.5cm} \mathbf{F} &=&-\mathcal{L} D\nabla{E} \ ,  \label{nb:Fc}
\end{eqnarray}
where the diffusion coefficient $D$ is given by
\begin{equation}
D = \frac{\lambda c}{3} \ , \nonumber
\end{equation} 
and  $\mathcal{L}$ denotes a flux limiter, a common device~\cite{nb:587,nb:261,nb:629,nb:736,LP81} invoked to limit the diffusion flux so that it does not exceed the streaming limit, $cE$.
  In RAGE this limiter, if active, depends on the dimensionless quantity $\lambda |\nabla(E)|/E$, among other things.  RAGE uses the Levermore-Pomraning flux limiter~\cite{LP81}; others have been tried at various times (by us and others~\cite{nb:629}) and not found to make much difference. 

The material momentum equation, to ${\mathcal O}(u/c)$, is
\begin{eqnarray} \label{nb:matter-mom}
\frac{\partial}{\partial t}( \rho \mathbf{u} ) + \nabla\cdot \left[ \rho \mathbf{u} \mathbf{u} + p \right] = \frac{ \mathbf{F} } {\lambda c} \ , 
\end{eqnarray}
and the material energy equation, to the same order,  is
\begin{flalign} \label{nb:matter-rel}
\frac{\partial}{\partial t} \left(\rho e +\frac{1}{2}\rho u^{2} \right) 
+\nabla \cdot \left[ \left( \rho e +\frac{1}{2}\rho u^{2} +p \right) \mathbf{u} \right] + \nabla\cdot\mathbf{F}_m &=& c\mu \left( {E - \Phi } \right) 
 + \frac{1}{\lambda c} \mathbf{u}\cdot \mathbf{F} \ ,
\end{flalign}
with the material heat flux $\mathbf{F}_m$, given by
\begin{eqnarray}
\mathbf{F}_m = - \chi \nabla \theta\ . \nonumber
\end{eqnarray}
The material conductivity, $\chi$, can be read from SESAME tables or calculated from simple analytic expressions.

\subsubsection{Split Equations}
At each timestep RAGE splits the time integration into three separately performed phases: hydro, material heat conduction, and radiation.

The various relativistic terms are treated differently in RAGE. The advection term, line 2 of \Eq~\eref{nb:uoverc}, is calculated explicitly in the hydro phase and the updated result becomes the initial  value ${E_{-}}$, for the radiation phase. The energy and momentum changes wrought  by the radiation pressure and the flux, line 3, are explicitly treated at  the beginning of the  radiation phase. Line 1 is of course solved implicitly as the final step of the radiation phase.
Since $\mathbf{u}$ and $\rho$ are regarded as time-independent functions of position during the radiation phase, the matter equation during this part of the operator split simplifies, leading to the following coupled equations,
\begin{eqnarray}
\frac{\partial E(t)}{\partial t} & - \nabla \cdot \mathcal{L}(E) D(t;E)\nabla E(t) & 
                                             = - c\mu(t) ( {E(t) - \Phi (t)} ) \ ,  \label{nb:eq01} \\
 \rho\frac{\partial e(t) } {\partial t} &   &= +c\mu(t) ( {E(t) - \Phi (t)} ) ,  \label{nb:matter} 
\end{eqnarray}
where material properties ($D$, $\mu$, $C_{V}$, etc.) depend on $t$ because of their dependence on $\theta$, which itself depends on the time dependent internal energy $e$. Of course these properties depend on $\rho$ too; we shall not bother to indicate this.

Adding \Eqs~\eref{nb:eq01} and \eref{nb:matter} gives the basic diffusion equation to be solved in this phase:
\begin{eqnarray} \label{nb:dif-basicdn}
\frac{\partial (E+\rho e)}{\partial t} - \nabla \cdot \mathcal{L} D\nabla E = 0 \ .
\end{eqnarray}

\subsubsection{Special Case: Equilibrium Diffusion}\label{nb:xxx}
If the coupling coefficient $\mu$ is very large, in a sense which we will make precise below, \Eq~\eref{nb:matter} becomes stiff and implies
\begin{eqnarray} \label{nb:equilibrium}
E(t) - \Phi( t)=0\ .
\end{eqnarray}
From the definition of $\Phi$, \Eq~\eref{nb:Phidef}, we have
\begin{eqnarray} \label{nb:thetaofPhi}
\theta(t)=\sqrt[4]{\Phi(t)/a}\ ,
\end{eqnarray}
and the  equation of state, $\mathcal{E}$,  provides the material specific energy $e$, and specific heat $C_{V}$:
\begin{eqnarray} 
e (t) &= \mathcal{E}(\theta(t)) \ , \nonumber \\
C_{V}(\theta)&=\left. \frac{\partial\mathcal{E}}{\partial \theta}\right|_{\rho} \ . \nonumber
\end{eqnarray}
We see that \Eq~\eref{nb:dif-basicdn} requires $\partial (\rho e) /
\partial t$, which we can evaluate as
\begin{eqnarray} \label{nb:drhoedt}
\frac{\partial (\rho e) }{\partial t} &= \rho \ \frac{\partial e }{\partial \theta} \ \frac{\mathrm{d} \theta}{\mathrm{d} \Phi } \ \frac{\partial \Phi }{\partial t} \ ,  \nonumber \\
&=\rho \ C_{V} \ \frac{1}{4a\theta^{3}} \ \frac{\partial E }{\partial t}\ ,
\end{eqnarray}
where the last line follows from the result of the large $\mu$, \Eq~\eref{nb:equilibrium} (remember that $\rho$ is fixed in this phase).
Inserting this into \Eq~\eref{nb:dif-basicdn} produces the equation of equilibrium diffusion
\begin{eqnarray} \label{nb:equilibriumdiffusion}
 (1+{\rho C_{V}}/{4a\theta^{3}})\frac{\partial E}{\partial t}  - \nabla \cdot \mathcal{L} D\nabla E = 0 \ .
\end{eqnarray}
(This is often expressed as an equation for $\theta$: $ (\rho C_V  + 4a \theta^3)\frac{\partial \theta }{\partial t} = \nabla 4a\mathcal{L}D\theta^3\nabla \theta$, and may be referred to as 1-T radiation conduction.)  In general, \eref{nb:equilibriumdiffusion} is an implicit and nonlinear equation.  (The exception is the unphysical case in which $D$ is constant and $C_{V}$ is proportional to $\theta^{3}$, introduced by Pomraning~\cite{nb:813} specifically for the purpose of generating a simple test problem and so used by Su and Olson~\cite{nb:319,SuO97}, and others~\cite{nb:736,nb:756,nb:814}.) 
It is implicit because the material properties $C_{V}$ and $D$ are related to $\theta$ through the equation of state and opacity function, and the flux limiter $\mathcal{L}$ (if active) is explicitly nonlinear.
It is nonlinear in addition because $\theta^{3}$ appears, and $\theta$ is a function of $E$, not just a parameter: $\theta =  \sqrt[4]{\frac{E}{a}}$.

\subsubsection{Special Case: Linear Diffusion}
If $\mu$ is very small, the matter and radiation decouple,  \Eq~\eref{nb:eq01} becomes
\begin{equation} \label{nb:lineardiffusion}
\frac{{\partial E(t)}}{{\partial t}} - \nabla \cdot \mathcal{L} D\nabla E(t) = 0 \ ,
\end{equation}
and the material temperature never changes. Unless the flux limiter or other phenomenon ({\it e.g.}, two-temperature opacities, see appendix~\ref{mrc:alternative}) causes $\mathcal{L} D$ to vary with $E$, this is a linear equation and can be solved by standard methods.

\subsection{Time Differencing}

We now return to the case of non-equilibrium diffusion.
One of the unique features of RAGE's treatment is the asymmetric temporal treatment of the radiation energy equation and the matter energy equation.  As we will see, the radiation energy (diffusion) equation is handled with a typical backward difference approach, but the matter equation is rearranged, treated as an ordinary differential equation (ODE), and integrated exactly, leading to what we term exponential  differencing.

\subsubsection{Solving for $E$ Using Backward Differencing}
At every timestep RAGE integrates the diffusion equation from a starting time $t_{-}$ to a finishing time $t_{+}$, an interval $\Delta t$.
It assumes that $E(t)$ throughout the timestep can be approximated by its final value, $E_{+}$. This is the backward difference approximation, only first-order accurate in time but stable and very robust. Thus integration of \Eq~\eref{nb:dif-basicdn} over a timestep results in
\begin{eqnarray} \label{nb:radiation}  
(E_{+}+\rho e_{+} )- (E_{-}+\rho e_{-} ) - \nabla \cdot \overline{\mathcal{L} D} \nabla E_{+} \ \Delta t = 0 \ ,
\end{eqnarray}
where $\overline{\mathcal{L} D}$ is
\begin{equation}
\overline{\mathcal{L} D} \equiv   \frac{1}{\Delta t} \int_{t-}^{t+}{\mathcal{L} D\, {d}t}  \ , \nonumber
\end{equation}
and the evaluation of $\overline{\mathcal{L} D}$ is left ambiguous at this point.
This is not a closed system, because we do not yet have a value for $e_{+}$, but with the $E$ variation now fixed the matter energy equation, \eref{nb:matter}, becomes
\begin{equation}
\rho\frac{\partial e (t) } {\partial t} = c\mu(t) ( {E_{+} - \Phi ( t)} ) \ , \nonumber
\end{equation}
or equivalently,
\begin{eqnarray} \label{nb:E-matter} 
\rho C_V \frac{\partial \theta(t) } {\partial t} = c\mu(t) ( {E_{+} - \Phi ( t)} ) \ .
\end{eqnarray}
Integration of \Eq~\eref{nb:E-matter} will enable us to calculate  $e_{+}$ needed in \eref{nb:radiation}, effectively allowing us to solve those two equations  simultaneously.

\subsubsection{Solving for $e_{+}$ Using Backward Differencing}
A very popular approach to finding $\ e_{+}$ is to use a backward difference scheme for the matter equation also. This approach is used by Freeman~\cite{BF89}, Knoll {\it et al.}~\cite{nb:562}, Shestakov {\it et al.}~\cite{nb:736}, and many others. It leads to a replacement of \eref{nb:E-matter} by
\begin{eqnarray} \label{nb:eback}
\rho  C_V  (\theta_{+}- \theta_{-})=c \overline{ \mu} (E_{+} - \Phi_{+}) \, \Delta t\ .
\end{eqnarray}
\Eqs~\eref{nb:radiation} and \eref{nb:eback} form a closed system once $\overline{\mathcal{L} D}$ and $\overline{ \mu}$ are specified. It is implicit and non-linear for all the same reasons as the equilibrium diffusion system, \Eq~\eref{nb:equilibriumdiffusion}, and in addition involves the implicit (via $\theta$ and the opacity function) and (likely) nonlinear variation of $\mu$.

\subsubsection{Solving for $\ e_{+}$ Using Exponential Differencing}
We choose not to use the backward difference scheme on the material energy equation, \eref{nb:E-matter}, because it can result in predicting noticeable temperature differences between matter and radiation in cases where the coupling is large enough that these should be in, or close to, equilibrium, as we shall see below.  
We are able to avoid this problem by treating \Eq~\eref{nb:E-matter} as an ODE -- possible because the coordinate $\mathbf x$ is there  only as a parameter; no spatial derivatives appear. This will let us find the actual time dependence of $\Phi$ (and $\theta$ and the variables which depend on it) during the timestep interval $\Delta t$, allowing a more accurate evaluation of the energy transfered during the timestep.

If we define a variable with the dimension of time,
\begin{eqnarray} \label{nb:taudef}
\tau \equiv \frac{{\rho {\kern 1pt} C_V}}{{4a c\,\mu\,\theta ^{3} }} \ ,
\end{eqnarray}
then \Eq~\eref{nb:E-matter} implies
\begin{eqnarray} \label{nb:Phidot}
\frac{d\Phi( t)} {dt} = \frac{ {E_{+} - \Phi( t)}} {{\tau( {\Phi( t)})}} \ .
\end{eqnarray}
From this we see that $\tau$ is the scale time for radiation-matter coupling (strictly speaking, $\tau$ is the relaxation time for $\Phi$ and $E$ coupling in a gradient-free medium)
and it is the proper measure by which we can quantify our earlier statements about $\mu$ being large, leading to \Eq~\eref{nb:equilibrium} or $\mu$ being small, leading to \Eq~\eref{nb:lineardiffusion}  ({\it i.e.}, $\tau$ small compared to the time of interest or {\it vice versa}).

Viewing $t$ as the dependent variable and $E_{+}$ as a simple fixed parameter we can rearrange \Eq~\eref{nb:Phidot}:
\begin{eqnarray} \label{nb:dt}
{dt} = \tau ( \Phi )\frac{d \Phi }{ E_{+} - \Phi } \ .
\end{eqnarray}
This implies that as the interval time increases monotonically from zero toward infinity, the difference between $E_{+}$ and $\Phi$ relaxes from its initial value, ($E_{+}-\Phi_-$), towards $0$.
$\tau$, which is positive and finite on physical grounds, affects only the rate at which the equilibrium is approached. Equation~\eref{nb:dt} has the formal solution,
\begin{eqnarray} \label{nb:eq08}
\Delta t \equiv t_{+}-t_{-} = \int_{\Phi _- }^{ \Phi _+ } {\tau ( {\Phi } )\frac{d\Phi }
{ E_{+} - \Phi }} \ ,
\end{eqnarray}
and inversion of this equation gives $\Phi_{+}$ as a function of $E_{+}$. This in turn gives the temperature $\theta_{+}$ at the end of the timestep and, {\it via} the equation of state, $ e_{+} $. Thus we have in principle:
\begin{eqnarray} \label{nb:emplus}
\ e_{+}= e (E_{+}) \ .
\end{eqnarray}

\subsection{Inverting $\Delta t (\Phi_{+})$ }
The method RAGE uses to invert \Eq~\eref{nb:eq08} is the subject of the next few sections. We begin with an illuminating special case.

\subsubsection{An Ideal Case}
\begin{figure}[htbp]
\begin{center}
\includegraphics[width=416 pt,height=421 pt,angle=0,scale=1.]{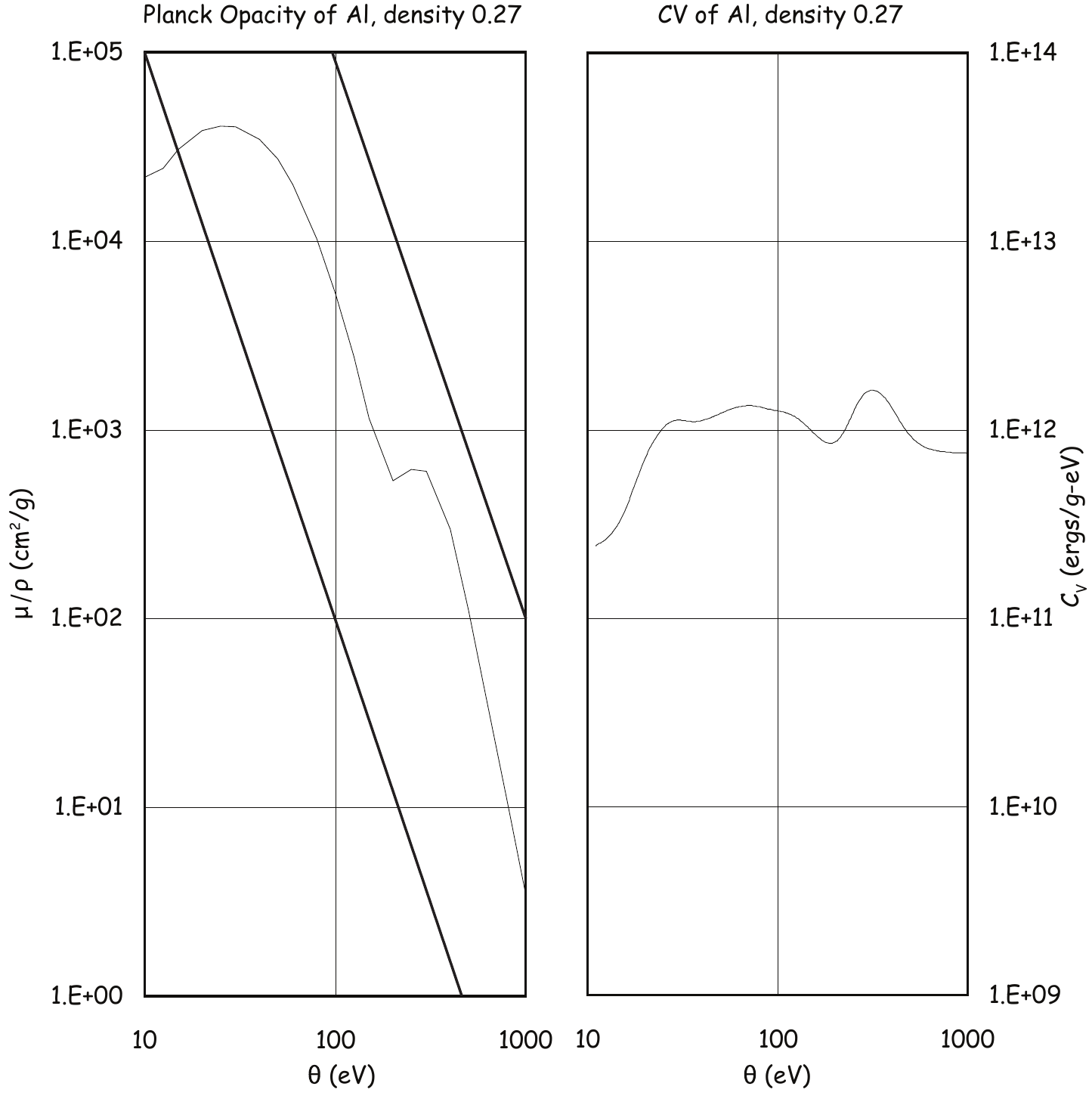}
\caption{Left: Planck opacity of Al at $\rho= 0.27$ g/cm$^3$. The heavy lines illustrate $\theta^{-3}$ behavior. Right: Specific heat, $C_{V}$,  for Al at $\rho= 0.27$ g/cm$^3$. }\label{nb:oneovert3}
\end{center} 
\end{figure}

It is very common that $\mu$ (for ionized materials) varies approximately as the inverse cube of the temperature $\theta$ while $C_{V}$ varies slowly by comparison. Figure~\ref{nb:oneovert3} shows, at left, a log-log plot of the Planck absorption opacity $\mu$ of aluminum at one tenth normal density, $\rho= 0.27$ g/cm$^3$, as calculated by the TOPS opacity code at LANL~\cite{TOPS85};  at right is a plot of the specific heat $C_{V}$ of the same material, calculated  using a Saha-based equation of state~\cite{CWC78}. These plots suggest, at least in this case, a) that a $1/\theta^{3}$ approximation is roughly correct and b) $C_{V}$ is constant by comparison.

Let us idealize this behavior, for the sake of an example, to the case where $\mu$ varies as $\theta^{-3}$ and $C_{V}$ is constant. Then $\tau$ is also constant, and the integration and inversion of \Eq~\eref{nb:dt} becomes trivial:
\begin{eqnarray}
\Phi_{+} -E_{+} = e^{ - \Delta t / \tau}\ (\Phi _{-}- E_{+})\ , \label{nb:exp-ideal}
\end{eqnarray}
explicitly displaying the above-mentioned relaxation toward $E_{+}$.  (The same favorable circumstance of constant $\tau$ occurs for gray Pomeranium, a (mythical) element with a material specific heat  $C_V \propto \theta^{3}$, but with a constant opacity~\cite{nb:813}.)

The fully backward differenced matter equation, \eref{nb:eback},  with the same assumptions, would replace \Eq~\eref{nb:Phidot} with
\begin{eqnarray}
 \frac{d \Phi}{d t} = - \frac{\Phi_{+}-E_{+} }{\tau} \ , \nonumber
\end{eqnarray}
giving the solution,
\begin{eqnarray} \label{nb:fullback}
\Phi_{+} -E_{+} = \frac{1}{1+ \Delta t / \tau}\ ( \Phi_{-}-E_{+})  \ .
\end{eqnarray}
If we recall that the equilibrium diffusion model, \Eq~\eref{nb:equilibriumdiffusion}, gives
\begin{eqnarray} \label{nb:eqdif}
\Phi_{+} -E_{+} = 0 \ ,
\end{eqnarray}
we see that all three schemes' solution can be written as
\begin{eqnarray}\label{nb:all3}
\Phi_{+} = \alpha \Phi _- + (1-\alpha) E_{+} \ ,
\end{eqnarray}
with $\alpha=0$, $\alpha = \frac{1}{1+ \Delta t / \tau}$, or $\alpha = \ e^{ - \Delta t / \tau} $, for equilibrium diffusion, fully backward differencing, or exponential differencing,  respectively. 

\begin{figure}[!h]
\begin{center}
\includegraphics[width=340 pt,height=260 pt,angle=0,scale=1.]{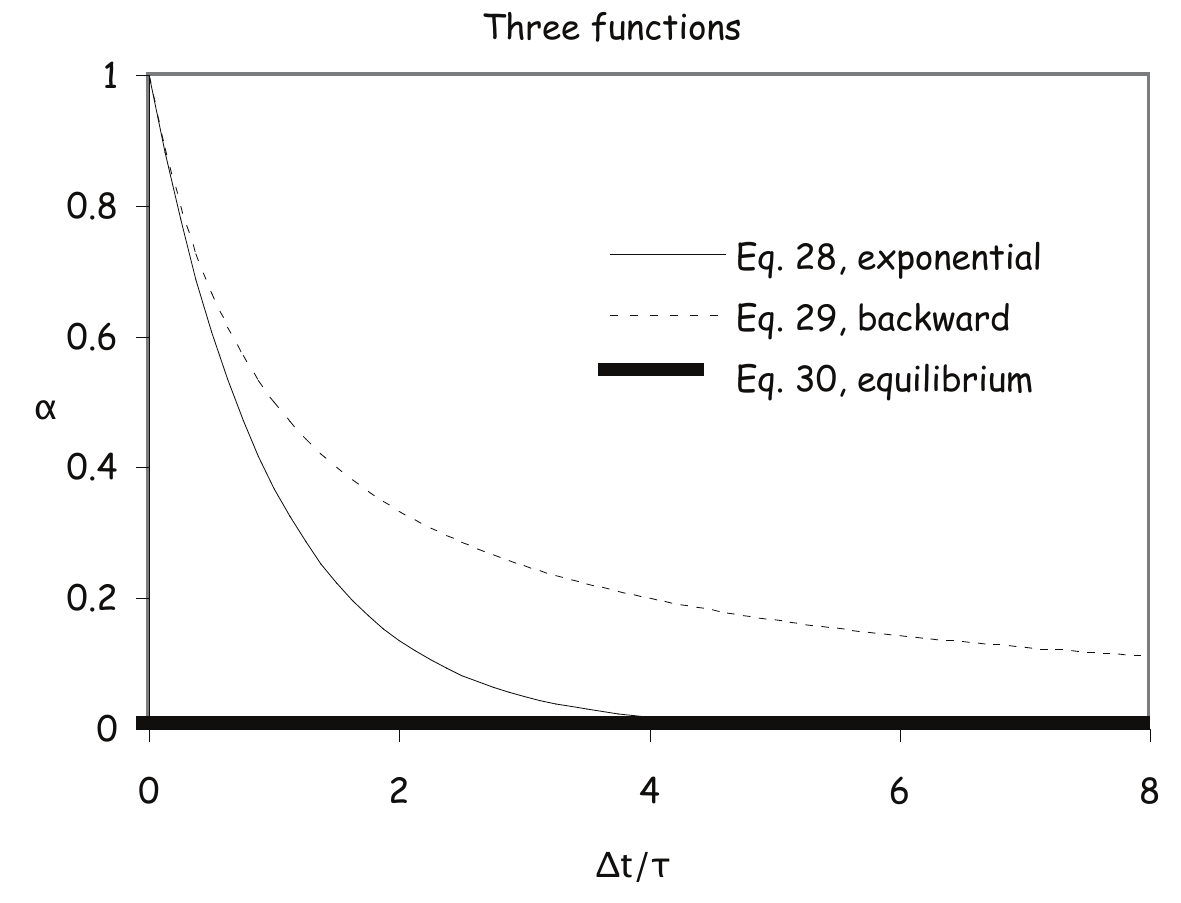}
\caption{Three models for $\alpha$ }\label{nb:funks}
\end{center} 
\end{figure} 

To repeat, the decay toward equilibrium in our scheme proceeds as $\exp{(-\Delta t/\tau)}$ whereas the backward scheme goes like $1/(1+\Delta t/\tau)$. This means that, although both agree to first order in $\Delta t/\tau$, the exponential scheme always transfers more energy between matter and radiation. Although it does not insist on equilibrium, it gets very close as $\Delta t / \tau$ gets large. We expect our method to do better for intermediate timesteps where $\Delta t \stackrel{>}{\sim} \tau$. Figure~\ref{nb:funks} shows these differing solutions over a range of a few time constants;  
Appendix \ref{nb:powerlaw} discusses the application of this technique to the case when the opacity has a power-law behavior other than $\mu\sim \theta^{-3}$, and  we discuss arbitrary opacity behavior in the next section.

\subsubsection{$\Phi$ in the General Case}\label{secn:genl-phi}

As can be seen from our example opacity plot, \Fig~\ref{nb:oneovert3} (left), the inverse cube behavior of $\mu$ is common but not a law, nor, as seen on the right, is $C_{V}$ always a constant. For aluminum in the temperature range 10-30 eV,  both actually {\em  increase} with $\theta$, and $\tau$ turns out to vary approximately as $\theta^{-5/2}$.

RAGE's strategy for handling such cases 
 consists of a more careful treatment of \Eq~\eref{nb:eq08}. It is invoked if we find that the constant-$\tau$ approximation fails, as given by this criterion: evaluate $\tau$ at $\Phi_-$, use it with $E_{-}$ in \Eq~\eref{nb:exp-ideal} to find $\Phi_{+}$ and hence $\theta_{+}$ (and thence $C_{V}$ and $\mu$) and reevaluate $\tau$. If the two values of $\tau $ are substantially different, then the timestep $\Delta t$ is broken into $M$ subdivisions $\Delta t^{(m)}$ small enough that $\tau$ is reasonably constant over each one. 
    Equation~\eref{nb:eq08} is evaluated piece by piece with $\tau$ time lagged in the integral on each subinterval,
\begin{equation} \label{nb:eq12}
\Delta t^{\left( m \right)} = - \tau ^{\left( {m - 1} \right)} 
 \int_{E_{+} - \Phi ^{\left( {m - 1} \right)} }^{E_{+} - \Phi ^{\left( m \right)} } 
 {\frac{{d\left( {E_{+} - \Phi '} \right)}}{{\left( {E_{+} - \Phi '} \right)}}} \ ,
\end{equation}
so that  we can do the integration and inversion for that subinterval:
\begin{eqnarray} 
\alpha ^{(m)} &=& \exp \left( { - {{\Delta t^{(m)} } \mathord{\left/
 {\vphantom {{\Delta t^{(m)} } {\tau ^{(m - 1)} }}} \right.
 \kern-\nulldelimiterspace} {\tau ^{(m - 1)} }}} \right) \ , \label{nb:eq13} \\
\Rightarrow \left( {E_{+} - \Phi ^{(m)} } \right) &=& 
\alpha ^{\left( m \right)} \left( {E_{+} - \Phi ^{(m- 1)} } \right) \ .
\end{eqnarray}
Expanding $\Phi^{(m-1)}$ in terms of $\Phi^{(m-2)}$, {\it etc.},  then  gives
\begin{eqnarray} \label{nb:eq15}
\Phi ^{(m)} &=& \alpha ^{\left( m \right)} \Phi ^{(m- 1)} + \left( {1 - \alpha ^{\left( m \right)} } \right)E_{+} \ , \\
 &=& \alpha^{(m)}\alpha^{(m-1)}\Phi^{(m-2)} + \left( \alpha^{(m)}(1-\alpha^{(m-1)}) + (1-\alpha^{(m)} \right) E_{+} \ , \nonumber \\
 &=& \alpha^{(m)}\cdots\alpha^{(1)} \Phi_{-} + \left( 1 - \alpha^{(1)}\cdots \alpha^{(m)}\right) E_{+} \ , \label{nb:eq64}
\end{eqnarray}
which lets us find the new time constant $\tau^{(m)}$:
\begin{equation} \label{nb:eq16}
\tau^{(m)} = \tau(\Phi^{(m)}) \ ,
\end{equation}
and we repeat until $\Delta t$ is reached. We emphasize that $\Delta t^{(m)}$ is chosen first, then $\Phi^{(m)}$ is found from it.  Defining an unscripted $\alpha$,
\begin{equation} 
\alpha = \alpha ^{\left( 1 \right)} \cdot \alpha ^{\left( 2 \right)} \cdot \alpha ^{\left( 3 \right)} \cdot \ldots \cdot \alpha ^{\left( M \right)} \ , \label{nb:eq17}
\end{equation}
and using \Eq~\eref{nb:eq64} we have, as before,
\begin{eqnarray}\label{nb:eq11}
\Phi_{+} = \alpha \Phi _- + ( {1 - \alpha } )E_{+} \ ,
\end{eqnarray}
giving the desired $\Phi_{+}$ in terms of known quantities, even though $\alpha$ is no longer a simple exponential.

This subcycling need be done only for those cells that are undergoing large material property changes in a given timestep, for instance when the opacity does not approximate the common inverse cube law or the specific heat increases suddenly as the result of beginning to burn out an ionization stage. In practice this is a rare event. The overall cost in time and load balance is small anyway compared to the cost of a diffusion solve, at least in 2-D or 3-D, since inversion of the diffusion operator is not involved in this procedure. We also reap the benefit of treating non-ideal cells accurately without slaving the timestep of the entire mesh to the small values required by a few stiff zones.  For example,  a non-subcycled backward difference code might require a timestep based on a relative temperature change of 5-10\%, while we can allow an order of magnitude change without adverse effect (if this is the  only stiff effect).

Notice that we have separated the problem of handling material properties that have \emph{any}, possibly tricky, variation with temperature from the problem of dealing with the so-called ratio of specific heats.
This latter problem involves only the treatment of the behavior of $\theta^{3}$, simpler in principle than dealing with the vagaries of temperature variation of opacity or specific heat, and is with us even in the case of equilibrium diffusion, \Eq~\eref{nb:equilibriumdiffusion}.
We treat it in section~\ref{secn:ratiocv}.


Note that $\alpha$ is always between 0 and 1. For the simple case of constant $\tau$ this is obvious because $\alpha$ is an exponential of a negative real ($\alpha = e^{ - \Delta t / \tau}$). For the more complicated case of general $\tau$, $\alpha$ is a product of $M$ factors, each of which is bounded by 0 and 1.
We can test the accuracy of the approximation that $\tau$ is a constant (over the subdivision) by comparing successive values of $\tau^{(m)}$, and adapting the subcycle timesteps appropriately (e.g., given the  logarithms of the ratios of the two $\tau$'s and the two $\theta$'s, we can parameterize $\tau$ as a power law in temperature ($\tau \sim \theta^p$). Requiring $\frac{\delta \tau}{\tau}< f \Rightarrow  p \frac{\delta \theta}{\theta}<f \Rightarrow  \frac{p}{4} \frac{\delta \Phi}{\Phi} = \frac{p( E_{-}-\Phi_{-})}{4\Phi_{-}}(1-e^{\-\Delta t^{(m)}/\tau^{(m-1)}})< f$.  Using Newton-Raphson, one can solve for $\Delta t^{(m)}$. )

\subsubsection{A Practical Illustration  of the Iterated $\Phi$}

\begin{figure}[h!]
\begin{center}
 \begin{minipage}[t]{0.8\linewidth}
   \includegraphics[scale=0.45]{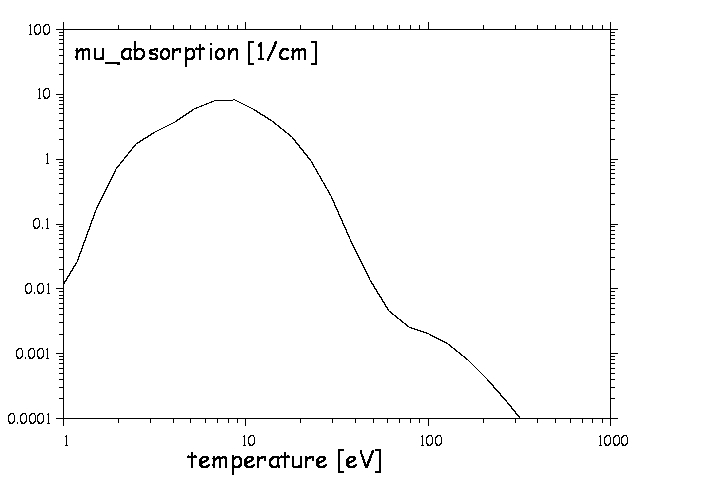}  
    \caption{Absorption coefficient $\mu_{abs}(\theta)$, for oxygen ($\rho = 1.3\,10^{-4}$ g/cm$^3$).} \label{nb:td12opc}
  \end{minipage}
  \end{center}
      \begin{minipage}[bl]{.5\linewidth}
    \includegraphics[scale=.31]{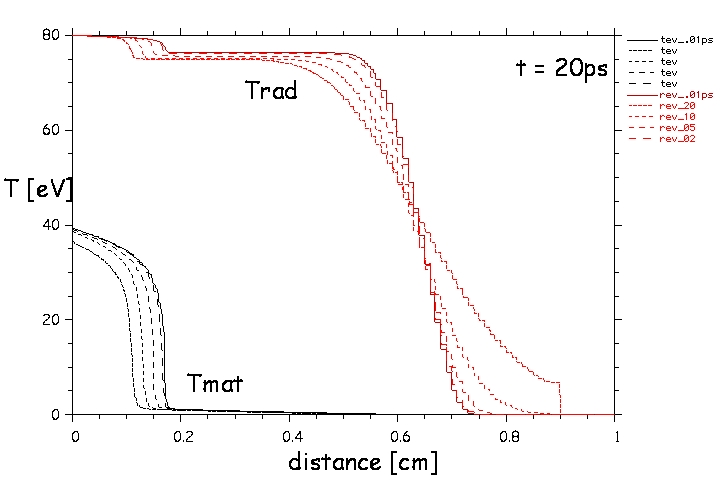}
    \caption{Material and radiation temperature profiles in oxygen at 20 ps. No subcycling of opacity occurs.  Dashed curves used timestep  controlled by $\Delta \theta/\theta= 20, 10, 5, 2$\%.  The solid curves used 2000 timesteps ($\Delta t = 10^{-14}$ s) to converge.} \label{nb:td12bad}
  \end{minipage}
  \begin{minipage}[bc]{.01\linewidth}
 \hspace{1pt}
  \end{minipage}
  \begin{minipage}[br]{.45\linewidth}
   \includegraphics[scale=.29]{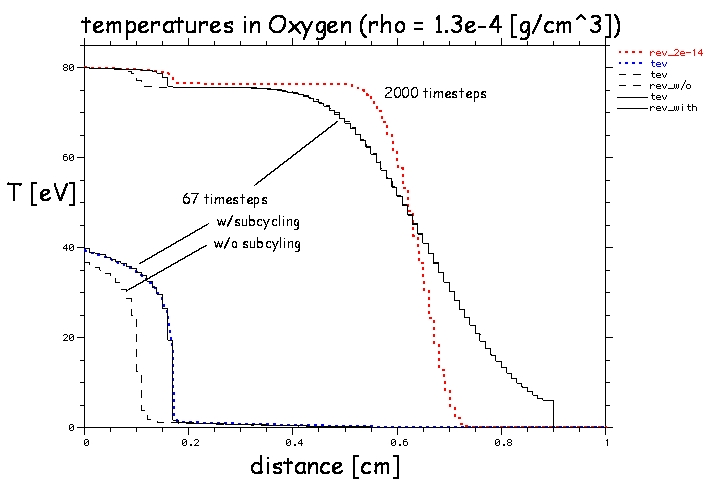}
   \caption{Material and radiation temperature profiles in oxygen at 20 ps.  The solid black curves subcycled opacity, the long-dashed curves didn't; both used $\Delta t=3\cdot10^{-13}$ s.  The heavy-dotted curves used $\Delta t = 10^{-14}$ s.} \label{nb:td12good}
   \end{minipage}
\end{figure}

A valuable side effect of this approach is that it has the capability to handle cases in which the opacity has a local maximum, and this approach can actually provide the correct solution even if some zones start the timestep below the opacity peak peak and end above the peak: one merely marches through the difficult region using \Eqs~\eref{nb:eq12}-\eref{nb:eq16}.  An example is given by the gray opacity 
 of oxygen at $\rho= 1.3\cdot10^{-4}$ g/cm$^3$, as it responds to an 80 eV radiation front.  Figure~\ref{nb:td12opc} shows that  the opacity increases by almost $10^3$ between 1 and 10 eV before it drops back over the next decade of temperature (SESAME's EOS 5010 has $C_V$  constant over this interval).

Without subcycling, one has to hope that $\tau$ is constant for $\Phi$ to be correct, and when $\mu$ is not proportional to $\theta^{-3}$, this means limiting relative changes in $\tau$ to a small value, say 10\%.  However, as \Fig~\ref{nb:td12opc} shows, for temperatures between 1 and 10 eV, $\mu \sim \theta^{+3}$, so $\tau \sim \theta^{-6}$, and a 10\% change in $\tau$ requires no more than a 1.66\% relative change in temperature.  Figure~\ref{nb:td12bad} shows a non-equilibrium Marshak wave in relatively low-density oxygen (idealized from the gas-filled hohlraum that raised this issue).  The oxygen sits in a (relatively) high temperature radiation bath until enough ionization occurs to allow it to rapidly heat up.  The solid black ($\theta(x)$) and solid red ($T_R(x)$) curves shows a converged calculation, run with 2000 timesteps to a final time of 20 picoseconds.  The various dashed curves show profiles of $T_R(x)$ and $\theta(x)$ at $t=20$ ps, when RAGE was altered to {\em not} do any opacity subcycling but to run on a timestep control that kept the relative temperature change below 20, 10, 5 and 2\%  (requiring 122, 254, 554 and 1493 timesteps, respectively).  Only the last calculation has converged the matter temperature.

When RAGE has adapted an interface in order to get the correct energy deposition into a wall, say between air and gold in a hohlraum, it has to adapt in the other direction(s) as well.  For the case depicted by \Fig~\ref{nb:td12bad}, RAGE has very small zones in the direction that the Marshak wave is propagating, even though it does not {\em need} them in that direction, and would prefer to spend as little time as possible calculating the flow in that direction.  Contextually, ``little time'' means flowing radiation  on a radiation Courant timestep, $\Delta t \approx \Delta x/c$,  taking 1 or 2 timesteps to cross each zone. Since the level 1 mesh size is 0.01 cm in this example, this means $\Delta t \approx 3\cdot10^{-13}$ s,  requiring 67 timesteps to get to the displayed time of 20 ps. 
If one used traditional timestep controls, allowing 2\% relative temperature changes from 1 eV (when opacity begins to change) to 40 eV, one would require 186 timesteps to cross each zone.  That becomes prohibitive given the size of 2D and 3D AMR matrices and the time required to invert them.
Figure~\ref{nb:td12good} compares the results with and without the opacity subcycling when the timestep is forced to be the radiative Courant timestep, and both are compared to the 2000 timestep converged result. The width of the leading edge of the radiation temperature reflects the $\mathcal{O}(\Delta t)$ accuracy of the (flux-limited) diffusion solution; it can only be improved with smaller timesteps, and even $c\Delta t/\Delta x = 1/2$ will be much better.
  If RAGE had not been able to subcycle and thereby dispense with traditional timestep controls, it would not have been able to run this type of problem.

\subsubsection{$\mathrm{d}\Phi_{+}/\mathrm{d}E_{+}$}\label{secn:dphide}

Recalling that the equation we need to solve involves $\partial (\rho e)/\partial t$, and using the chain rule, we can now  write
\begin{eqnarray}
\frac{\partial (\rho e) }{\partial t} &= \rho \ \frac{\partial e }{\partial \theta} \ \frac{d\theta}{d \Phi} \ \frac{d\Phi }{dE}  \frac{\partial E}{\partial t}\  
 = \frac{ \rho C_V}{4a\theta^3} \ \frac{d\Phi }{dE}  \frac{\partial E}{\partial t}\  . \nonumber
\end{eqnarray}
As we will see presently, our solution scheme will require the time-advanced $d\Phi_{+} /dE_{+}$.
For $1/\theta^3$ opacities and constant specific heat,  $\tau$ is constant,  $\alpha = e^{-\Delta t/\tau}$, and  $d\Phi_{+} /dE_{+}$ results trivially from \Eq~\eref{nb:all3}:
\begin{equation} \label{nb:eq68}
\frac{d\Phi_{+} }{dE_{+}} = 1 - \alpha \ .
\end{equation}
In the general case we still have \Eq~\eref{nb:eq11}, but since $\alpha$ depends upon $E_{+}$, we must compute the derivative in a more roundabout way. By taking the derivative with respect to $E_{+}$ of \Eq~\eref{nb:eq08} and doing a little manipulation, we have
\begin{equation} \label{nb:eq27}
\frac{d\Phi_{+} }{dE_{+}} = \frac{E_{+} - \Phi_{+} }{\tau_{+} }\int_{\Phi _{-} }^{\Phi_{+}} {\frac{{\tau \left( {\Phi '} \right)}}{{\left( {E_{+} - \Phi '} \right)^2 }}\, d\Phi '} \ .
\end{equation}

Again we break the integral into smaller pieces. The added cost is small because we can recycle the partial results from \Eqs~\eref{nb:eq12}-\eref{nb:eq16}, building
\begin{eqnarray}\label{nb:dphidE}
 \frac{d\Phi_{+} }{dE_{+}} 
 =& \frac{1}{\tau_{+}}  [ \ \ \ \tau^{(0)}(1 - \alpha ^{(1)}) \cdot \alpha^{(2)} \cdot \alpha^{(3)} \cdots \alpha ^{\left( M \right)} \hfill \nonumber \\
 & + \tau^{(1)}(1 - \alpha^{(2)}) \cdot \alpha^{(3)} \cdots \alpha^{(M)} \hfill \nonumber \\
 & + \tau^{(2)}(1 - \alpha ^{(3)}) \cdot \alpha^{(4)} \cdots \alpha^{(M)} \hfill \nonumber \\
 & + \cdots \nonumber \\
 & + \tau^{(M - 1)}(1 - \alpha ^{(M)})]  \ , \nonumber \\
 =& \frac{1}{\tau_{+}}  \sum_0^{M-1} \tau^{(m)} W^{(m)}  \ , 
\end{eqnarray}
and calculate this at the same time we calculate $\alpha$.
By working out the telescoping sums of the weights, we see that $\sum_m W^{(m)} = 1-\alpha$, and rescaling to create normalized weights $w^{m}$ that sum to unity,
\begin{eqnarray}
w^{m} &\equiv W^{m} / (1-\alpha) \nonumber \\
& = \left(1-\alpha^{(m+1)} \right) \cdot \alpha^{(m+2)} \cdot \alpha^{(m+
3)} \cdots \alpha ^{\left( M \right)} / (1-\alpha) \nonumber \ ,
\end{eqnarray}
we can define an average $\langle \tau \rangle$,
\begin{equation}
\langle \tau \rangle = \sum\limits_0^{M - 1}{\tau^{(m)} w^{(m)} } \ , \nonumber
\end{equation}
enabling us to rewrite \Eq~\eref{nb:dphidE} as
\begin{equation} 
\frac{d\Phi_{+} }{dE_{+}} = (1 - \alpha)\frac{\langle \tau \rangle}{\tau_{+}} \ , \nonumber
\end{equation}
with $\alpha$ given by \Eq~\eref{nb:eq17}. This means that in the general case we have  the result for a constant $\tau$, \Eq~\eref{nb:eq68},  multiplied by the ratio of the average of $\tau$ over the timestep to its value at the end of the timestep. In all but the most perverse cases this ratio will be of order unity (RAGE does not depend on this supposition, using \Eq~\eref{nb:dphidE} directly).

\subsection{The Equations for $E_{+}$ and $e_{+}$.}

Equation~\eref{nb:eq11} tells us that $\Phi_{+}$ can always be written as $\alpha \Phi_{-} + (1-\alpha)E_{+}$,   regardless of the behavior of $\mu$ and $C_{V }$. What about $e_{+}$, which is required to solve for the radiation energy at the new time, as seen in \Eq~\eref{nb:radiation}?

We use exactly the same logic as in the case of equilibrium diffusion, described above in section~\eref{nb:xxx}. The equation of state $\mathcal{E}(\theta)$ gives
\begin{eqnarray} \label{nb:e-of-theta}
\ e_{+} =\mathcal{E}\left(\sqrt[4]{\Phi_{+}/a}\right) \ ,
\end{eqnarray}
where $\Phi_{+}$ has been used to find $\theta_{+}$, as per  \Eq~\eref{nb:Phidef}. Hence we have 
  $e_{+}$ as a function of $\Phi_{+}$. 


\subsubsection{The Equations to be Solved.}

Finally, we specify $\overline{\mathcal{L}D}$ as evaluated at the forward time. Our equations then become the non-linear coupled set, for which we invert a matrix equation for $E_+$,
\begin{eqnarray}
E_{+} + \rho e_{+} - \nabla \cdot (\mathcal{L}D)_{+} \nabla E_{+} \ \Delta t &=& E_{-} +\rho e_{-} \ ,  \label{nb:radiation1a}  
\end{eqnarray}
and then solve for the remainder,
\begin{eqnarray}
\Phi_{+}&=&\alpha \Phi_{-} +(1-\alpha)E_{+} \ ,  \label{nb:radiation1b} \\
\theta_{+}&=&\sqrt[4]{\Phi_{+}/a} \ ,                     \label{nb:radiation1c}  \\
\ e_{+}&=&\mathcal{E} \left( \theta_{+} \right) \ ,  \label{nb:radiation1d} 
\end{eqnarray}
where \Eq~\eref{nb:radiation1a} represents the diffusion equation \eref{nb:radiation} and  \Eqs~\eref{nb:radiation1b} \ldots \eref{nb:radiation1d} result from the exponential treatment of the radiation-matter heat exchange, \Eq~\eref{nb:E-matter}. This set is to be solved at every timestep for every space point, thus advancing $E$ and $e$ from time $t_{-}$ to  time $t_{+}$. Their subordinate variables can then also be advanced.

\subsection{Spatial Differencing}
All cells in RAGE have rectilinear cross sections with unit aspect ratio  in every direction, and all faces are orthogonal to the appropriate coordinate axes.   Therefore only one component of a gradient will contribute to a divergence on any face in any RAGE geometry, and spatial differencing becomes quite simple. For example, if an interface normal $\hat{n}$ points in the $\hat{x}$ direction, then for any scalar $s$:
 \begin{equation}
\overrightarrow{\nabla} s \cdot \hat{n} \rightarrow \frac{\partial s}{\partial x} \ , \nonumber
\end{equation}
and this $\mathcal{O}(\Delta x)$  expression for the face gradient suffices to build $\mathcal{O}(\Delta x^2)$ zone-centered divergences.

\subsubsection{Touching Cells}

There are two possible configurations of the geometry of touching cells: either both cells are the same size or else one cell is half the size of the other (a situation we refer to as T-cells, given the form of the interface ([ $\vdash$ ], as in \Fig~\ref{nb:Splat}). In 1D, adaption does not affect the size of the interface, but in 2D or 3D the larger cell communicates through two or four interfaces with the  two or four smaller cells.

\subsubsection{Differencing for Same Size Interface Cells} 

Same size interface cells in RAGE have the same size in the dimension(s) perpendicular to the line joining cell centers. In this sense {\em all}  1D cells qualify, as do all other geometries {\em except} across interfaces where cell adaption has occurred.

RAGE's difference stencil is simple. 
For every interface not subject to a boundary condition, we require that the flux calculated between the cell centroid and the interface on one side be equal to the flux between the interface and cell centroid on the other side, and that they be equal to a net flux between the two centroids: $F_L = F_R = F$, where
\begin{eqnarray}
F_L   =& -D_L \frac{E_{face} - E_L}{\Delta s_{Lf}}  &\equiv -\frac{c}{3}\frac{E_{face}-E_L}{\Delta \tau_L } \ , \nonumber \\
F_R   =& -D_R \frac{E_R - E_{face}}{\Delta s_{fR}} &\equiv  -\frac{c}{3}\frac{E_R - E_{face}}{\Delta \tau_R} \ , \nonumber \\
F =& - D_{face}\frac{E_R - E_L}{\Delta s_{Lf}+ \Delta s_{fR}} &\equiv -\frac{c}{3}\frac{E_R - E_L}{\Delta \tau_{LR}} \ , \label{nb:flux}
\end{eqnarray}
 $\Delta s_{Zf}$ and $\Delta s_{fZ}$ are the distances between the appropriate cell centroid ($Z=L$  or $R$) and the interface $f$ (half the zone size along Cartesian coordinates), and the $\tau$ are optical depths ($D=\lambda c/3 \Rightarrow \Delta \tau = \Delta s /\lambda$).  Solving these equations gives the face quantities:
\begin{eqnarray}
E_{face} &=& \frac{\Delta \tau_L E_R + \Delta \tau_R E_L}{  \Delta\tau_L+  \Delta\tau_R} \ , \label{mrc:fluxeq} \\
D_{face} &=& \frac{c}{3}\frac{\Delta s_{Lf}+ \Delta s_{Rf}}{  \Delta\tau_L+  \Delta\tau_R} \ , \label{mrc:tausum}
\end{eqnarray}
implying that optical depths add: $ \Delta \tau_{LR} =  \Delta\tau_L+  \Delta\tau_R$.

\subsubsection{Temperatures on a Face}

Experience has shown that if the two diffusion coefficients ($D_{L}, D_{R}$) are evaluated at the zone-center temperatures $(\theta_L$, $\theta_R$), it is next to impossible to use \Eq~\eref{nb:flux} to calculate properly the self-similar solution of a Marshak~\cite{Mar58} wave.  The cold material has a very small $D_Z$ (very large $\Delta \tau_{Zf}$), hence  ${\Delta \tau}_{RL}$ is also large and thus very little radiation flows  -- according to the difference equations. The solution to this problem has been to find a common face temperature,  use it to define $D_L(\theta_{face})$ and $D_R(\theta_{face})$, and then compute the $\Delta \tau_L$, $\Delta \tau_R$ and  $\Delta \tau_{LR}$ from those $D$'s. Even though the temperature(s) are now the same, these $D$'s will differ if the densities or constituents of the two cells differ.

Flux equality requires that $E$ vary linearly in optical depth, \Eq~\eref{mrc:fluxeq}, and we note that if radiation and matter were in thermal equilibrium ($E=\Phi$) then   $\theta^4$ would vary similarly. We adopt this {\it ansatz} in all cases, using a two step approximation:  First we calculate $D$'s using the current cell-center temperatures. This lets us define optical depths and so find $\theta^{4}_{face}$ by interpolation: 
\begin{eqnarray}\label{nb:odwgt-t4}
\theta^{4}_{face} &\equiv \frac{\Delta \tau_L\, \theta_R^4+ \Delta \tau_R \, \theta_L^4}{  \Delta\tau_L+  \Delta\tau_R}  \ .
\end{eqnarray}
We then use this value of $\theta_{face}$ for the ``real'' lookup of opacities, $D$'s, and $\Delta \tau$'s.  Some alternate ways of estimating $\theta_{face}$ are discussed in Appendix~\ref{mrc:alternative}; we have not found them to make enough of a difference to offer users a choice in the matter.

\subsubsection{Differencing for Unequal Size Interface Cells}

\begin{figure}[htbp]
\begin{center}
\includegraphics[scale=1]{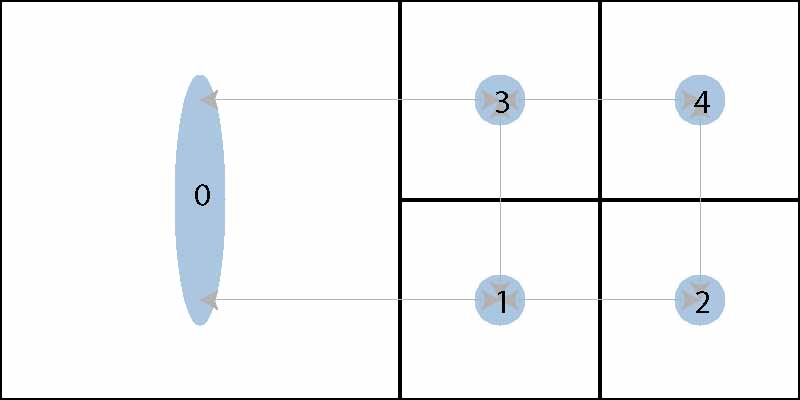}
\caption{RAGE Difference Star for Non-conforming Cells.}\label{nb:Splat}
\end{center} 
\end{figure}

In the case where adaption has occurred, as depicted in \Fig~\ref{nb:Splat}, we ``splat'' (extend central values across the bigger cells) and continue to use \Eq~\eref{nb:flux},  for $F_{03}$ and $F_{01}$. This leads to a formal loss of spatial accuracy, but is mitigated by the fact that in our AMR implementation the gradients are relatively small at any place in the mesh where such scale size changes occur. Splatting has the advantage that the resulting matrix couples only cells which share an interface.

Edwards~\cite{nb:615} introduced a fix for this loss of accuracy, in the context of petroleum extraction, that he called the ``Flux Continuous Approximation" (FCA). Lipkinov {\it et al.} and Olson {\it et al.}~\cite{LMS04,nb:650}, re-derived it in the context of support-operators (such that {\tt div} and {\tt grad} are consistent) for diffusion. The argument is summarized briefly in Appendix~\ref{nb:Shashkov}.

FCA was coded for RAGE in 1996 for 2D geometries. Unfortunately in 3D FCA requires a coupling between cells that touch only at an edge. This was not allowed by RAGE's data structure at that time. In order that the 3D code reproduce 2D results where expected, RAGE chose not to use this  technique in 2D either.

This is not such a blemish on the code as might be expected from the resulting formal loss of accuracy. For one thing, RAGE's AMR logic places such scale changes only in regions with small gradients (gradients are the trigger for adaption, after all).  In the case of  isotropic coefficients and a 2:1 split in cell sizes -- both always the case in RAGE -- simple splatting, observed Edwards\cite{nb:615}, ``can produce extremely good results''.

In any event, recent changes to the data structure  should now allow such couplings, and we expect the FCA treatment to appear in the next major revision of RAGE.

\subsubsection{Boundary Conditions}

RAGE  users have the option of four different types of boundary conditions for radiation: Neuman, Dirichlet, and two forms of Robin.

 The default condition is to assume a Neuman (mirror)  condition, $\nabla E = 0$,  at all edges of the mesh. (This is  reasonable for heat  transfer mediated by ions and electrons that one does not expect to leave the mesh, but less so for radiation, since one generally does expect photons to leave the problem if they get to the boundary.)

All other boundary conditions are specified by nonzero values of temperatures  on the appropriate face.  The user can choose to impose either a Dirichlet or a  Robin  boundary condition on all  boundaries with non-zero temperatures  ({\it e.g.}, in 3D, one could have all six boundaries at six different temperatures, as long as they all use  the same  condition).  The Dirichlet condition holds the temperature to a fixed value, $T_f$ at face $f$; of the two types of Robin condition, the Marshak/Milne holds it at that temperature but 2/3 of an optical depth beyond the face, and the Spillman variant~\cite{SPBC} of Marshak/Milne holds it to that temperature at a variable optical depth beyond the face. The Marshak/Milne boundary condition is designed to reproduce the effect for a single energy equation that the transport equation would accomplish with two equations; it requires optically thin boundary zones to succeed. Spillman's variant is conceptually similar to, but considerably simpler to implement than, the Levermore-Pomraning boundary condition~\cite{LP81} that enabled their flux-limited diffusion to recover transport solutions in thin and thick limits.  We compare and contrast the Robin conditions in Appendix~\ref{mrc:robin}.

\subsection{Solution Techniques for Non-Linear Systems }\label{secn:solution}

As mentioned above, \Eqs~\eref{nb:radiation1a}\ldots \eref{nb:radiation1d} are very similar to the result of a backward difference scheme, so it is of interest to examine techniques for solving that scheme. Shestakov \emph{et al.}~\cite{nb:736} discuss the use of pseudo-transient continuation to stabilize a Picard-Newton series of linearizations and iterations, and Knoll {\it et al.}~\cite{nb:562} discuss wrapping a Newton-Krylov iteration around an inner iteration. We might well profit from the use of such methods, but at present RAGE uses a different (nameless) method, also involving linearization and iteration.

\subsubsection{RAGE's Technique\label{secn:ratiocv}}

 At each timestep we execute an iteration process, with iterates  $\{E_{+}^{(n)},e_{+}^{(n)}\} $ $\left( n=1,2,3,\cdots \right)$ for the time-advanced solution of \Eqs~\eref{nb:radiation1a}\ldots\eref{nb:radiation1d}.  Each iteration consists of three conceptually distinct steps. 

The first step consists of making an initial estimate $\{E_{+,est}^{(n)}, e_{+,est}^{(n)}\} $ of the $n^{th}$ iterate based on the previous iterate $\{ E_{+}^{\left( n-1\right)}, e_{+}^{\left( n-1\right) }\} .$ Currently we simply take%
\begin{equation}
\{ E_{+,est}^{(n)},e_{+,est}^{(n)}\}
  =   \{ E_{+}^{\left( n-1\right) },e_{+}^{\left( n-1\right) }\} , \nonumber
\end{equation}
where in particular the $0^{th}$ iterate consists of the initial values at the beginning of the timestep:%
\begin{equation}
\{ E_{+}^{\left( 0\right) },e_{+}^{\left( 0\right) }\} =\{ E_{-},e_{-}\} . \nonumber
\end{equation}%
We identify this as a separate step in the iteration process because more elaborate methods for this step are currently under study, having the flavor of the prediction step in a predictor-corrector technique.

The second step consists of solving a linearization of the radiation diffusion equation, in which the coefficients  are evaluated in terms of the estimated values $\left\{ E_{+,est}^{(n)},e_{+,est}^{(n) }\right\} $ from the first step. More precisely,  \Eq~\eref{nb:radiation1a},
\begin{equation}
E_{+}+\rho e_{+}-\nabla \cdot \left( \mathcal{L}D\right) _{+}\nabla E_{+}\Delta t=E_{-}+\rho e_{-},  \nonumber
\end{equation}
is linearized as%
\begin{equation}
E_{+,lin}^{(n)}+\rho e_{+,lin}^{(n)}-\nabla \cdot  (\mathcal{L}D)_{+,est}^{(n)}\nabla E_{+,lin}^{(n) }\Delta t=E_{-}+\rho e_{-},  \label{tlb:eq79}
\end{equation}%
where the coefficient $(\mathcal{L}D)_{+,est}^{(n)}$ is evaluated at the values  $\{ E_{+,est}^{(n) },\theta _{+,est}^{(n)} \} $ with $\theta _{+,est}^{(n) }$  given in terms of $e_ {+,est}^{(n)}$ by the equation of state $e_{+,est}^{(n)}=\mathcal{E}\left( \theta _{+,est}^{(n)}\right) $,  and where $e_{+,lin}^{(n)}$ is given by%
\begin{equation}
e_{+,lin}^{(n)}=e_{+}\left\vert_{E_{+}=E_{+,est}^{(n) }}\right. +\left( \frac{de_{+}}{dE_{+}}\right)
_{E_{+}=E_{+,est}^{(n)}}\left( E_{+,lin}^{(n)}-E_{+,est}^{(n)}\right) .  \label{tlb:80a}
\end{equation}
The RHS of \Eq~\eref{tlb:80a} is in fact the first two terms of a Taylor series expansion of $e_{+}$ as a function of $E_{+}$ about the linearization point $E_{+,est}^{(n)}$, with $e_{+}$ and $\left( \frac{de_{+}}{dE_{+}}\right) = \frac{de_{+}}{d\theta_{+}} \frac{d\theta_{+}}{d\Phi_{+}}\frac{d\Phi_{+}}{dE_{+}}$ given by \Eqs~(\ref{nb:radiation1b}-\ref{nb:radiation1d}), reproduced here:
\begin{eqnarray}
\Phi_{+}^{(n)}&=\alpha \Phi_{-} +(1-\alpha)E_{{+}}^{(n)}  \ , \label{nb:nonlinb} \\
\theta_{+}^{(n)}&=\sqrt[4]{\Phi_{+}^{(n)}/a}                      \ ,  \label{nb:nonlinc}   \\
\ e_{+}^{(n)}&=\mathcal{E} \left( \theta_{+}^{(n)} \right)  \ ,    \label{nb:nonlind} 
\end{eqnarray}
where $\alpha $ and $d\Phi/dE$ are evaluated as described in sections~\ref{secn:genl-phi} and \ref{secn:dphide}. Having solved \Eq~\eref{tlb:eq79} for $E_{+,lin}^{(n)}$, we compute $e_{+,lin}^{(n)}$ by substituting $E_{+,lin}^{(n) }$ into \Eq~\eref{tlb:80a}. We note that the solution to the second step, $\{ E_{+,lin}^{(n)},e_{+,lin}^{(n)}\}$,  conserves energy exactly, subject only to the accuracy of the equation of state inversion and of the linear solver used to solve the system \Eq~\eref{tlb:eq79}. 

Early versions of RAGE used a conjugate gradient linear solver with a point Jacobi (diagonal scaling) preconditioner. We have now converted to an algebraic multigrid preconditioner (LAMG) developed at LANL by Joubert~\cite{JC04}; this preconditioner has typically given, for moderate to large 2D and 3D meshes, a factor of five or better in speed-up of the linear solver, relative to the point Jacobi preconditioner.

Owing to the linearization in the second step, the values $\{E_{+,lin}^{(n)}, e_{+,lin}^{(n)}\} $ will not, in general, simultaneously solve the material energy equation, \Eq~\eref{nb:nonlind}. Hence we do a third (corrective) step to bring the solution into agreement with the material energy equation for each cell, without, however, changing the total energy in each cell from the value obtained in the second step. This step consists of solving, for each cell, the energy balance equation%
\begin{equation}
E_{+}^{(n)}+\rho e_{+}^{(n)}=E_{+,lin}^{(n) }+\rho e_{+,lin}^{(n)},  \nonumber
\end{equation}%
for $\{ E_{+}^{(n)},e_{+}^{(n)}\} $ in which the RHS is the total cell energy from the second step, and $e_{+}^{(n)}$ on the LHS is given self-consistently as a function of $E_{+}^{(n)}$ by \Eqs~(\ref{nb:nonlinb}-\ref{nb:nonlind}).

This completes the specification of a single iteration, giving the $n^{th}$ iterate $\{ E_{+}^{(n)},e_{+}^{(n)}\} $
in terms of the $( n-1)^{th}$ iterate $\{ E_{+}^{(n-1) },e_{+}^{( n-1) }\} $.  In order to decide when to terminate the iterative process, we (currently) look at the total energy in the cell, exiting when changes in that quantity becomes sufficiently small. We are also investigating the more usual method of looking at the nonlinear residual. Normally, one or two iterations suffice. 

Our experience has been that, for small to moderate timesteps, the dominant nonlinearities come from the radiative heating and cooling associated with the material energy equation. For large timesteps, the nonlinearity associated with the diffusion coefficient and flux limiter can become more important. In fact, for large enough timesteps $\Delta t$, we may encounter diffusion fronts that move faster than $\Delta x/\Delta t$, where $\Delta x$ is the size of a computational cell; in this case it is necessary to execute
several iterations in order for the front to traverse the required number of cells.

The great bulk of the cpu time (at least for 2D and 3D meshes) is normally spent in executing the linear solve in the second step of the iteration (and possibly in the first step in future versions of the code). No other computations require inter-cell  communication; they execute relatively quickly, particularly in a parallel environment.

The current version of the code makes the simplifying approximation that the specific heat $C_{V}$ is held fixed during the radiation diffusion computations. This simplification eliminates a number of equation of state calls, replacing \Eq~\eref{nb:nonlind} with the simple linear relation   $e_{+}=e_{-}+C_{V}\left( \theta _{+}-\theta _{-}\right) $. Future versions of the code will include as an option the elimination of this simplification.

\subsection{Comparisons to Other Codes and Standard Problems}

\begin{figure}[htbp]
\begin{center} 
\includegraphics[scale=.5]{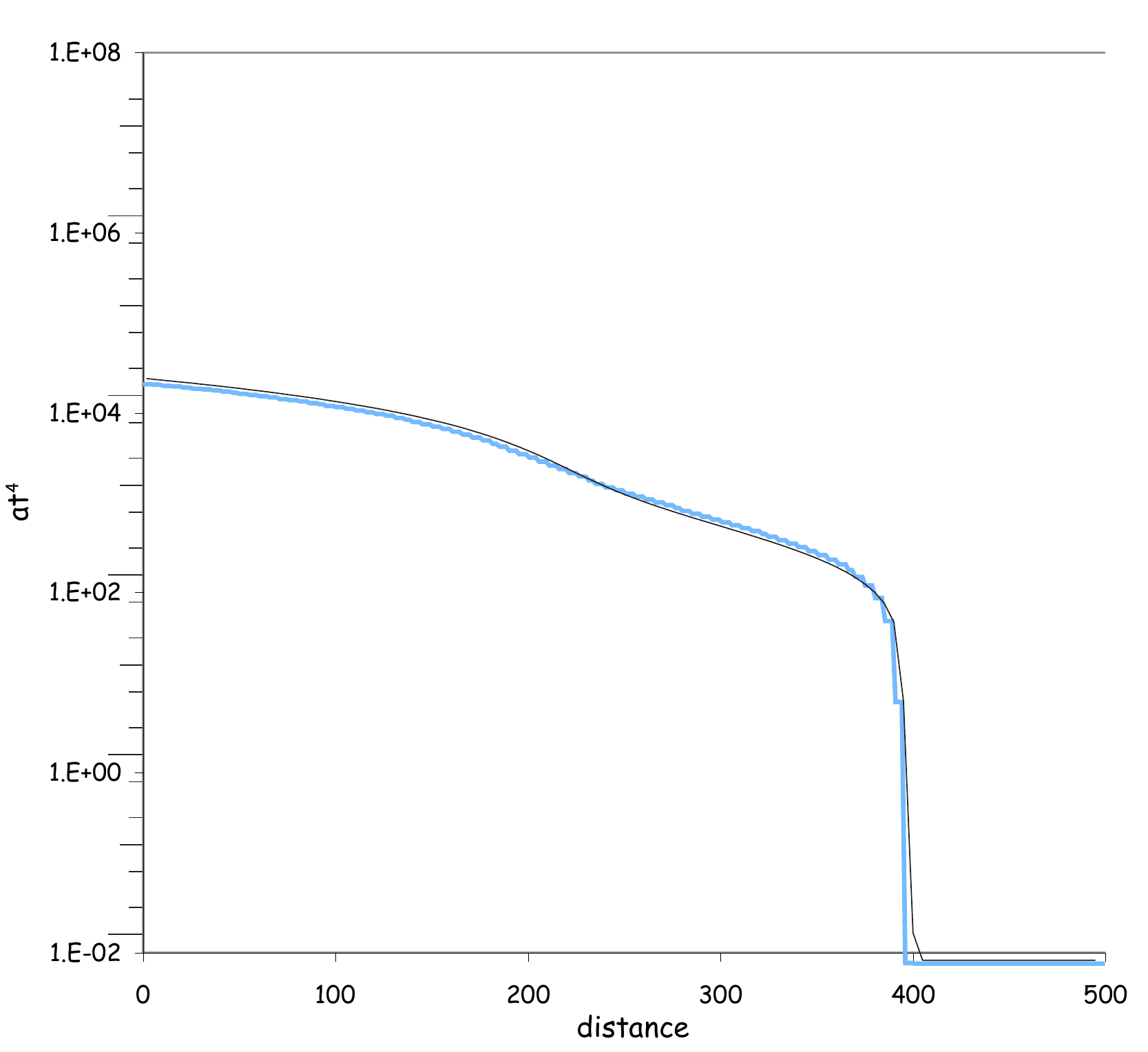}
\caption{Comparison of  $aT^4$ spatial profiles at 20 ns  from RAGE (thin black histogram) and $\Psi$tc~\cite{nb:736} (smooth blue curve).} \label{nb:ASfig3}
\end{center} 
\end{figure}

Figure~\ref{nb:ASfig3} shows a comparison of the results of a test problem taken from Shestakov \emph{et al.} \cite{nb:736}.
 The RAGE results for $\Phi$ $(a\theta^4)$ at 20 ns  are shown  as histograms and the original calculation using pseudo-transient  continuation ($\Psi$tc) as a curve through cell centers. As one can see, the temperatures  (or $\Phi$'s) calculated by the two codes (with the same zoning and timesteps) are in very close agreement. For this problem RAGE was modified to use the same hydrogenic Saha equation of state used by Shestakov \emph{et al.} rather than SESAME tables, in part because the SESAME EOS includes the effect of dissociation of molecular $\mathrm{H}_2$, not just ionization of atomic hydrogen (as specified by the test problem).  RAGE was also modified to use the $n=1$ Larsen flux limiter~\cite{nb:629} (RAGE's Levermore limiter agreed  with the $n=2$ Larsen limiter, and both disagreed with the $n=1$ limiter by a little more than the thickness of the lines in \Fig~\ref{nb:ASfig3}).
 
\begin{figure}[htbp]
\begin{center} 
  \begin{minipage}[t]{.9\linewidth}
\includegraphics[scale=.5]{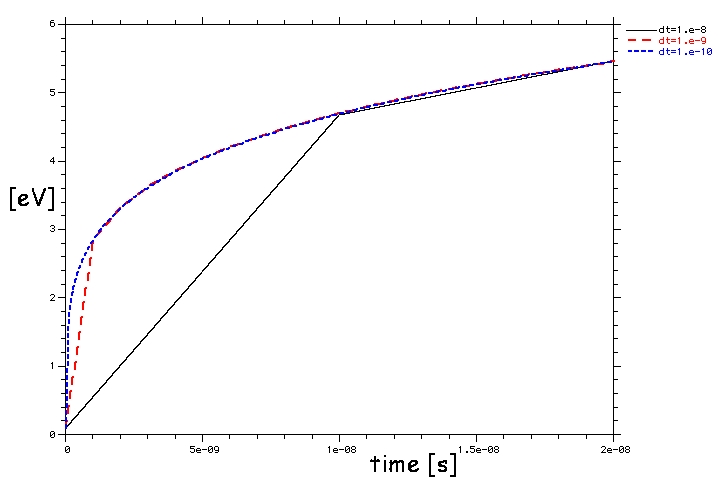}
\caption{$\theta(t)$ temperature relaxation in an infinite medium's  27.2 eV radiation bath. $C_V$ is constant, $\mu_a \sim T^{-7/2}$, and $\tau$ is not constant.} \label{nb:ASinf}
\end{minipage}
\begin{minipage}[c]{0.9\linewidth}
\end{minipage}
 \begin{minipage}[b]{0.9\linewidth}
\includegraphics[scale=.5]{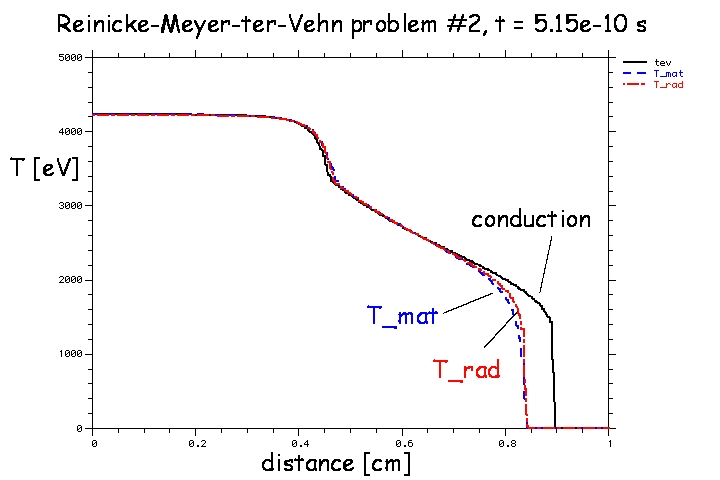}
\caption{Temperature profiles when  thermal wave has reached $r=0.9$ ($t=5.15\,10^{-10}$ s).  Black solid: heat conduction only; blue dashed: material temperature, red dot-dashed: radiation temperature from (2-T) radiation diffusion.} \label{fig:rmv2}
\end{minipage}
\end{center} 
\end{figure}

Figure~\ref{nb:ASinf} shows the temporal behavior of a related problem. The configuration is changed to an infinite medium with fixed $C_V$  but the same $\mu_a \propto T^{-7/2}$ opacity, so we can focus on the coupling of matter and radiation only ({\it i.e.} without diffusion complications). The plot shows the electron temperature as a function of time, up to 20 ns. The problem has been run three times, varying the integration timestep $\Delta t$. The smoothest curve was generated by using 200 timesteps, each of size $10^{-10}$ s. Another curve used only 20 timesteps of $10^{-9}$ s, and finally the solid curve used only two(!) timesteps of $10^{-8}$ s. Clearly one loses information as to what happens in the intervals between such monster timesteps, but we call attention to the fact that RAGE's exponential difference scheme \emph{gets the right answer at a given time regardless of the timestep} --  as long as opacity subcycling is allowed.

Figure~\ref{fig:rmv2} shows temperature profiles for a self-similar problem due to Reinicke and Meyer-ter-Vehn~\cite{RMV91} and popularized by Shestakov~\cite[his figure 4, ``E0=235'']{AS99}.  A thermal wave and a hydrodynamic shock are created whose radii maintain a fixed ratio, generated by a point explosion in cold material with $\rho \sim r^{-19/9}$. The solid black curve of the figure shows the resultant temperature profile using the heat conduction package in RAGE (with the appropriate  analytic conductivity, $\kappa = 31.62 \cdot\rho^{-2}\, T^{6.5}$); the red dot-dashed curve shows the radiation temperature $T_{rad}$ and the blue dashed curve the material temperature $T_{mat}$ calculated by RAGE's radiation diffusion package, all at a time of $5.15\cdot 10^{-10}$ s.   $T_{mat}$ and $T_{rad}$ do not overlay the conduction solver's temperature profile because the self-similar problem was defined with a constant $C_V$~\cite{RMV91}; unlike the conduction package, the radiation diffusion package cannot help including an extra $4a\theta^3$ in the total specific heat, a significant component at these temperatures that destroys the self-similarity of the solution.  
 The more interesting point is that RAGE's exponential differencing has kept the two temperatures in equilibrium (at an early time of $1\cdot10^{-10}$ s, there is a slight disequilibrium near the origin); we expect standard backward differencing methods would have trouble keeping the temperatures clamped together (recall \Fig~\ref{nb:funks}).   We did, however, assume the freedom to define a Planck-weighted material absorption opacity $\kappa_{abs}= 5.433\cdot 10^{12}\rho T^{-3.5}$, 31.32 times larger than the presumed Rosseland total opacity $\kappa_{tot} = 1.735\cdot10^{11}\rho T^{-3.5}$, making relaxation times $\tau\sim 10^{-12}-10^{-11}$ s.  Without this, the two temperatures were out of equilibrium everywhere, and the thermal front was another half millimeter retarded ($\tau \sim 10^{-11} - 10^{-10}$ s).

Finally, we present an example of the ability of RAGE to deliver reasonable results even under unreasonable conditions. In particular, we display solutions of a Marshak-like problem, with a left boundary held at 1000 eV. Cell 1 is initialized to that same temperature, but the rest of the (1D) grid is started at room temperature. We use a dummy material with a constant specific heat, $C_V = 10^{12}$ erg/g/eV, constant density, $\rho=1$ g/cm$^3$, and variable opacity,
\begin{equation}
\kappa=\left(\frac{1000}{\theta}\right)^3 \ \ \mbox{cm$^2$/g} \ . \nonumber
\end{equation}
Each cell is 0.02 cm wide. We run to $10^{-11}$ s, at which time a simple constant-flux-approximation argument suggests that a wave would have burned through about $6\frac{1}{2}$ cells. RAGE normally chooses its timesteps fairly carefully, beginning with a very small one and inching up as seems appropriate, but for this series we forced it to use a constant timestep. We ran the problem three times, taking one, two, and fifty timesteps ($\frac{c\Delta t}{\Delta x} \approx 15, 7, 1/3$) to get to the final time; the results are shown in \Fig~\ref{nb:1,2,50rt} (radiation temperature \emph{vs.} space) and \Fig~\ref{nb:1,2,50et} (electron temperature \emph{vs.} space).

\begin{figure}[htbp]
  \begin{center} 
    \begin{minipage}[t]{.9\linewidth}
    \includegraphics[scale=.4]{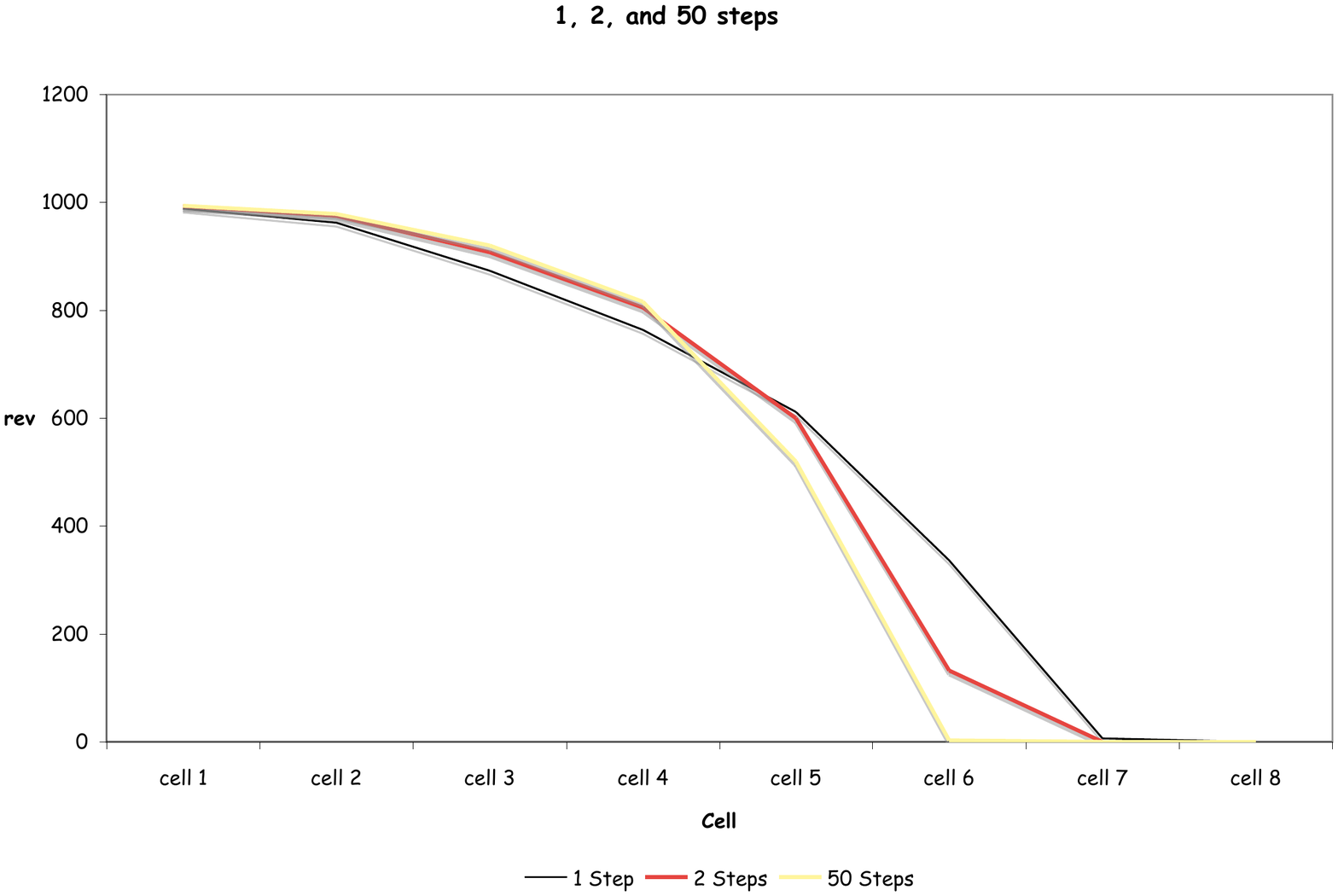}
    \caption{Radiation temperature profiles for a non-equilibrium Marshak wave using one, two, and fifty timesteps ({\it i.e.}, 15, 18, and 122 matrix solves).} \label{nb:1,2,50rt}
  \end{minipage}
  \begin{minipage}[b]{.9\linewidth}
   \includegraphics[scale=.4]{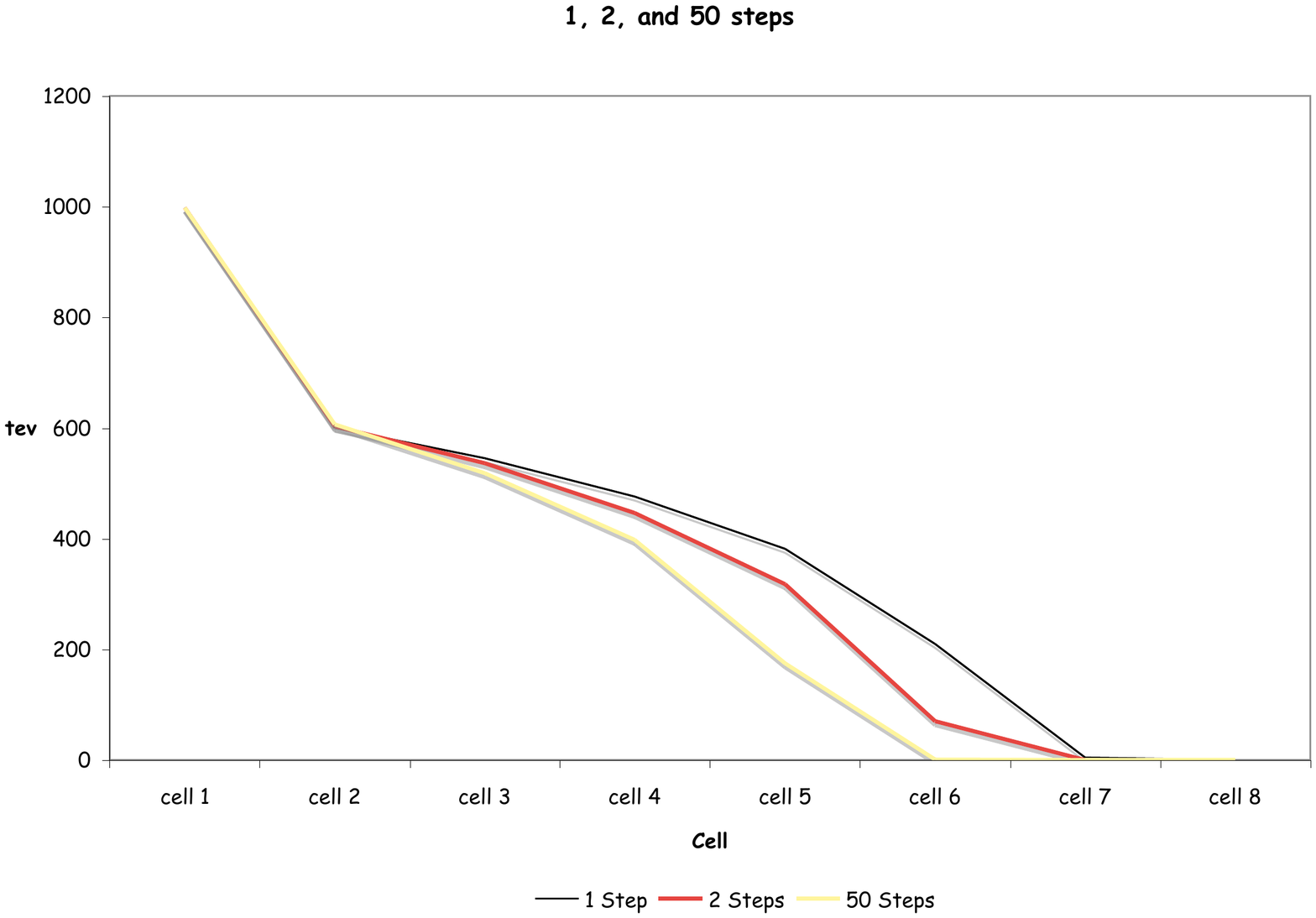}
   \caption{Matter temperature profiles for a non-equilibrium Marshak wave using one, two, and fifty timesteps ({\it i.e.}, 15, 18, and 122 matrix solves).} \label{nb:1,2,50et}
   \end{minipage}
 \end{center} 
\end{figure}  

We do not claim that the results obtained with the ridiculously large timestep are accurate  (after all, the backward difference scheme used for the diffusion equation is only first-order accurate), rather that they are not unreasonable; in particular, since we have no theorem that proves our iteration method will converge, these results are encouraging in that they do not diverge.  Moreover, they show that when multiple (expensive) physics packages are turned on ({\it e.g.} the gravitational Poisson solver), we can iterate the matrix-solves in the radiation package without requiring the rest of the code to run on the smaller effective timestep.   

\subsection{Radiation Summary}

We have described just a subset of RAGE's radiation capability: gray diffusion. RAGE also contains a multifrequency diffusion package, in which the spectrum of the radiation energy field is calculated by the multifrequency-gray method~\cite{nb:526}, and of which this algorithm is a subset, and a gray $P_1$ package which contains a number of novel features. The gray diffusion algorithm described here is of note primarily because of the exponential differencing of the material energy equation and of the novelties in the iteration scheme. Exponential differencing is particularly useful in a class of problems in which radiation floods a region of space and serves to heat a contained body. This situation occurs frequently in inertial confinement systems, where the radiation source is produced by e-beam or laser deposition in a hohlraum wall and the goal is to heat and compress a capsule containing the nuclear fuel.



\section{Verification and Validation for SAGE and RAGE}
\label{vandv}      
                   
During the course of the ASCI (now ASC) Project, substantial time has been devoted at LANL, either directly or indirectly, to the verification and validation (V\&V) of the Project codes.
Direct V\&V results from a Project team member running specific simulations to quantitatively compare to analytic results (verification) or to results from experimental data (validation). 
Indirect validation results from the use of the Project codes by end-users for their own purposes, either in designing a new AGEX (Above Ground EXperiment) experiment,  in the analysis of a previous experiment, or merely as a tool for analysis of a physical problem.

\subsection{Verification}

There exists a large number of verification test problems in our community, including the Sedov exploding blast wave~\cite[pp. 393-396]{LL59}, the Guderley converging shock,\cite{L81},~\cite[p. 793]{ZR02}, anthologies of hydrodynamic~\cite{C91} and 
 radiation-hydrodynamic problems~\cite{CA86}, both LTE and non-LTE~\cite{Mar58,Pet60,SuO97},   Marshak radiation wave problems, and many more  (non-LTE in this context means that $T_r \ne T_e$).
A detailed report on the suite of verification problems is being prepared (see e.g. Ð\cite{KR98,KRV99}), as well as work on automating such ongoing testing~\cite{TSH05}.

\begin{figure}
  \begin{minipage}[t]{.95\linewidth}
\includegraphics[width=5.5in,height=2.5in]{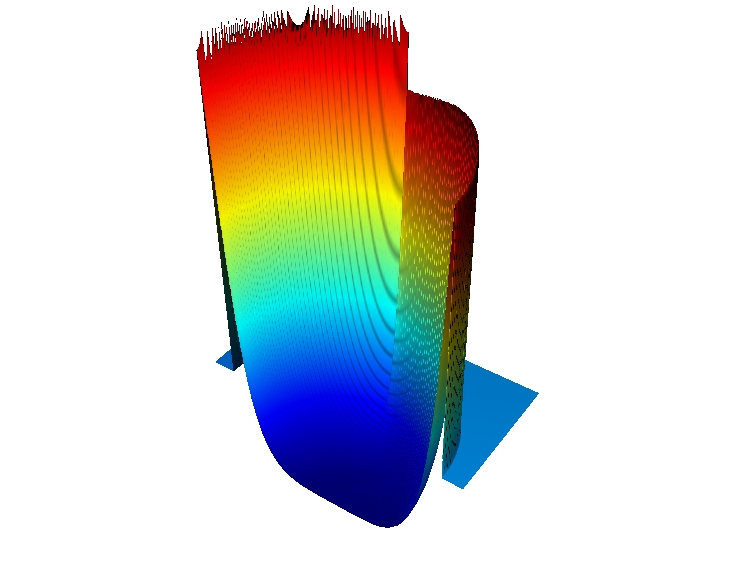}
   \caption{2D (rz) Spherical Sedov problem. Density is plotted vertically and colored by density; the analytic solution ($\rho_{max}=4$) is reflected in the left quadrant for comparison to the calculation in the right quadrant (numerically diffused $\rho_{max}\sim 3.5$). } 
   \label{fig:vandv_sedov2d}
   \end{minipage}
\end{figure}

One example of such a verification problem is the Sedov problem, with unit density and zero initial velocity but  $4.936 \cdot 10^{15}$ erg in a delta function at the origin ({\it i.e.} in one zone at the origin, so that $r_{shock}=1$ cm at $t$=10 ns).  The results from the (2d) cylindrical Sedov problem are shown in Figure~\ref{fig:vandv_sedov2d}, comparing the SAGE adaptive mesh hydrodynamics to the analytic result, and some error norms are provided in table~\ref{tab:l1errs}, which incidentally shows that the same accuracy can be achieved in less time and fewer zones using adaption.  The ratio of $L_1(\rho, u, p)$  errors for the fixed meshes is consistent with those errors being proportional to $\Delta x^{0.6 - 0.75}$, typical for most shock problems we have looked at (shockless problems without discontinuities do behave like $\Delta x^{2}$).

\begin{table}[h!]
\begin{tabular}{|| l | l | l | l | l | l | l | l ||} \hline
level & $L_1(\rho)$  & $L_1(p)$ & $L_1(u)$ & $N_{cycles}$ & $N_{zones}$  & cpu-time[s] & cc/s/pe \\ \hline
1      & 0.13              & 3.99e13   & 1.80e6    & 289       & 900      & 13.2           & 1.97e4 \\
2      & 0.0869          & 2.59e13   &  1.23e6    & 1016     & 3600    & 139.3         & 2.63e4 \\
3      & 0.0555          & 1.67e13   &  7.87e5    & 3406     & 14400  & 2020.3       & 2.43e4 \\
4      &  0.0328         & 1.00e13   &  4.74e5    & 10867   & 57600  & 24512.        & 2.55e4 \\
A1-4 & 0.0327         & 1.00e13    & 4.72e5    & 8052     & 30921   & 8268.7       & 1.90e4 \\ \hline
\end{tabular}
\caption{Comparisons of $L_1$ error norms for density, pressure and velocity on 4 fixed size meshes as well as a mesh that adapts (``A1-4'') to the finest level.  Also shown are measures of the time required to calculate the answer (on a Mac G5).\label{tab:l1errs}}
\end{table}

In general we are very pleased with the success of the SAGE and RAGE algorithms in comparing to analytic and semi-analytic results (see e.g., \cite{WGB97} for an example of comparisons to semi-analytic problems and Section~\ref{secn:hydrovandv} for four comparisons to analytic solutions). 
Many of these verification problems are included in a regression test suite (run on the LANL machines) as part of the software quality assurance process.

\subsection{Validation}
\begin{figure}
  \begin{minipage}[t]{.9\linewidth}
  \includegraphics[width=4.5in,height=3.0in]{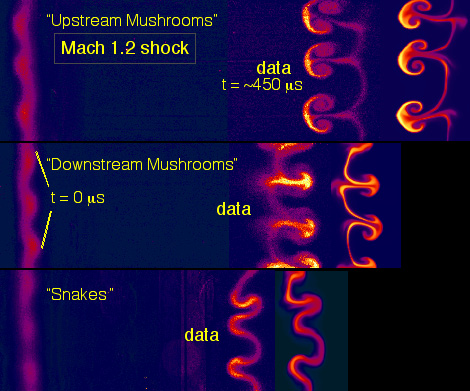}
   \caption{Comparison of experimental and calculated density profiles for a gas curtain shock-tube experiment. $t=0$ profiles provided initial conditions.} 
   \label{fig:vandv_balt1}
   \end{minipage}
\end{figure}

Our first validation effort,~\cite{BGWBB96}, ``The Simulation of Shock Generated Instabilities''  compared detailed code simulations of shock tube experiments  to the experimental results in a quantitative manner;  the results from this (``gas curtain'') work are summarized (in the ``vu-graph'' norm) in figure~\ref{fig:vandv_balt1}.  In order to agree with the data at $t\sim 450$ ms, the code was initialized with the density profile measured at $t=0$ ms (after background subtraction and some smoothing).  Further Richtmyer-Meshkov validation was done using  shocks generated by lasers~\cite{HDF99} and pulsed-power~\cite{KAB02}.  RAGE has also been used to model more general laser-induced shock effects~\cite{GCW99,GBC00,LMB06} and more recently, 3-D oceanic disturbances generated by asteroid impacts and tsunamis~\cite{GWMG03,GWGM04,GWG06,GWG06a}.

\begin{figure}[htbp]
  \begin{minipage}[t]{1.0\linewidth}
 \hspace{1.0cm}   \includegraphics[width=5.5in,height=2.5in]{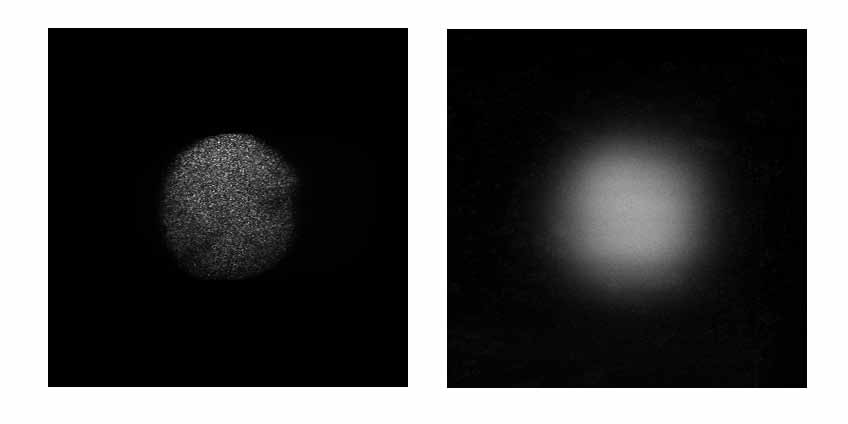}
   \caption{Experimental images of a gas cylinder, using ``fog'' tracer particles on the left, and  Rayleigh-scattering on the right.} 
   \label{fig:vandv_cz1}
   \end{minipage}
  \begin{minipage}[bl]{0.55\linewidth}
 \includegraphics[width=3.5in,height=4.0in]{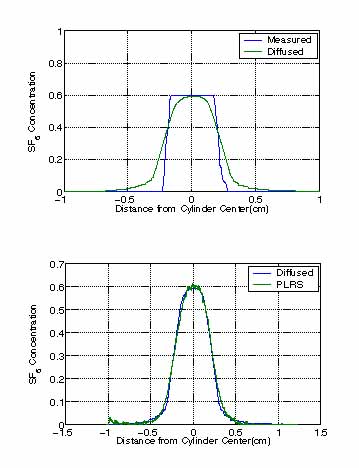}
   \caption{Top: original (measured) and  ``fitted'' (diffused) initial density profiles for a gas cylinder experiment; the original was an (angularly) smoothed measured fog distribution.  Bottom: comparison of that ``fitted'' profile to the measured Rayleigh-scattering profile ({\it n.b.}, scales have changed).} 
   \label{fig:vandv_cz2}
  \end{minipage}\hspace{0.2cm}
\begin{minipage}[c]{0.45\linewidth}
\hspace{1.0cm}  \includegraphics[width=4.5in,height=3.9in, angle=90.]{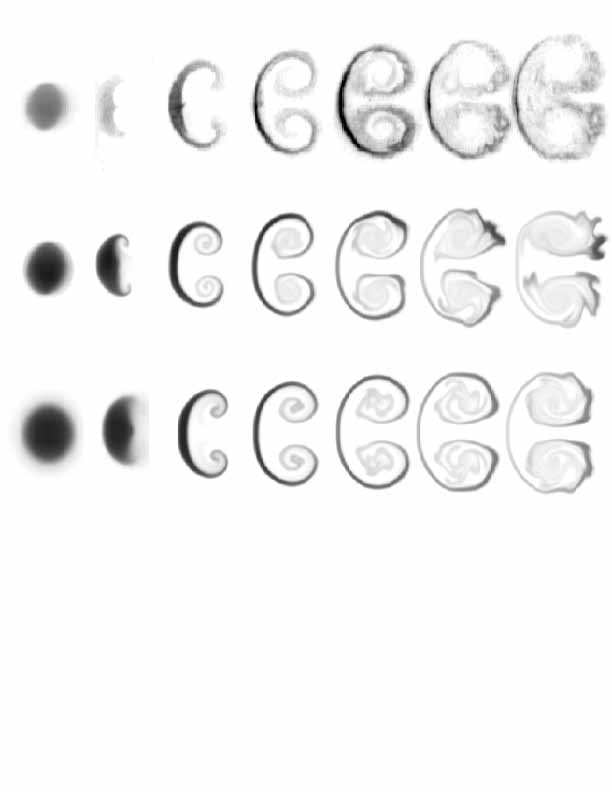}
   \caption{Experimental images of a gas cylinder, using ``fog'' tracer particles on the left; calculated images using sharp (middle) and diffuse (right) profiles from \Fig~\ref{fig:vandv_cz2} top.} 
   \label{fig:vandv_cz3}
\end{minipage}
\end{figure}

Code validation also formed the basis for a Ph.D. thesis for a SUNY Stony Brook student~\cite{Z02,F02}.
In this work, code simulations were done for a similar shock tube environment but in this case the Mach 1.2 shock interacted with a dense cylinder of gas, instead of a gas curtain.  Both the code and the experimental techniques had changed by the time of this comparison (the code was rewritten from Fortran77 with a Lagrange plus Remap hydro to Fortran90 with the Direct Eulerian hydro described in this paper;  the  original experiments~\cite{BGWBB96} used Rayleigh scattering as a diagnostic while the new ones used ``fog'' tracer particles, which were thought to be equivalent to Rayleigh images and were essential in order to perform particle velocity measurements)  and the agreement between the two was not as good as it was in the earlier work.  Having determined that changing the hydro algorithm didn't change the discrepancy with experiment, changes to the initial conditions were tried -- it was hypothesized that the SF$_6$ diffused faster than the fog particles before the first image was taken, leading to incorrect initial density profiles -- and much better agreement was obtained. The top part of figure~\ref{fig:vandv_cz2} shows how much the initial RAGE density profile had to be diffused in order to match the (late-time) data (shown in figure~\ref{fig:vandv_cz3}), compared to the original profile based on the ``fog'' tracer particles (left half, figure~\ref{fig:vandv_cz1}).   This prompted the experimenters to re-image their initial profiles using a Rayleigh scattering technique (right half, figure~\ref{fig:vandv_cz1}).    The lower part of  figure~\ref{fig:vandv_cz2} shows how RAGE's diffused density profile (which gave good late-time agreement) compared to a (later) Planar Laser Rayleigh Scattering (PLRS) measurement.\footnote{The agreement between the profile used to model one experiment and the PLRS profile from another shot shows how reproducible the experimental conditions were.} 
 In this case, one could say that the code validated the experimental technique.  On the other hand, the advantage of the fog tracer particles is that they enable the experimenters to measure the velocity field in the gas cylinder (from two closely spaced images).  When the calculated and simulated images of the density agreed (with the diffused density profiles), the velocity data were also found to agree, to with 10-15 percent, and one can say that the experiment validated the code in this case.


Another example of validation is the work being done as a part of a multi-lab, multinational,  collaboration on the simulation of supersonic jets and shock interactions \cite{FWR02}, where detailed quantitative inter-comparisons have been done among several computational tools (including RAGE) and the actual experimental data obtained from a series of AGEX (``Above Ground EXperiments'') experiments.
All the code techniques, including CAMR RAGE and ALE codes from both Livermore and AWE (the UK Atomic Weapons Establishment), show good agreement with the data.  In this and other cases~\cite{BHW03}, results of code simulations are used to design as well as  predict future experiments.

These examples touch on the breadth of validation efforts completed and underway in the on-going effort to understand the range of capability for the CAMR Eulerian-based hydrodynamics and radiation energy flow as coded in the massively-parallel, modular version of the RAGE code. More detailed documentation efforts are underway. 
Two of the areas of active research in the Crestone Project currently are the treatment of interfaces between materials and the improvement of the material strength package.



\section{Summary}

We have presented a description of the SAIC/LANL Radiation Adaptive Grid Eulerian code, RAGE, describing some of  its data structures, its parallelization strategy and performance, its hydrodynamic algorithm(s), and its (gray) radiation diffusion algorithm.  We have also shown that all packages and the code as a whole are subject to a considerable amount of verification and validation efforts.

The Godunov-based hydrodynamics uses a second-order alternating-direction explicit (ADX) method, and this algorithm determines the maximum timestep the code can take. The iterative nature of the anti-cavitation logic handles stiffnesses in the equation of state   in an implicit manner, meaning that we do not have to control the timestep based on, for example, relative changes in density or mass fraction.  

The gray radiation package uses a temporally first-order and spatially second-order backward Euler method to solve the diffusion equation.  The material energy equation is integrated exactly for opacities proportional to $T^{-3}$, and to any specified accuracy for all other behaviors by subcycling the numerical integration.  This solution is then linearized to build the coupled total energy diffusion equation, and a subsequent Newton-Raphson step rebalances the material and radiation energy to correct for any linearization errors.  Finally, an outer iteration is performed until the total radiation plus material energy in each zone is relatively unchanged between two iterations.

Because the radiation package has taken great pains to treat (1) stiff material coupling terms ($\mu_{abs}$ or $C_V$) by subcycling, and (2)  linearization errors by Newton-Raphson fixups, it can be considered to be maximally implicit.  This means that its accuracy is completely determined by the relative changes in radiation energy, independent of the changes in material temperature.

RAGE by default attempts to take relatively large timesteps, based on controlling just the relative change in total energy over the entire timestep, $\delta (\rho E_M + E) /(\rho E_M + E)$,  to some tolerance, {\it e.g.}, $\mathtt{sie\_pct}=0.2$.  It is also possible to control the material temperature change on {\em each} step of the operator splitting to a tolerance ($\mathtt{tev\_pct}$), but this is a legacy of Lagrangian technology.  We intend to update the splitting logic to instead control relative changes of material and radiation energy on each step of the operator split, and may couple this with an overall control on the non-linear residual.

The implementation of these implicit treatments allow our AMR code to run on timesteps that are considerably larger than would be allowed by the traditional small-percentage-changes in material and radiation temperature on each step of an operator split, and this means that our AMR code can run more highly-resolved problems than might otherwise have been possible, within time and resource constraints.

\section*{Acknowledgments}
We wish to thank Bill Archer and the ASC Crestone Team at LANL for their on-site support as well as the efforts of many of our colleagues at Los Alamos National Laboratory who have worked with us to develop RAGE into the ASCI code that it is, including 
C. Zoldi of X-2 for permission to use figures from her thesis; W. Joubert, of CCS-2 and Xiylon Software for his work developing the Algebraic Multigrid Matrix solvers we now use; and to Darren Kerbyson and Adolfy Hoisie at the Performance and Architecture Laboratory (PAL), CCS-3, for their work benchmarking RAGE on multiple architectures and for providing the results to us.   We also wish to thank the anonymous referees whose comments significantly improved this paper.

 This work  was supported by the University of California, operator of the Los Alamos National Laboratory  under the auspices of the US Department of Energy at LANL under contract number W-7405-ENG-36,  and  at  SAIC under subcontract number  96558-001-04 4t. 


\begin{appendix}

\section{ A Less Ideal Opacity}\label{nb:powerlaw}
The case of constant $\tau$ is but one example of a whole class of situations in which exponential differencing \eref{nb:eq08} can be treated analytically: that for which $\tau$ has a power-law dependence on $\theta$, and thus $\Phi$. Given that
\begin{eqnarray}\label{nb:power-tau}
\tau = \tau_{*} \times {\left(\frac{\Phi_{\  }}{\Phi_{*}}\right)}^{\lambda} \ ,
\end{eqnarray}
\Eq~\eref{nb:eq08} becomes
\begin{eqnarray}\label{nb:here-is-hyper}
\Delta t & = &-\tau_{*}{ \left( \frac{\phi_{-}}{ \phi_{*}} \right) }^{\lambda} \ \frac{1}{\phi_{-}^{\lambda}} \int_{\phi _- }^{ \phi _+ } {\phi^{\lambda}\frac{\mathrm{d}\phi }
{ \phi-1 }} \ , \nonumber \\
\frac{\Delta t }{\tau_{*}}{ \left( \frac{\Phi_{*}}{ \Phi_{-}} \right) }^{\lambda} & = & - \frac{1}{\phi_{-}^{\lambda}} \int_{\phi _- }^{ \phi _+ } {\phi^{\lambda}\frac{\mathrm{d}\phi }
{ \phi-1 }} \ ,  \nonumber \\
&= & \frac{1}{\phi_{-}^{\lambda}} \ \left. \frac{\phi^{\lambda+1} \, _2F_1(\lambda+1,1;\lambda+2;\phi)}{\lambda+1}\right|_{\phi_{-}}^{\phi_{+}}  \ ,
\end{eqnarray}
where any $\phi$ represents the corresponding $\Phi$ scaled to $E_{+}$:
\begin{equation}
\phi_{x} \equiv \frac{\Phi_{x}}{E_{+}} \ . \nonumber
\end{equation}
 $ _2F_1(\lambda+1,1;\lambda+2;\phi)$ is of course the confluent hypergeometric function of the second kind~\cite{wolframweb}.  

This is a very nice form: since $\Phi_{-}$, $\lambda$, $\Phi_{*}$, and $\tau_{*}$ are all known, the left side of \Eq~\eref{nb:here-is-hyper} is just a constant multiple of $\Delta t$ (also known). The right side, on the other hand, is a function of both $E$ and $\Phi_{+}$. This then provides all we need to find $\Phi_{+}$ as a function of $E$; if need be we can resort to a numerical root-finding routine to extract the information.

This chore can be rather intimidating in general, but is much easier when $\lambda$ is a multiple of 1/4, because then the hypergeometric function can be expressed in terms of standard simple functions.

For example, consider the case $\lambda=-\frac{1}{2}$, as occurs with a constant specific heat and $1/\theta$ opacity. Then since
\begin{eqnarray*}
{\left. \, _2F_1(l+1,1;l+2;\phi) \right|}_{l=-1/2}= \frac{\tanh ^{-1}\left(\sqrt{\phi}\right)}{\sqrt{\phi}} \ ,
\end{eqnarray*}
\Eq~\eref{nb:here-is-hyper} becomes:
\begin{eqnarray*}
\frac{\Delta t }{\tau_{*}}{ \sqrt\frac{\Phi_{-}}{ \Phi_{*}}  } = -\sqrt{\phi_{-}} \   \ln\left[  \left( \frac{1-\sqrt{\phi_{+}}}{1+\sqrt{\phi_{+}}} \right)  \left( \frac{1+\sqrt{\phi_{-}}}{1-\sqrt{\phi_{-}}} \right)  \right] \ .
\end{eqnarray*}

Using this we can explicitly display $\phi^{+}$ in terms of a simple parameter, $X$,
\begin{eqnarray*}
\phi_{+}= \left( \frac{X-1}{X+1}\right) ^{2 }\ ,
\end{eqnarray*}
where $X$ is
\begin{eqnarray*}
X &= \exp\left[{-\frac{\Delta t}{\tau_{E}}}\right] \left( \frac{1-\sqrt{\phi_{-}}}{1+\sqrt{\phi_{-}}}   \right) \ .
\end{eqnarray*}
Note that we used \Eq~\eref{nb:power-tau}, the scaling law for $\tau$, to evaluate $\tau_{E}$.

Earlier we discussed three different simple models for the parameter $\alpha$, which governs the energy transferred between matter and radiation during a timestep, \Eq~ \eref{nb:all3}. It is of interest to examine $\alpha$ for this new case:
\begin{eqnarray*}
\alpha &= \frac{1-\phi_{+}}{1-\phi_{-}} 
             =\frac{4 X}{(1+X)^2} \frac{1}{1-\phi_{-}}  \ .
\end{eqnarray*}

\begin{figure}[!hbtp]
\begin{center}
\includegraphics[width=340 pt,height=260 pt,angle=0,scale=0.7]{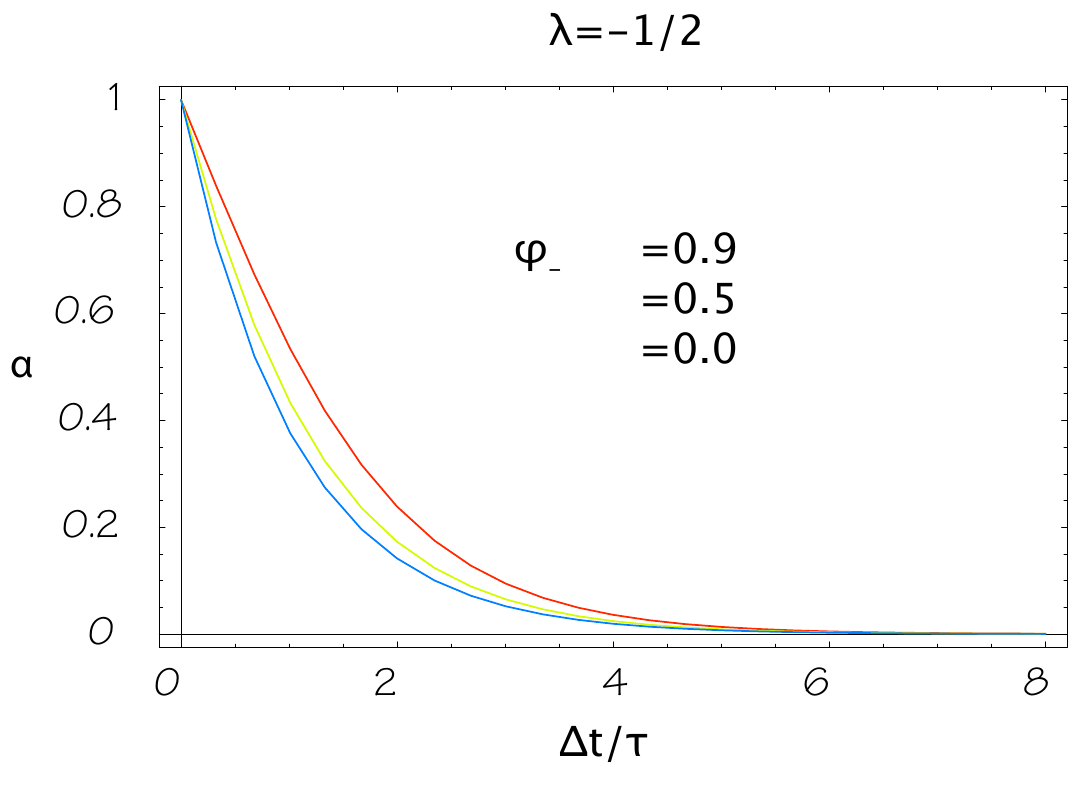}
\caption{A fourth model for $\alpha$ }\label{nb:lambda-1-2}
\end{center} 
\end{figure}

Figure~\ref{nb:lambda-1-2} shows the results for three different values of $\phi_{-}$, a parameter heretofore irrelevant. The bottom curve, for $\phi_{-}=0$, is the same as the curve labeled ``exponential'' in \Fig~\ref{nb:funks}. Again this more sophisticated treatment yields seemingly more reasonable results.

\section{ Alternative Face Temperature Calculations}\label{mrc:alternative}
Instead of assuming that $\theta^4$ varies linearly with optical depth across an interface, 
\begin{eqnarray}
\theta^4_{lin} = \frac{\tau_L \theta^4_R + \tau_R \theta^4_L}{\tau_L  + \tau_R} \ , \label{eq:odwat4face} 
\end{eqnarray}

another choice would be to make the minimalist assumption that the zone-centered temperatures vary linearly between zones and interpolate a material temperature on the face,
\begin{eqnarray}
\theta_{lin} = \frac{\Delta x_L \theta_R + \Delta x_R \theta_L}{\Delta x_L  + \Delta x_R} \ , \label{eq:lintevface} 
\end{eqnarray}
unless the problem is in radiative equilibrium, in which case we might expect $E$ and $\theta^4$ to vary linearly in space, leading to
\begin{eqnarray}
\theta^4_{lin} = \frac{\Delta x_L \theta^4_R + \Delta x_R \theta^4_L}{\Delta x_L  + \Delta x_R} \ . \label{eq:linat4face} 
\end{eqnarray}

If one knows that if the problem is in a completely ionized regime, so that the opacity is dominated by free-free (Bremsstrahlung) physics, the mono-frequency opacity goes like $T_e^{-1/2}/\nu^3$.  When this opacity is averaged over frequency using as a weight the Rosseland at the radiation temperature (since we are talking about gray diffusion, the only questions are whether the flux is a Planckian or Rosseland, and whether it is at $T_e$ or $T_r$; we make the latter choices), we end up with a ``two-temperature'' opacity  with mean free path $\lambda \sim T_e^{1/2} T_r^3$ (a two-temperature opacity is the result of using a non-equilibrium Planckian or Rosseland  radiation field to calculate a gray average: {\it e.g.},  $\kappa(\theta_e,T_r) \sim \int_0^{\infty} \kappa_{\nu}(\theta_e) B_{\nu}(T_r) d\nu$).  Ignoring the weak dependence on material (electron) temperature, we can write this flux as $F \sim T_r^3\nabla E \sim E^{3/4}\nabla E$.   If we ignore quarter powers of things, we can say that the flux behaves approximately as $F \propto \nabla E^2$, whose finite difference approximation would be
\begin{eqnarray*}
F  &\propto  \nabla E^2 \\
 &\propto  \frac{E_R^2 - E_L^2}{\Delta x_L  + \Delta x_R} \\
  & \sim   \left( \frac{E_R + E_L}{2}\right) \frac{E_R - E_L}{\Delta x_L  + \Delta x_R} \ . 
\end{eqnarray*}
(For electron conduction, the Spitzer conductivity varies as $\theta_e^{5/2}$, leading to $F\sim \theta_e^{5/2}\nabla \theta_e \approx \theta_e^2\nabla \theta_e$.  Since  $\nabla \theta_e^3/3 =  \frac{1}{3}(\theta_L^2 + \theta_L\theta_R + \theta_R^2) \nabla \theta_e$,  we have found that  the intermediate temperature, $\theta_{face}=\sqrt{ (\theta_L^2 + \theta_L\theta_R + \theta_R^2)/3}$,  in Spitzer's formula gives good results on this variant of a Marshak wave.)
From this point of view, one would argue that the face temperature ought to be related to the average of the two energy densities {\em without} any linear  interpolation (in space {\em or} optical depth):
\begin{eqnarray}
\theta^4_{face} &= \left( \frac{E_L + E_R}{2a} \right) \ . \label{eq:selfsimface}
\end{eqnarray}

It has been our experience that whether one uses the  assumption of linearity  in optical depth space for $\theta^4$ (\Eq~\ref{eq:odwat4face}), or either of the linear in space methods (\Eqs~\ref{eq:lintevface},~\ref{eq:linat4face}),  all Marshak-wave types of problems can be well modeled.  Although we have not looked at the last averaging method, \Eq~\eref{eq:selfsimface},  in any detail,  it ought to  give results similar to the linear-in-space average of $\theta^4$ in most cases, and might be the best approach to  use with two-temperature opacities in general.

\section{ Improved Difference Scheme for T-Cells}\label{nb:Shashkov}


\begin{figure}[htbp] 
\begin{center} 
\includegraphics[scale=1]{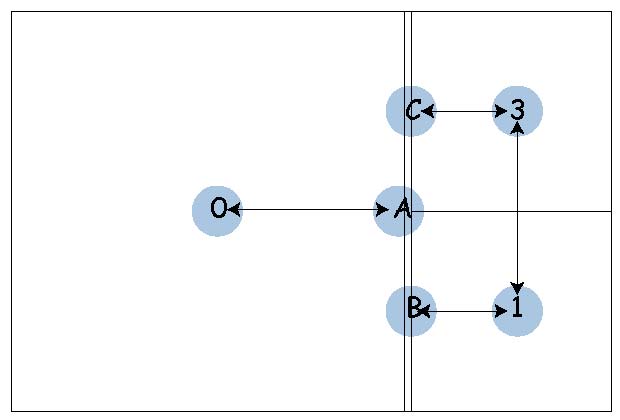}
\caption{Edwards' difference stencil.} \label{nb: Shashkov}
\end{center} 
\end{figure} 

When adjacent zones are equal-sized, as for the case of vertical flow between zones ``1'' and ``3'' in \Fig~\ref{nb: Shashkov}, the assumption of piecewise continuity of $E$ along with the requirement of continuity of flux from zone 1 to the (13) face and from that face to zone 3 gave the two equations that defined $E_{face}$ as the optical-depth weighted average of the zonal $E$'s  and $\tau_{face}$ as the sum of the zonal optical depths.  (In principle, $E_{13}^{above} \ne E_{13}^{below}$, but piecewise continuity means that $\int dV \nabla E \rightarrow 0$ around a tiny volume straddling the (13) interface  as $dV \rightarrow 0$, and that forces $E_{13}^{above} = E_{13}^{below} = E_{face}$.)

For T-cells (reverting to the horizontal direction of flow in \Fig~\ref{nb: Shashkov}),  piecewise continuity of $E$ means that as $\int dV \nabla E = \oint dA_f E_f \rightarrow 0$,  the three non-coincident values of $E$ at the respective faces of the zones are related by:  $A_0 \cdot E_{A_{face}} = A_3\cdot E_{C_{face}} + A_1 \cdot E_{B_{face}}$, where $A_0$ is the area of the large zone's face at the T-cell interface, and $A_1$ and $A_3$ are the areas of the interface as seen by zones 1 and 3 respectively. 

 In 2-D Cartesian geometry, the $A_i =A_0/2$; in 3-D, they are a  $A_0/4$.  In 2-D cylindrical geometry, for face-normals in the $\hat{r}$ direction, the $A_i$ are also $A_0/2$, but for normals in the $\hat{z}$ direction, they vary because $dA = r dr$, and $r$ varies (the smaller radius face varies from $\frac{1}{4}A_0$ at the origin to $\frac{1}{2}A_0$ approaching infinity).  Let us define $a_i \equiv A_i / A_0$.
 
 By requiring continuity of the fluxes at T-cells  and using the relation $E_A = a_1 E_B + a_3 E_C$, we have the following two constraints:
 \begin{eqnarray}
 F_L = - \frac{c}{3}\frac{E_A - E_0}{\Delta \tau_0} 
    & = & - \frac{c}{3} \frac{E_3 - E_C}{\Delta \tau_3}  = F_3\ , \nonumber \\
 F_L = -\frac{c}{3} \frac{E_A - E_0}{\Delta \tau_0} 
     & = & - \frac{c}{3} \frac{E_1 - E_B}{\Delta \tau_1} = F_1  \ , \nonumber 
\end{eqnarray}
Once one has solved for the uninteresting $E_B$ and $E_C$,  one can calculate the common face flux:
\begin{eqnarray}
 F_{face} &=& -\frac{c}{3}\frac{a_1 E_1 + a_3 E_3 - E_0}{a_1 \Delta \tau_1 + a_3 \Delta \tau_3 + \Delta \tau_0 } 
  =  -\frac{c}{3}\frac{a_1 (E_1-E_0)  + a_3( E_3 - E_0) }{a_1( \Delta \tau_1+ \Delta \tau_0)  + a_3(  \Delta \tau_3 + \Delta \tau_0) }  \ . \nonumber
 \end{eqnarray}
Note that this flux could be regarded as the weighted average of the {\it naive} fluxes, $F^{naive}_i = -\frac{c}{3}\frac{(E_i - E_0)}{\Delta\tau_i + \Delta \tau_0}$, with the weights $a_i(\Delta\tau_i + \Delta \tau_0)$.  On the other hand, we can define an optical depth for the faces,
 \begin{eqnarray}
 \tau_{face} &=& \Delta \tau_0 + a_1 \Delta \tau_1 + a_3 \Delta \tau_3  =  a_1 (\Delta \tau_0 +  \Delta \tau_1 )+ a_3( \Delta \tau_0 +\Delta \tau_3) \ , \nonumber
\end{eqnarray}
in which case the fluxes can be written as the area-weighted average of less-{\it naive} fluxes:
\begin{eqnarray}
 F_{face} &=& -\frac{c}{3}\frac{a_1 E_1 + a_3 E_3 - E_0}{ \tau_{face} } 
     =   a_1\left[ -\frac{c}{3} \frac{ (E_1 -E_0)}{ \tau_{face} } \right] 
      +  a_3\left[ - \frac{c}{3}\frac{ (E_3 - E_0)}{ \tau_{face} } \right] \ . \nonumber
 \end{eqnarray}
  
  In 3D, we assume zone $0$ abuts 4 smaller zones, $1,2,3$ and $4$.  In that case,
  \begin{eqnarray}
 \tau_{face} &=& \Delta \tau_0 + a_1 \Delta \tau_1 + a_2 \Delta \tau_2 + a_3 \Delta \tau_3+ a_4 \Delta \tau_4  \ , \nonumber \\
    &=&   a_1 (\Delta \tau_0 +  \Delta \tau_1 )+ a_2( \Delta \tau_0 +\Delta \tau_2) +   a_3 (\Delta \tau_0 +  \Delta \tau_3 )+ a_4( \Delta \tau_0 +\Delta \tau_4)\ , \nonumber
\end{eqnarray}
and the mean flux is similarly given by
 \begin{eqnarray}
 F_{face} &=& -\frac{c}{3}\frac{a_1 E_1 + a_2 E_2 +  a_3 E_3 + a_4 E_4- E_0}{ \tau_{face} } \ , \nonumber \\
     &=& +   a_1\left[ -\frac{c}{3} \frac{ (E_1 -E_0)}{ \tau_{face} } \right] 
      +  a_2\left[ -\frac{c}{3} \frac{ (E_2 -E_0)}{ \tau_{face} } \right] \nonumber \\
    & &   + a_3\left[ -\frac{c}{3} \frac{ (E_3 -E_0)}{ \tau_{face} } \right] 
       +  a_4\left[ - \frac{c}{3}\frac{ (E_4 - E_0)}{ \tau_{face} } \right] \ . \nonumber
 \end{eqnarray}

 One can define a simple test problem, with a linear temperature distribution ({\it e.g.}, 1000 eV at $x=10$ to 2000 eV at $x=20$ independent of $y$).  In Cartesian geometry, the $L_{\infty}$ error in temperature (assuming both $C_V$ and $\kappa$ are constant in the material heat conduction equation) is approximately half of one percent and behaves like $\mathcal{O}(\Delta x^1)$ on refinement.  Including these face effects in $\tau$ and $\nabla\cdot F$, the $L_{\infty}$ error can be reduced to machine accuracy ($10^{-13}$), even {\em without} modifying the matrix (nor does including the modifications in the matrix change the perfect result). 
  
  In cylindrical geometry, the uncorrected  $L_{\infty}$ error is again about a half percent and consistent with  $\mathcal{O}(\Delta x^1)$.  It is reduced to about a part per thousand with the full ($\tau$, $\nabla\cdot F$, and matrix) correction, now consistent with $\mathcal{O}(\Delta x^2)$, since the steady state solution has $T \propto \ln(r)$.  In this case, correcting the face $\tau$ and $\nabla\cdot F$ without correcting the matrix generates $L_{\infty}$ errors that are comparable to the temperature ({\it i.e.}, greater than 100 percent). 
  
   For this reason, it was and is better to do nothing than try to do a partial fix, and until we are certain that our current implementation runs as well on multiple processors as on one processor, we will continue to do nothing.

\section{Spillman's variant of the Marshak/Milne  Boundary Condition}\label{mrc:robin}

When constructing divergences of fluxes, the elemental building block is the contribution of each face to the divergence, and this quantity, 
\begin{equation}
(\mathbf{F}\cdot d\mathbf{S})_{face} = -\frac{ cA_{face}}{\Delta\tau_L + \Delta\tau_R} (E_R - E_L)  \ , \nonumber
\end{equation}
 is used to calculate divergences of the flux for the matrix source term, edits, {\it etc}.  
 At boundaries, one of the $\tau$'s obviously fails, leaving only the other,  $\Delta\tau_Z$. In that case, the Dirichlet condition sets the contribution to
 \begin{eqnarray}
(\mathbf{F}\cdot d\mathbf{S})_{face} &=& -\frac{ c A_{face}}{\Delta\tau_Z } \cdot (E_R - aT_{face}^4 ) \ , \nonumber \\
 \mbox{or}\  (\mathbf{F}\cdot d\mathbf{S})_{face} &=& -\frac{ c A_{face}}{\Delta\tau_Z } \cdot (aT_{face}^4 - E_L ) \ , \nonumber
 \end{eqnarray}
 depending on the boundary.  Henceforth, we will concentrate on the Left boundary and zone $R$.  The Marshak/Milne condition sets the contribution to
 \begin{equation}
(\mathbf{F}\cdot d\mathbf{S})_{face} = - \frac{ c A_{face}}{ 2/3 +\Delta \tau_Z }\cdot (E_R - aT_{face}^4 ) \ . \nonumber
 \end{equation}
  The Spillman variant~\cite{SPBC} is an attempt to cover more bases.  Assume that the user specifies a (dimensional) vacuum boundary length, $\lambda_V$ (default = 1 cm), and recall that $\lambda_Z$ was used to calculate the optical depth: $\Delta\tau_Z =   \Delta s_{Zf}/\lambda_Z$.   Defining the ratio,
  \begin{equation}
 \xi = \frac{ 2\lambda_V + 3\lambda_Z}{\lambda_V + 2 \lambda_Z} \ , \nonumber 
 \end{equation}
Spillman sets the contribution to
 \begin{equation}
(\mathbf{F}\cdot d\mathbf{S})_{face} = - \frac{c A_{face}}{\frac{2}{3}(\xi - e^{-3\Delta\tau_Z/2})}\cdot (E_R - aT_{face}^4 )  \ , \nonumber
\end{equation}
so that 
when $\lambda_Z$ is large compared to both $\lambda_V$ and $\Delta x$,  $\xi \rightarrow \frac{3}{2}$ and the denominator goes to $\frac{1}{3}+\Delta\tau_Z$. When $\lambda_Z$ is small compared to $\lambda_V$ but still large compared to $\Delta x$,  $\xi \rightarrow 2$ and the denominator goes to $\frac{2}{3}+\Delta\tau_Z$, the standard Milne result.  Even if the user makes a poor choice of $\lambda_V$, in optically thin zones, Spillman will always give Milne-ish behavior.

However, if the user intended to have inflow via a Milne condition, but made the zone abutting the boundary so large that its optical depth was enormous, Milne's $2/3+\Delta \tau$ is indistinguishable from Dirichlet's $\tau$,  and the flux $\rightarrow 0$.  In this case the numerics allows essentially no flow across the boundary, contrary to what would have been allowed if the zone had been thinner, and what {\em physics}  presumably always requires.
 
In such a case of large optical depth ($ \Delta s_{Zf} \gg \lambda_Z $), Spillman's exponential term vanishes and the denominator varies between unity and $\frac{4}{3}$ (depending on the exact value of $\lambda_V$ relative to $\lambda_Z$).  The numerical fore-factor ($\sim \frac{3}{4}\cdots 1$)  will now be within a factor of two of the optically thin Milne result ($\frac{3}{2}$),  instead of vanishing.
These contributions from boundary faces do not contribute to any off-diagonal elements of the matrix to be solved, but the factor multiplying $(E-aT^4)$ does add (actually, subtract) from the diagonal element of the matrix for the zone adjacent to the face.  This coefficient also contributes to calculating the source term of the matrix equation (the ``vacuum'' ``zone'' shows up as an explicit source for the adjacent real zones ).  In the case of thermal conduction, one calculates the divergence twice: once to form the source for the matrix solver, and a second time afterwards to calculate a conservative energy transfer between all faces based on $\kappa\nabla \theta$ fluxes; for radiation, $E$ is a conserved quantity already, so only the first, matrix-source divergence is necessary.

Thus we can see that the optically thick limit of Spillman's variant allows the vacuum to radiate like a black body (more correctly, it radiates like the difference of two black bodies). 
 $F\cdot dS \approx \pm \frac{3}{4}A_{face}c (E_Z-aT_b^4)$, and compares well  to the optically thin Milne result, $\pm \frac{3}{2}A_{face}c  (E_Z-aT_b^4)$, and disagrees with  the optically thick Marshak/Milne (or Dirichlet) result, $ \pm A_{face}c (E_Z-aT_b^4)/ {\tau_Z} \rightarrow 0$.

 \end{appendix} 




\bibliographystyle{unsrt}
\bibliography{RAGE}




\end{document}